\PassOptionsToPackage{unicode=true}{hyperref} % options for packages loaded elsewhere
\PassOptionsToPackage{hyphens}{url}
\documentclass[11pt,a4paper]{article}
\usepackage{jheppub,multirow}
\usepackage{float}
\usepackage{graphicx}
\usepackage{url}
\usepackage{cancel}
\usepackage{slashed}
\usepackage{caption}
\usepackage{placeins}
\usepackage{feynman}
\usepackage{soul} % For strikeout, using  \st{Hellow world}
\usepackage{bm}
\usepackage{amsmath}
\usepackage{amssymb}
\usepackage{listings}
\usepackage[toc, page]{appendix}
\usepackage{color}
\usepackage{booktabs} % for tables
\usepackage{pgf}
\usepackage{empheq}
\usepackage{tikz-feynman}
\usepackage{tikz}

\newcommand{\mcscales}{{\tt MCscales} }

\def\G1{{\bf \gamma^{(1)}_N}}

\def\gsim{\gtrsim}

\definecolor{violet}{cmyk}{0,1,0,0.2}
\definecolor{red}{cmyk}{0,1,1,0}

\usepackage[colorlinks=true, linkcolor=black!50!blue, urlcolor=blue, citecolor=blue, anchorcolor=blue]{hyperref}
\usepackage[font=small,labelfont=bf,labelsep=period]{caption}
\usepackage{tabularx}
\newcolumntype{C}[1]{>{\centering\arraybackslash}p{#1}}

\newcommand{\be}{\begin{equation}}
\newcommand{\ee}{\end{equation}}
\newcommand{\bea}{\begin{eqnarray}}
\newcommand{\eea}{\end{eqnarray}}
\newcommand{\bi}{\begin{itemize}}
\newcommand{\ei}{\end{itemize}}
\newcommand{\ben}{\begin{enumerate}}
\newcommand{\een}{\end{enumerate}}

\numberwithin{equation}{section}
\numberwithin{figure}{section}
\numberwithin{table}{section}
\title{Parton distributions with scale uncertainties: a Monte Carlo sampling approach}
\author[a]{Zahari Kassabov,}
\author[a]{Maria Ubiali,}
\author[b]{Cameron Voisey}
\affiliation[a]{DAMTP, University of Cambridge, Wilberforce Road, Cambridge, CB3 0WA, United Kingdom}
\affiliation[b]{Cavendish Laboratory (HEP), JJ Thomson Avenue, Cambridge, CB3 0HE, United Kingdom}
\emailAdd{zk261@cam.ac.uk}

\abstract{We present the {\tt MCscales} approach for incorporating scale
uncertainties in parton distribution functions (PDFs). The new methodology
builds on the Monte Carlo sampling for propagating experimental uncertainties
into the PDF space that underlies the NNPDF approach, but it extends it to the
space of factorisation and renomalisation scales.  A {\it prior} probability is
assigned to each scale combinations set in the theoretical predictions used to
obtain each PDF replica in the Monte Carlo ensemble and a {\it posterior}
probability is obtained by selecting replicas that satisfy fit-quality criteria.
Our approach allows one to exactly match the scale variations in the PDFs with
those in the computation of the partonic cross sections, thus accounting for the
full correlations between the two. We illustrate the opportunities for
phenomenological exploration made possible by our methodology for a variety of
LHC observables.  Sets of PDFs enriched with scale information are provided,
along with a set of tools to use them.}

\keywords{Parton Distribution Functions, Monte Carlo, Theory Uncertainties.}
\begin{document}

\maketitle
\flushbottom

\hypertarget{introduction}{%
\section{Introduction}\label{introduction}}

Parton distribution functions (PDFs) are one of the
crucial ingredients of any theoretical predictions at the Large Hadron
Collider (LHC).
At present, the uncertainty associated to state-of-the-art
NNLO PDF sets~\cite{Ball:2017nwa,Bailey:2020ooq,Hou:2019efy,nnpdf40,Ball:2022hsh} only accounts for the
statistical and systematic uncertainties of the experimental data that
are fitted in order to extract PDFs. Thanks to the addition of a large number of
precise experimental measurements, PDF uncertainties have
decreased to as little as 1-2\% for several key processes at the
LHC~\cite{Amoroso:2022eow,Ball:2022hsh}.
Once data from the High-Luminosity LHC are included in their
determination, PDF uncertainties are likely to further
decrease~\cite{Khalek:2018mdn}.

Given such a level of precision, it is paramount that all relevant sources of
uncertainty, including the methodological and theoretical uncertainties that
until recently were considered subdominant, are properly accounted for in PDF
determinations, and are propagated to all predictions involving PDFs.
Uncertainties related to the fitting methodology can be kept under control using
statistical closure tests~\cite{Ball:2014uwa,DelDebbio:2021whr}, as well as
``future tests"~\cite{Cruz-Martinez:2021rgy} which validate   uncertainties in
extrapolation regions, not constrained by data. In terms of theory,
state-of-the-art PDF fits include fixed-order calculations computed at
next-to-leading order (NLO) in QCD via fast interpolating
grids~\cite{Carli2010,Bertone:2016lga,Kluge:2006xs}, augmented by
pre-computed local next-to-next-to-leading order (NNLO) $K$-factors, that make the
calculations accurate at NNLO in QCD. Electroweak corrections could also be
easily incorporated in a PDF fits thanks to fast interpolation
grids~\cite{Frederix:2018nkq,Carrazza:2020gss}. Recently in
~\cite{McGowan:2022nag} an approximate N$^3$LO PDF set has been
presented along with a formalism for the inclusion of theoretical
uncertaintie into a PDF fit.

One of the main sources of theoretical uncertainty are scale uncertainties, which stem from the fact that
fixed-order calculations depend on unphysical parameters, the factorisation
and renormalisation scales. The dependence upon these scales would completely
disappear if we were able to compute cross sections to all orders and it typically
decreases as higher perturbative orders are added.
For this reason, scale uncertainties are typically used to estimate missing higher-order uncertainties (MHOUs), which is a more general concept
involving all of the uncertainty accrued from the fact that calculations are
truncated at a given perturbative order.  
%Although scale vairations are a somewhat {\it ad hoc} method without
%probabilistic interpretation, it is still the most popular way to estimate MHOUs.
Alternative approaches to estimate MHOUs based on Bayesian frameworks have been proposed,
which might improve over the somewhat ad-hoc 7- or 9-point envelope
scale variations~\cite{Cacciari:2011ze,David:2013gaa,Bonvini:2020xeo,Duhr:2021mfd}. 
%We point out that scale uncertainties are related to the more general concept of a Missing Higher
%Order Uncertainty (MHOU). This refers to the uncertainty induced by neglecting
%terms of higher order than the fixed order expansions being considered. Scale
%uncertainties are often used to estimate such uncertainties.
At the moment, it is unclear to what extent these Bayesian approaches to the estimate of
MHOUs can be implemented in a PDF fit.
%, without however imposing any symmetry in the prior probability.
%This is particularly important
%given that the recent analysis of~\cite{Duhr:2021mfd} suggests that
%the prior probability in the space of factorisation and
%renormalisation scales does not have to be necessarily symmetric. 

The inclusion of scale variations in a state-of-the-art NLO PDF
fit was presented for the first time in~\cite{AbdulKhalek:2019bux,AbdulKhalek:2019ihb}.
The approach is built upon the construction of a theory covariance matrix that 
describes the scale variations of the processes included in a PDF fit and
models their theoretical correlations. It can then be added to the
experimental covariance matrix, whereafter the data are fitted using a total covariance matrix,
which is the sum of the two contributions. The method has been shown to
improve the accuracy of calculations with NLO PDFs for a range of LHC
standard candle processes, as the inclusion of scale uncertainties moves the
NLO result towards the NNLO result. A similar method has been also been
applied to nuclear uncertainties, which are relevant in the description of
data based on nuclear targets~\cite{Ball:2020xqw,Pearson:2019upi}.

In this work we take a somewhat different point of view: % on scale uncertainties:
instead of viewing scale variations primarily as proxies for guessing
MHOUs, here we consider the renormalisation and factorisation
scales as free parameters of the fixed-order theory, that 
induce an uncertainty on the theory predictions included in a PDF fit,
which needs to be propagated.
These parameters then have to be treated like other parameters that cannot be
obtained from first principles, indeed like the PDFs themselves. It follows
that, in this picture, PDFs and scale parameters should be chosen jointly so
as to produce theory predictions that are compatible with the experimental data
included in a PDF fit. Scales choices should, however, be made in a way that favours
perturbative convergence, by choosing values close to the physical scale of
the process, which, whenever it can be well defined, minimises higher-order
terms containing logarithms of the ratio of the scale of the process to the
unphysical scales, and follow consistent criteria for similar theory
predictions. These requirements can be implemented by specifying a suitable
prior probability distribution of all possible scale choices.

We present such a methodology to determine jointly PDFs and scales parameters,
which we dub \mcscales. It extends the NNPDF fitting
methodology~\cite{Ball:2014uwa,Ball:2017nwa,nnpdf40}, wherein each Monte Carlo
sample of PDFs (known as a replica) is generated by providing one Monte Carlo
sample of experimental data, varied according the experimental uncertainties. We
additionally endow each PDF replica with a sample from the prior distribution of
scale choices for each of the theory predictions that enter the PDF fit. Then,
by exploiting another component of the NNPDF methodology, the post-fit
replica selection, we obtain a posterior distribution of scales. In this way, a
refined model with a reduced dependence on prior assumptions is obtained and a
statistical treatment of scale variations is achieved for the first time.
Indeed, the posterior distribution of scales constitutes a first comprehensive
benchmark of scale choices across all data sets included in a PDF fit.

We record the scale information associated to each replica during the fit and
make it available to the user so that, when scale variations for a given
theoretical prediction are computed, it becomes possible to associate each variation in
the partonic cross section with the corresponding variation in the
PDFs (given by a subset of \mcscales replicas). This allows one to
overcome a drawback of the theory covariance matrix approach:
the fact that within that method the scale uncertainties are fully integrated within the PDF set in such a
way that scale variations in the PDF fit cannot be matched with the scale
variation in the partonic processes when computing a theoretical
prediction~\cite{Harland-Lang:2018bxd}.
In~\cite{AbdulKhalek:2019bux,AbdulKhalek:2019ihb} we argued that neglecting
this correlation yields a small effect in most phenomenological applications.
In a recent work~\cite{Ball:2021icz} however it is claimed that the correlations might be not so negligible
for processes already included in a PDF fit.

The approach that we present here allows
% this limitation to be overcome and
for the scales to be correlated exactly, so that the size of the
correlations can be checked explicitly. We find that, although for most
processes that are not included in a PDF fit the effect of the
correlation is thought to be negligible, in the case of Higgs production via gluon fusion,
not accounting for correlation between the factorisation scale used in the fit
and the one used in the calculation of the partonic cross section would underestimate the
total theory uncertainty by a significant amount. On the other hand, for processes included in a PDF fit,
particularly in those that provide a strong constraint on the PDFs (such as $W$
and $Z$ boson production), the inclusion of the correlation reduces
the size of the joint PDF and scale uncertainties by nearly a factor of 2. This is
discussed in more details in Sect.~\ref{sec:pheno}.

A further advantage of the \mcscales method is that the correlation
model applied to the factorisation and renormalisation scales is transparent
so that the user has more freedom. They can choose asymmetric {\it
  prior} probabilities or choose over which scale variations they would like
to be included in the PDFs they use, i.e. the user can tailor the prior
probability according to their theoretical prejudice and the phenomenological
application at hand. We provide a set of tools that allows the user to
manipulate the prior distribution of scales, the only input being the
{\tt LHAPDF}~\cite{Buckley:2014ana} sets accompanying this publication, allowing
also the integration of \mcscales PDFs with any existing Monte Carlo event generators.

A possible criticism of the perspective presented here is that scale choices should
be independent from experimental data, as otherwise it may not be possible to find
potential new physics signals. We note, however, that theory and experiment are
already inextricably mixed in current PDF determinations, based on fitting
theory parameters to experimental data which is assumed to be compatible with
the Standard Model. Furthermore, an experimental deviation from theory
predictions can hardly be considered meaningful evidence of new physics if it
can be explained away by scale variations in the PDFs~\cite{Iranipour:2022iak,McCullough:2022hzr,Greljo:2021kvv}.
Moreover, in order to respond to a further possible criticism based on the impression
that the PDF sets that we provide here break PDF universality, we outline that, while
each replica is labelled by the scale multipliers associated with each process that enters
a PDF fit, the full ensemble of the \mcscales replicas does not depend on any
specific process, as all scale choices for all possible processes are represented in the
sample determined via the post-fit selection that determines the {\it a-posteriori} distribution. 

This paper is laid out as follows. In Sect.~\ref{sec:the-method} we describe our approach to
including scale uncertainties in PDFs . In Sect.~\ref{sec:data-driven} we determine the
probability distribution of the renormalisation and
factorisation scales entering the theoretical predictions in a PDF
fit. In
Sect.~\ref{sec:xsec} we describe how to compute cross sections with matched
scale variations using \mcscales
PDFs. In Sect.~\ref{sec:pdf-9pts} we compare our procedure with the earlier
theory covariance matrix approach~\cite{AbdulKhalek:2019bux,AbdulKhalek:2019ihb}
as well as with a similar procedure to study correlations of theory
uncertainties~\cite{Ball:2021icz}. In Sect.~\ref{sec:pheno} we apply our method
to several phenomenologically relevant processes at the LHC and
assess the size of the correlations between the scale variations in the PDF fits
and those in the partonic cross sections. In
Sect.~\ref{sec:delivery} we describe how the PDF sets augmented with
information on the scales are delivered and we present tools that can
be used when working with them. Finally, we conclude in Sect.~\ref{sec:conclusion}.

\section{The methodology}
\label{sec:the-method}

In this section we describe how the {\tt MCscales} method constructs a
joint {\it prior} probability in the space of PDFs and scales, and how
it determines the {\it posterior} probability  distribution. The method can be applied to a variety of prior specifications. In
Sect.~\ref{sec:the-sampling-model} we introduce the idea of sampling in
the joint space and list the generic assumptions we make, while in Sect.~\ref{sec:postfit} we describe how the {\it a
  posteriori} distribution is determined.

\subsection{Probability sampling}
\label{sec:the-sampling-model}

Within the NNPDF approach \cite{Ball:2014uwa,Ball:2017nwa,nnpdf40}, a Monte Carlo
sampling of the probability density in the functional space of PDFs is done
by exploiting the concept of importance sampling. By constructing a set of
pseudo-data replicas, which are found in accordance with the statistical
features of the experimental data going into the PDF fit, one finds
that a sample of 1000
replicas is large enough to reproduce central values, uncertainties and
correlations of the starting data to a few percent
accuracy~\cite{Giele:2001mr,Ball:2008by,Ball:2010de,Ball:2012cx,DelDebbio:2009zz,Ball:2011eq}.

In the \mcscales approach, we extend the Monte Carlo sampling
described above by adding scale fluctuations for each theoretical prediction on top of the
fluctuations on the input data that are used to propagate experimental
uncertainties. That is, each PDF replica is obtained by employing different
values of the scales entering each theoretical prediction in the
fit\footnote{We point out that here we use the 
definitions of factorisation and renormalisation scales discussed
in~\cite{AbdulKhalek:2019ihb}. Namely $\mu_r$ refers to the
renormalisation scale in the partonic cross sections, $\mu_f$ refers to
the factorisation scale in the PDF evolution and $\tilde{\mu}$ refers
to the scale of the process. }. Crucially, we record the scale information for each replica and make it available to the
user. This enables users to tailor the prior distribution of scales and, furthermore,
enables them to take into account correlations among scale variations in the
computation of theoretical predictions.

%In this way we
%allow for different correlation models to be externally specified, possibly taking into account
%scale variations in hard cross sections, while also offering the convenience of making it easy to
%obtain predictions with our default correlation model which we describe later.
%\subsection{The sampling model}

Incorporating scale variations within PDF fits can be viewed as the inclusion of different hypotheses
for the theory predictions corresponding to the input data, each with some probability. Each
hypothesis corresponds to a different set of scale choices. We therefore
need to construct a set of prior probabilities. These priors will subsequently be
refined into posterior distributions based on the
compatibility of each theory hypothesis with experimental data.
Given that both sets of hypotheses that we consider and the probability assigned
to each of them are necessarily dependent on assumptions, loosely based on
physical intuition, practical ease or convention, it is crucial to describe said
assumptions precisely.
In this section we list the general assumptions underpinning our method. Later,
in Sect.~\ref{sec:data-driven} we outline specific assumptions
relevant to the construction of the prior that we select here to
illustrate our results. 
%We also employ the same
%For the results presented in this work we use the very same fit settings as those used in
%Ref.~\cite{AbdulKhalek:2019ihb}.
% We use the following codes:
% \textsc{APFEL}~\cite{Bertone:2013vaa}, \textsc{APPLgrid}~\cite{Carli2010} and
% \textsc{APFELgrid}~\cite{Bertone:2016lga}. Our fits use a global dataset of 2819
% data points, which includes data from
% NMC~\cite{Arneodo:1996kd,Arneodo:1996qe},
% SLAC~\cite{Arneodo:1996kd,Arneodo:1996qe},
% BCDMS~\cite{Arneodo:1996kd,Arneodo:1996qe}, CHORUS~\cite{Onengut:2005kv},
% NuTEV~\cite{Goncharov:2001qe,Mason:2006qa},
% HERA~\cite{Abramowicz:2015mha,Abramowicz:1900rp,Aaron:2009af,Abramowicz:2014zub},
% E866~\cite{Webb:2003ps,Webb:2003bj,Towell:2001nh}, E605~\cite{Moreno:1990sf},
% CDF~\cite{Aaltonen:2010zza}, D0~\cite{Abazov:2007jy,Abazov:2013rja,D0:2014kma},
% ATLAS~\cite{Aad:2013iua,Aad:2014qja,Aad:2011dm,Aaboud:2016btc,Aad:2015auj,
% Aad:2011fc,Aad:2014kva,Aaboud:2016pbd,Aad:2015mbv}, CMS~\cite{Chatrchyan:2012xt,
% Chatrchyan:2013mza,Chatrchyan:2013tia,Khachatryan:2016pev,Khachatryan:2015oaa,
% Chatrchyan:2012bja,Khachatryan:2016mqs,Khachatryan:2015uqb,Khachatryan:2015oqa}
% and LHCb~\cite{Aaij:2012vn,Aaij:2012mda,Aaij:2015gna,Aaij:2015zlq}. We point out
% that the cuts and data set used are consistent across all of the PDF fits
% presented in this work, so that the data and cuts do not bias any comparisons
% made.
%

%\subsubsection{General assumptions}
%\label{sec:general-assumptions}
%\textcolor{red}{CV: should one of the assumptions be that the fact. scale is correlated across all processes?}

\begin{description}
\item[Scale variations by a factor of two.]
  For both the factorisation and renormalisation scales we only consider: a central scale,
  an upwards variation of twice the central scale
  and a downwards variation of half the scale as possible
  values. This is consistent with
  usual practice \cite{deFlorian:2016spz} and it simplifies notably the
  implementation in the NNPDF framework, given that it restricts the number
  of allowed scales that can be chosen for a given Monte Carlo replica.

\item[Factorisation scale correlated across all processes.]
  The factorisation scale is the same, i.e. fully correlated, across all processes.
  This is an approximation, which may not be accurate particularly for processes which
  depend on PDFs whose evolution is controlled by different anomalous dimensions (such as
  the isospin triplet and the singlet).

\item[Renormalisation scales correlated by process.]
  We select the same renormalisation scale for each data point belonging to a
  given process. The processes are the same as
  in~\cite{AbdulKhalek:2019bux,AbdulKhalek:2019ihb}. We assign data into one
  of the following five categories: charged-current DIS (`DIS CC'), i.e. DIS data where the mediator is
  a $W$ boson, neutral-current DIS (`DIS NC'), where the mediator is a $Z$
  boson or a photon, Drell-Yan (`DY'), jet production (`JET') and top-quark pair production
  (`TOP').
\end{description}

We thus allow for variations of \(N_{\text{p}}=5\) renormalisation
scales, each of which corresponds to one process, as well as one factorisation
scale variation per replica, since the factorisation scale is completely
correlated across all processes.

With these assumptions, we have to assign \(N_{\text{s}} = N_{\text{p}} + 1\)
scale variations to each PDF replica, where each variation is associated with a
factor of either \(\frac{1}{2}\), 1, or 2 times the nominal scale. Hence each
PDF replica (which we will consistently label with the index \(n\), with
$n=1,\cdots,N_{\rm rep}$)  acquires $N_{\text{s}}$ scale variations, \(\{k_f, k_{r_1},
\ldots, k_{r_{N_\text{p}}}\}\), associated with it, where \(k_f\) corresponds to
the factorisation scale variation and \(k_{r_p}\) corresponds to the
renormalisation scale variation for each process \(p\). Therefore, we can
parameterise a given theory hypothesis by a list of \(N_\text{s}\) elements from
the set
\begin{equation}
\label{eq:scalemultchoices}
\Xi = \bigg{\{}\frac{1}{2}, 1, 2\bigg{\}}\ .
\end{equation}
%corresponding to a multiplicative factor for the central
%choice of factorisation scale and \(N_\text{p}\) choices of factors for the
%renormalisation scales.
That is, each theory choice is one of the
\(3^{N_\text{s}}\) elements of
\begin{equation}
  \label{eq:omega}
\Omega=\{(\xi_{f},\xi_{1},\ldots,\xi_{N_\text{p}}) \; \forall \; \xi_{f},\xi_{1},\ldots,\xi_{N_\text{p}}\in\Xi\}\ .
\end{equation}
Each element of $\Omega$ determines the factorisation and renormalisation scales
in the theory predictions used in the fit that yields a given PDF replica. Each
replica is obtained as in a standard NNPDF fit, namely as the best-fit to
pseudo-data, but additionally by setting the scales of theoretical predictions
to those dictated by the given element of $\Omega$.
%Clearly there will be more than one PDF replica for each element $\omega$.  
%
To determine the prior probability is equivalent to determine the number of 
replicas generated for each of the elements $\omega\in\Omega$ corresponding to the particular set of
scale choices,
\begin{equation}
P\left(k_f=\xi_f,k_{r_1}=\xi_{1},\ldots,
  k_{r_{N_{\rm p}}}=\xi_{N_{\rm p}}\right) =
P(\omega) \, ,
\end{equation}
{\it i.e.} to determine how many replicas of the \mcscales prior
have the scales set to those corresponding to a given element of
$\Omega$. From now on, we will label $\omega^{(n)}$ the element of
$\Omega$ that sets the scales in the fit of the $n^{\rm th}$ replica of the Monte
Carlo ensemble. 
In Sect.~\ref{sec:data-driven} we present results for a
flat or uniform prior probability and for the simplest non-uniform
priors. In Appendix~\ref{app:abc} we present a more sophisticated
definition of non-uniform prior. Finally, in Sect.~\ref{sec:delivery} we
describe how each user can define their own prior probability and
select a subset of the {\tt MCscales} replicas accordingly.

\subsection{Post-fit selection of replicas} \label{sec:postfit}
Once a {\it prior} probability is generated, an {\it a posteriori}
probability is obtained by selecting the replicas based on the
data-theory agreement, which is defined by the $\chi^2$ statistic for each
replica,
\begin{equation}
  \label{eq:chi2k}
\chi^2_n=\frac{1}{N_{\rm dat}}\sum_{i,j=1}^{N_{\rm
    dat}}\left[D_i-T_i^{(n)}\right]\,({\rm cov}^{-1})_{ij}\, \left[D_j-T_j^{(n)}\right] \, ,
\end{equation}
where $N_{\rm dat}$ is the number of data included in the fit, $T_i^{(n)}$ is
the theoretical prediction for the $i^{\rm th}$ data point, computed with the
$n^{\rm th}$ PDF replica as input PDF and with the factorisation and
renormalisation scales set to $\omega^{(n)}$, and  $D_i$ are the corresponding central
values of the experimental data. The multiplicative uncertainties in the
experimental covariance matrix $({\rm cov}_{ij})$ are treated as explained
in~\cite{Ball:2009qv,Ball:2012wy}.
Throughout the rest of this work, we will be quoting the values of the $\chi^2$
statistic as normalised by the number of data points $N_{\rm dat}$.

The standard $\chi^2$ selection criterion adopted in the NNPDF global fits discards
replicas based on the difference between $\chi^2_n$ and a value
depending on the mean and standard deviation of the whole replica sample.
Specifically, the $n^{\rm th}$ replica is discarded if
\begin{equation} 
    \label{eq:nnpdf-veto}
    \chi^2_n > \left\langle  \chi^2 \right\rangle_n + 4 \, \text{std}(\chi^2)_n \, ,
\end{equation}
where the mean and standard deviation are taken over the whole replica sample. This
simple idea works well to remove a few isolated outliers, but it is likely to be
too lenient in the case that the replica sample contains many
outliers. Indeed, we observe that the fit quality (as measured by the $\chi^2$) deteriorates
substantially and unequally for certain scale combinations. This is a consequence of
some combinations of scales being disfavoured by the experimental data. As a result,
the mean $\chi^2$ will grow because of the presence of these inconsistent scale
choices and the standard deviation will grow because of the differences in fit
quality due to the different scale choices. Another option is to calculate the $\chi^2$ threshold
with only the replicas whose scale multipliers are all equal to one,
\begin{equation} \label{eq:mcscales-veto}
    \chi^2_n > \left\langle  \chi^2 \right\rangle_{n\,|\,\omega^{(n)} = \{1,\dots, 1\}} + 4 \, \text{std}(\chi^2)_{n\,|\,\omega^{(n)} = \{1,\dots, 1\}} \, ,
\end{equation}
which brings the threshold more in line with a standard NNPDF fit, which uses
central scales only. This is the choice we make.

Additionally, as is standard within the NNPDF framework, we apply vetoes related
to the arc-length and positivity of observables. The arc-length veto serves to remove
PDF replicas that are outliers in terms of their smoothness. For a typical
NNPDF fit, it takes the same form as the $\chi^2$ in
Eq.~\eqref{eq:nnpdf-veto}.
The positivity constraints are imposed as described in~\cite{Ball:2017nwa}.
For a replica to pass the postfit selection, it must yield positive
DIS structure functions at the scale $Q_{\rm pos}$ = 5 GeV in the large-$x$ region,
as well as positive pseudo-observables, such as tagged deep-inelastic structure functions $F_2^{u,d,s}$
and three flavour DY rapidity distributions $d\sigma_{u\bar{u}}^{\rm DY}/dy$,
$d\sigma_{d\bar{d}}^{\rm DY}/dy$ and $d\sigma_{s\bar{s}}^{\rm DY}/dy$.

After imposing the $\chi^2$, arc-length and positivity criteria, the replica sample that passes
the postfit cuts will have a different distribution of scale choices compared
to the prior distribution, precisely because some scale combinations will lead
more frequently to poor fits and consequently be discarded.
The distribution of surviving scale choices incorporates information on the
agreement with data to the prior sampling procedure described in
Sect.~\ref{sec:the-sampling-model}.
Specifically we assume a uniform distribution for
the replicas selected by postfit and zero for the replicas discarded
by it. The corresponding likelihood is
\begin{equation}
\label{eq:postfit}
    P(D|\omega^{(n)})\propto\begin{cases}
    1 & \text{Postfit passes}\\
    0 & \text{Otherwise}
    \end{cases}
    \ .
\end{equation}
Thus taking the replica sample after the postfit selection corresponds to
incorporating the posterior distribution over the scale choices.
%We henceforth call the second view a `data-driven approach'.

%\input{sec-settings.tex}
%\input{sec-sampling-model.tex}
\section{Implications for PDF fits} \label{sec:data-driven}

In this section we study the implications of the \mcscales approach on
the fit and the resulting PDFs. We start by choosing an uniform prior sampling of the theory hypotheses. 
The posterior sampling is determined according to criteria described
in Sect.~\ref{sec:postfit}. 

In Sect.~\ref{sec:fit-quality} we compare the fit quality of the prior set and the posterior set,
and in doing so we will ascertain the effect of the postfit criteria. We will also perform comparisons with a
baseline PDF set, which is equivalent to the NLO {\tt NNPDF3.1} set used in ~\cite{AbdulKhalek:2019bux,AbdulKhalek:2019ihb}.
%\footnote{except from some minor changes to the dataset and cuts, which are ..}.
%Note that this PDF set has had the same $\chi^2$
%and arclength vetoes applied as applied to MCscales with vetoes.
In comparing the {\tt MCscales} sets to those of {\tt NNPDF3.1}, the impact of scale variations included with our approach can be gauged.
%
%To compare these sets, we will first study the extent to which they are able to
%describe the data used in the fit by means of the $\chi^2$ estimator. This will
%form Sect.~\ref{sec:fit-quality}.
%
Then, in Sect.~\ref{sec:scale-dists} we will study the effect that the postfit criteria have on the scale choices of the
surviving PDF replicas. This will indicate the relative preference of the experimental data to particular scale choices.
In Sect.~\ref{sec:pdf-impact} we will look at these effects at the level of the
PDFs themselves 
%in Sect.~\ref{sec:pdf-9pts} we  compare the
%resulting PDFs with those obtained with the 9-point prescription of Refs~\cite{AbdulKhalek:2019bux,AbdulKhalek:2019ihb}.
while in Sect.~\ref{sec:theory-driven} we briefly discuss several examples of non-uniform prior samplings.

\subsection{Fit quality} \label{sec:fit-quality}
We begin with a set of 3000 PDF replicas, which we denote {\tt
  MCscales uniform prior}. We produce this set by generating a uniform
prior sampling over the set of theory hypotheses $\Omega$.  
We then apply the $\chi^2$, arclength and positivity vetoes described in Sect.~\ref{sec:postfit}, and end up
with a set of 823 replicas, which is dubbed {\tt MCscales with postfit}. First we will look at
the $\chi^2$ of individual replicas computed against two different data sets:
the data used to fit each replica (`training set') and the data used in the
cross-validation of each replica to prevent over-fitting (`validation set').
Note that these sets are different for each replica and that for each replica
the two sets contain entirely different data points that are randomly
selected.

%-------------------------------------------------------------------------------
\begin{figure}[!ht] \centering
\includegraphics[width=0.49\textwidth]{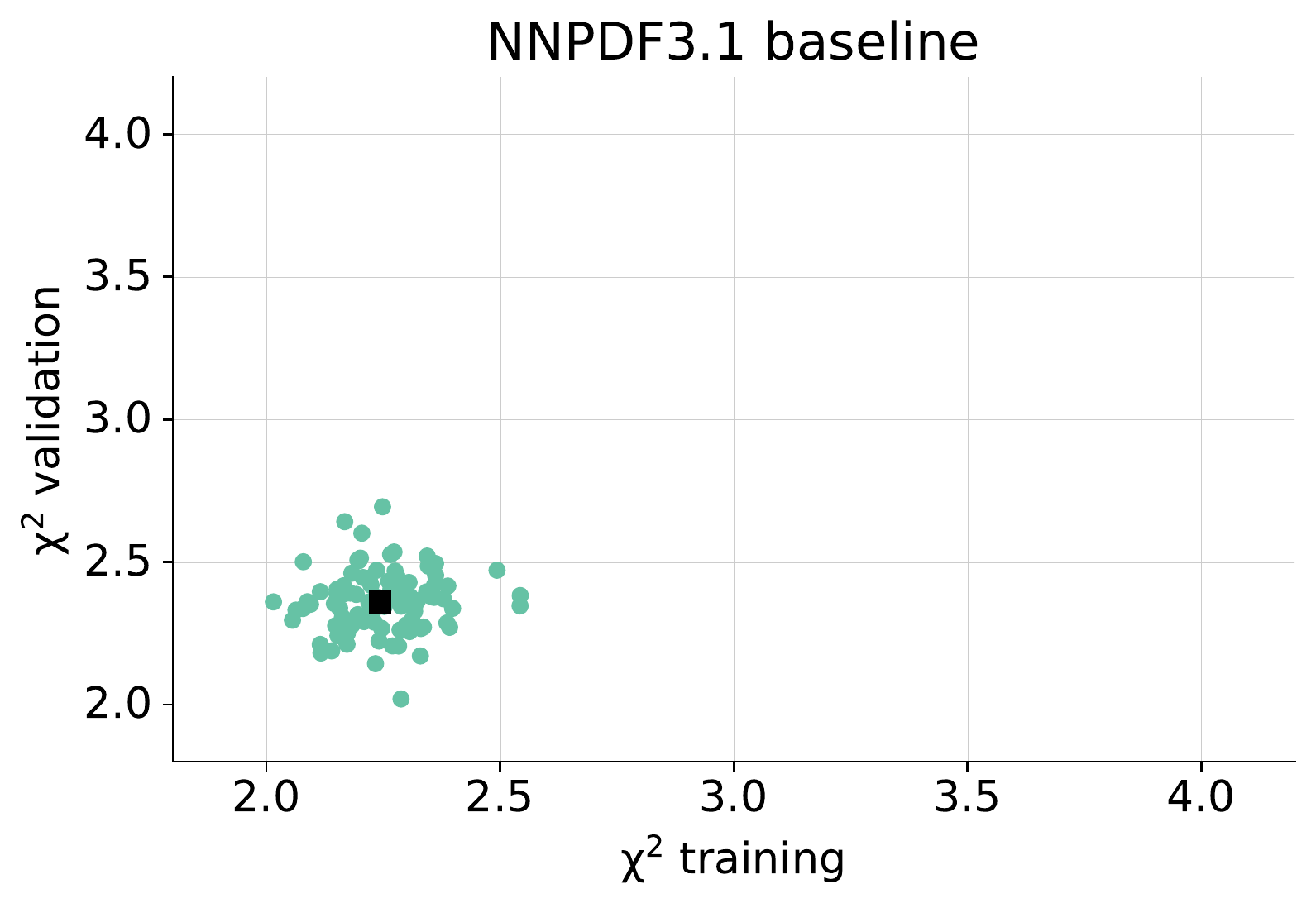}
\includegraphics[width=0.49\textwidth]{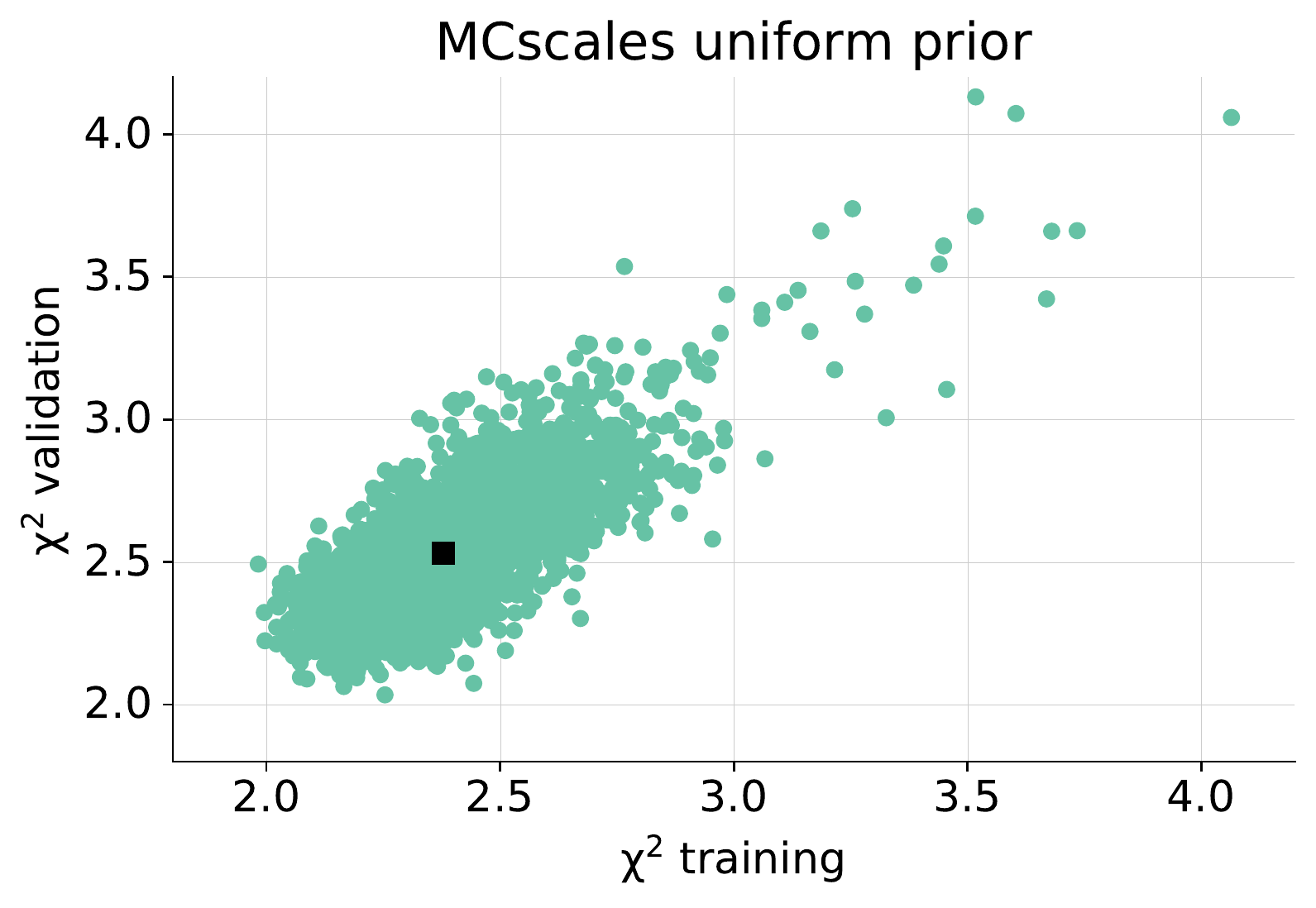}\\
\includegraphics[width=0.49\textwidth]{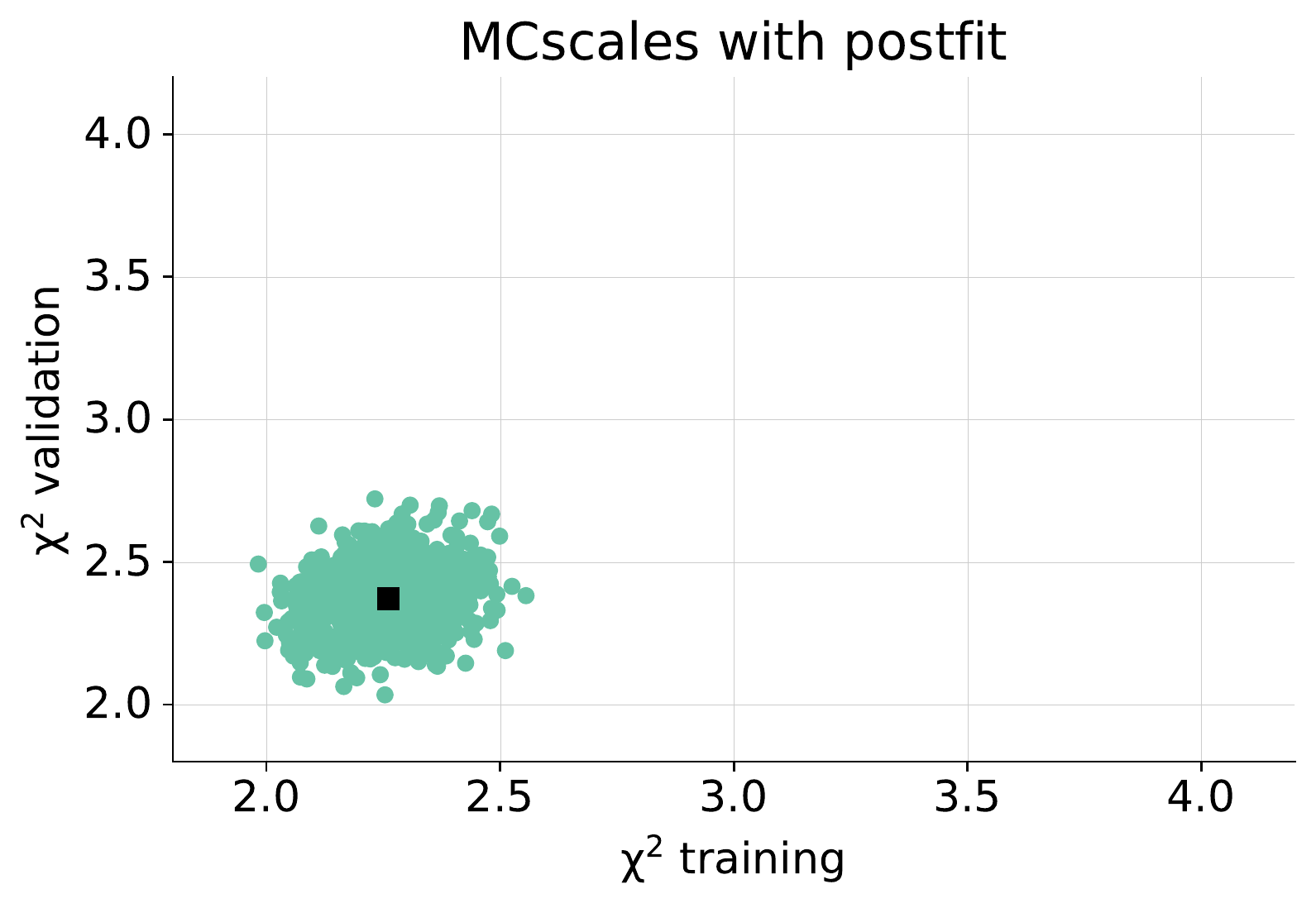}
\caption{The values of the training and validation $\chi^2$ per replica for the
  {\tt NNPDF3.1} baseline (top left), the \texttt{MCscales uniform prior}
  (top right) and the {\tt MCscales with postfit} set (bottom).
  The result for each replica is plotted in green, with the mean plotted with a black square.}
\label{fig:training-validation-comps}
\end{figure}
%-------------------------------------------------------------------------------
Fig.~\ref{fig:training-validation-comps} shows the $\chi^2$ to these two data
sets plotted for each replica. Three PDF sets are shown: the {\tt NNPDF3.1}
baseline,
the {\tt MCscales uniform prior} and the same set but after the
postfit vetoes have been applied to it. Table~\ref{tab:mean-std-training-validation}
shows the mean and standard deviation of each of these $\chi^2$ distributions.
From these we see that the {\tt MCscales} set with a uniform prior has many more outliers
than the other two sets, with these outliers exhibiting a poor agreement with
the experimental data. Such replicas drive sizeable increases in the mean and
standard deviation of the $\chi^2$ distributions versus the other two PDF sets.
This confirms the need of the modified selection procedure discussed in
Sect.~\ref{sec:postfit} and explicitly given in
Eq.~\eqref{eq:mcscales-veto} rather than the one of Eq.~\eqref{eq:nnpdf-veto}. Once it has been applied, the {\tt MCscales} set exhibits
very similar behaviour to the baseline {\tt NNPDF3.1} set, with the scatter plot
and the means and standard deviations being totally comparable. 

%-------------------------------------------------------------------------------
\begin{table}[!htbp]  \centering
\begin{tabular}{ |c|c|c|c|c|c| }
\hline
                             &         & Baseline & {\tt MCscales uniform prior} & {\tt MCscales with postfit} \\
\hline
$\chi^2_{\text{training}}$   & mean    & 2.24              & 2.38                     & 2.26                 \\
                             & std & 0.10              & 0.19                     & 0.10                 \\
\hline
$\chi^2_{\text{validation}}$ & mean    & 2.36              & 2.53                     & 2.37                 \\
                             & std & 0.11              & 0.22                     & 0.11                 \\
\hline
\end{tabular}
\caption{The mean values and standard deviations of the $\chi^2$ to the training
and validation data sets, computed from the distributions shown in
Fig.~\ref{fig:training-validation-comps}. Note that all $\chi^2$ values are per data
point.}
\label{tab:mean-std-training-validation}
\end{table}
%-------------------------------------------------------------------------------

\subsection{Distributions over scales} \label{sec:scale-dists}

\begin{figure}[!htbp] \centering
\includegraphics[width=\textwidth]{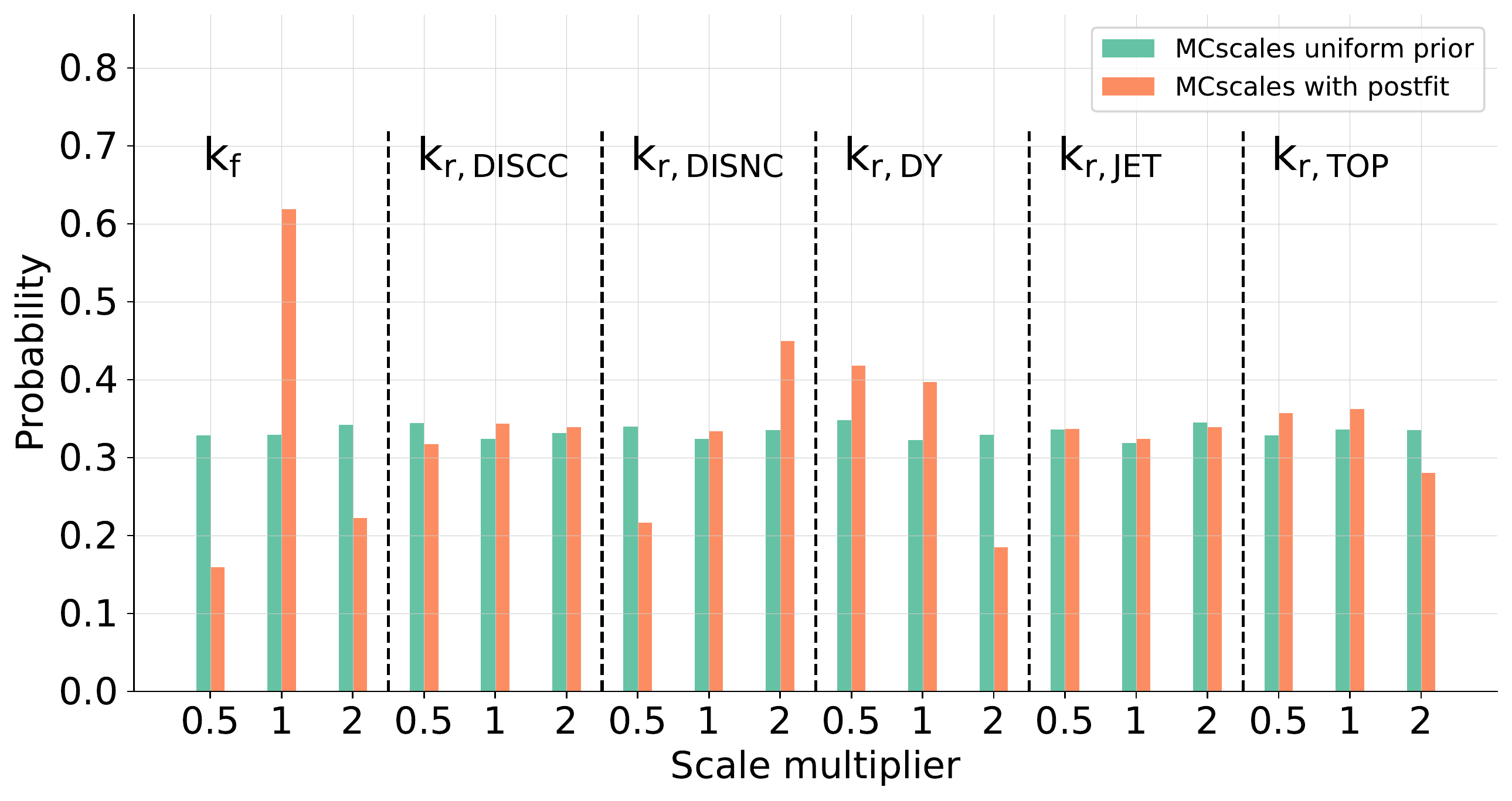}
\caption{The distribution of PDF replicas over each of the scales
in two cases: in green for the {\tt MCscales uniform prior} and in orange for the
distribution after postfit.}
\label{fig:data-driven-dists}
\end{figure}

We now turn our attention to how postfit affects the distribution of the scale
combinations in the {\tt MCscales} set.  To begin, in Fig.~\ref{fig:data-driven-dists} we look at the distribution
of each of the scales over replicas by comparing the flat distribution
corresponding to the {\tt MCscales} uniform prior to the distribution
after postfit selection. We observe that the 
distribution changes significantly after applying postfit. We observe a strong
preference for the central factorisation scale, $k_f=1$.
We also see that postfit affects each process in a
different way. As far as DIS CC or jets are concerned,
the data do not prefer any of the renormalisation scale choices, while for the
DIS NC the data strongly prefer $k_r=2$. DY data tend to prefer a
lower value of the renormalisation scale, with $k_r=1/2$ being
favoured over the other scale options, while the top data tends to disfavour
$k_r=2$. The differences over scale choices are driven by the $\chi^2$ postfit
criterion Eq.~\ref{eq:nnpdf-veto}, with little discernible correlation between the
positivity and arclength criteria with the choice of scale combination.

%-------------------------------------------------------------------------------
%------------------------------------------------------------------------------

\begin{figure}[!htbp] \centering
  \includegraphics[width=.31\textwidth]{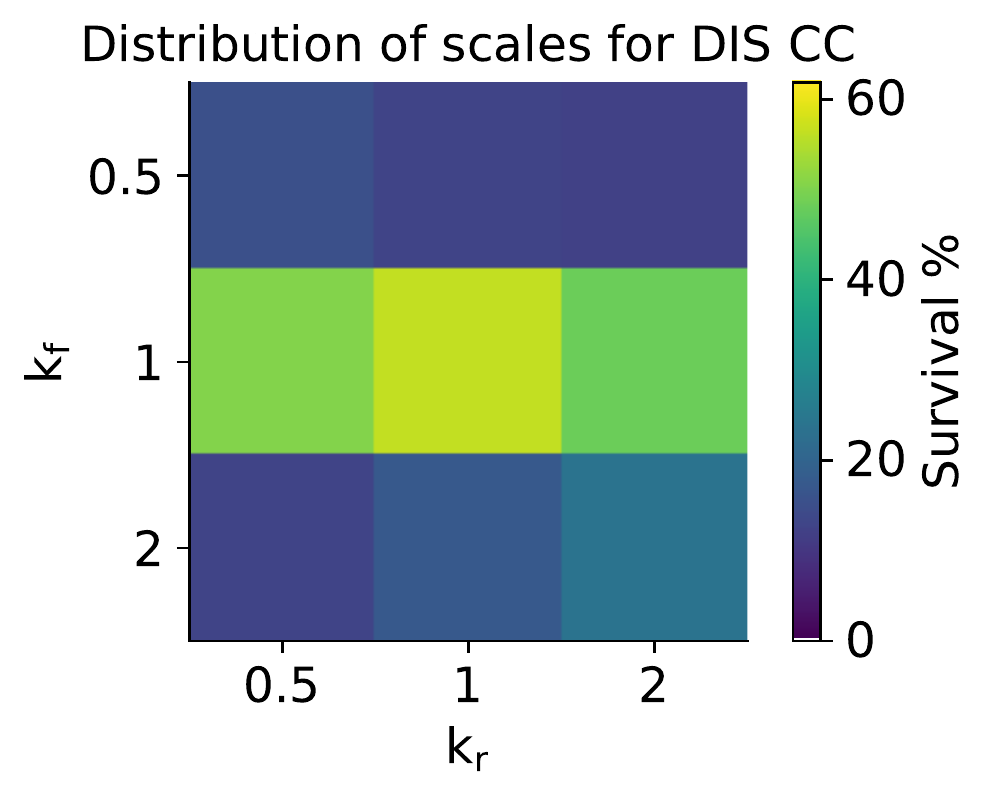}
  \includegraphics[width=.31\textwidth]{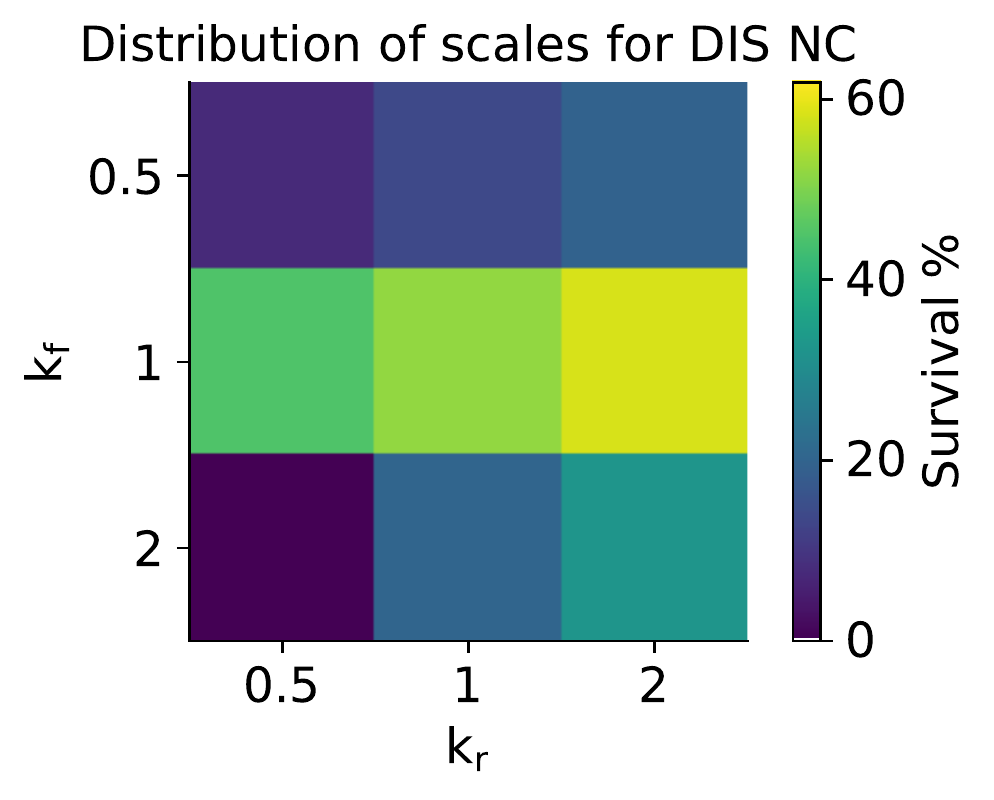}
  \includegraphics[width=.31\textwidth]{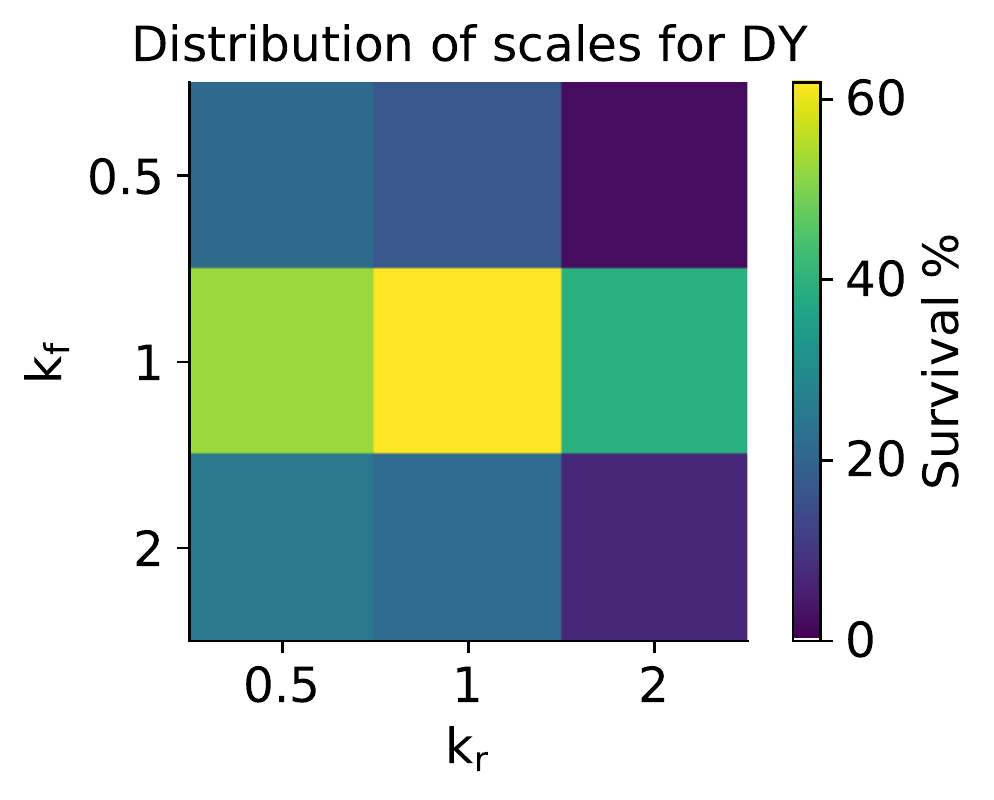}\\
  \includegraphics[width=.31\textwidth]{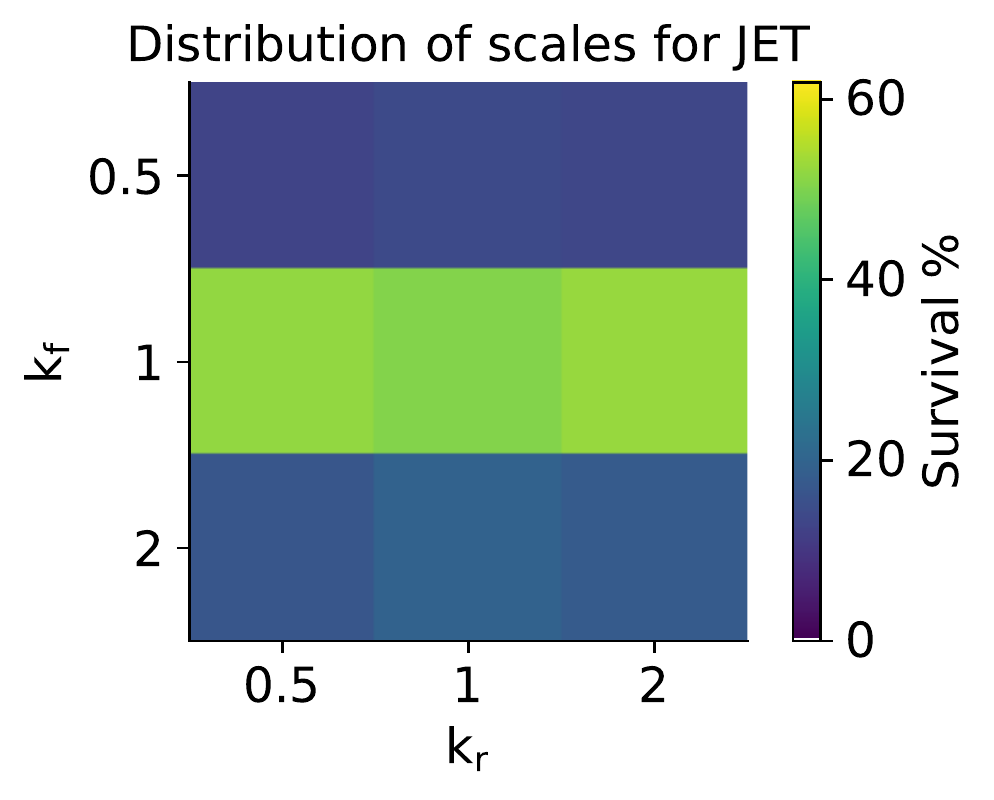}
  \includegraphics[width=.31\textwidth]{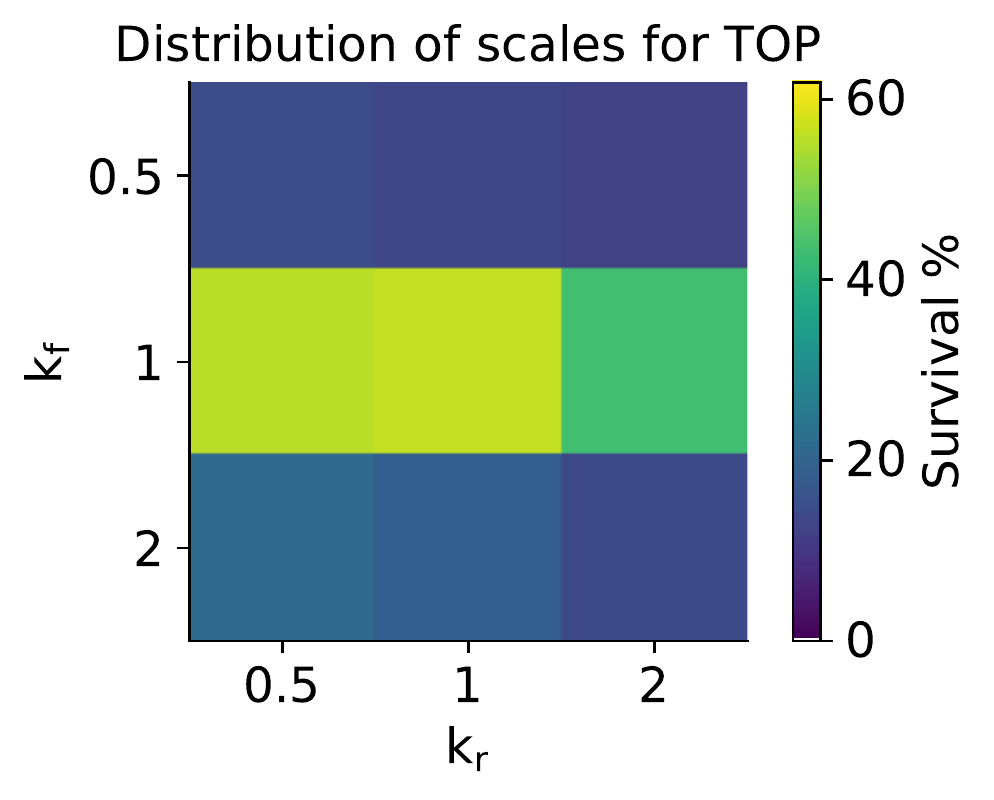}
  \caption{The percentage of surviving replicas after postfit relative to the {\tt MCscales uniform prior} for all scale combinations.}
\label{fig:data-driven-dists-comp}
\end{figure}

It is instructive to look at the distribution over scales in more fine-grained detail by
studying the effect on the renormalisation scale for each process in addition to the
factorisation scale. Since we assign one renormalisation scale to each process
and we vary the factorisation scale and the renormalisation scales
independently, there are nine scale combinations per process and therefore there
are 45 scale combinations in total. The percentage of
PDF replicas with a given scale combination after postfit
are shown in Fig.~\ref{fig:data-driven-dists-comp}. We observe that
across all processes the central value of the factorisation scale is
favoured by the data. In the case of DIS CC, the central value of the
renormalisation scale is also favoured, while in the case of DIS NC a
larger value is favoured by the data. As far as DY data are concerned,
a lower value of the renormalisation scale is favoured by the data
independently of the factorisation scale value.
Interestingly the survival percentage is larger for $(k_f,k_r^{\rm
  (DY)})=(0.5,0.5)$ and $(k_f,k_r^{\rm
  (DY)})=(0.5,2.0)$ than for $(k_f,k_r^{\rm
  (DY)})=(1.0,0.5)$ and $(k_f,k_r^{\rm
  (DY)})=(1.0,2.0)$. This compensates the larger survival percentage
for $(k_f,k_r^{\rm (DY)})=(1.0,1.0)$ compared to $(k_f,k_r^{\rm
  (DY)})=(1.0,0.5)$ and makes $k_r^{\rm(DY)}=0.5$ the scale favoured
by data.  
In the case of
inclusive jet data, there is basically no dependence on the
renormalisation scale, while in the case of top data a larger value of
the renormalisation scale is clearly disfavoured by the data.
%
%-------------------------------------------------------------------------------
%-------------------------------------------------------------------------------
%
We see that postfit has a strong impact on the distribution. As
we saw in Fig.~\ref{fig:data-driven-dists}, $k_f=1$ is preferred
to the non-central values.
We also observe that the most common scale combination after postfit has
been applied is not $(k_f, k_r) = (1, 1)$ for each process, as might be
expected.
A summary of the favourite scale combinations is given in
Table~\ref{tab:preferred-scale-combos}. In particular, we see that values of
$(1, 2)$ and $(1,\frac{1}{2})$ are preferred for the DIS NC and jets data,
respectively, while the other processes display a better agreement with the data
once central scales are selected.
%
%-------------------------------------------------------------------------------
\begin{table} [!htbp] \centering
\begin{tabular}{ |c|c|c| }
\hline
Scale multipliers & Process & Preferred values \\
\hline
$(k_f, k_r)$    & DIS CC  & $(1,1)$              \\
                & DIS NC  & $(1,2)$              \\
                & DY      & $(1,1)$              \\
                & Jets    & $(1,\frac{1}{2})$    \\
                & Top     & $(1,1)$              \\
\hline
\end{tabular}
\caption{The preferred scale combination for each process, calculated
  from the distributions given in Fig.~\ref{fig:data-driven-dists-comp}.}
\label{tab:preferred-scale-combos}
\end{table}
%-------------------------------------------------------------------------------

\subsection{Impact on PDFs} \label{sec:pdf-impact}

\begin{figure}[!htbp] \centering
\includegraphics[width=0.49\textwidth]{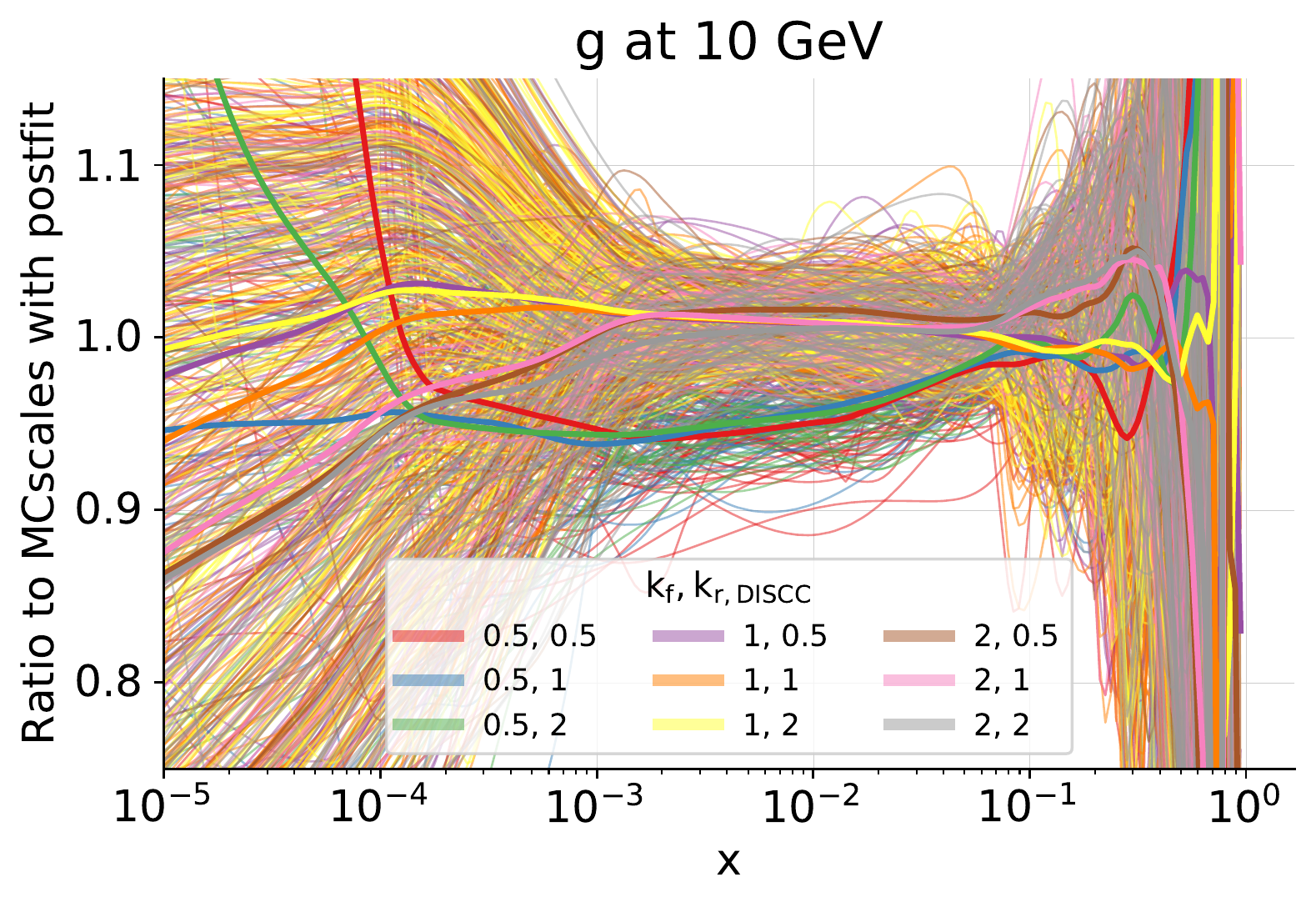}
\includegraphics[width=0.49\textwidth]{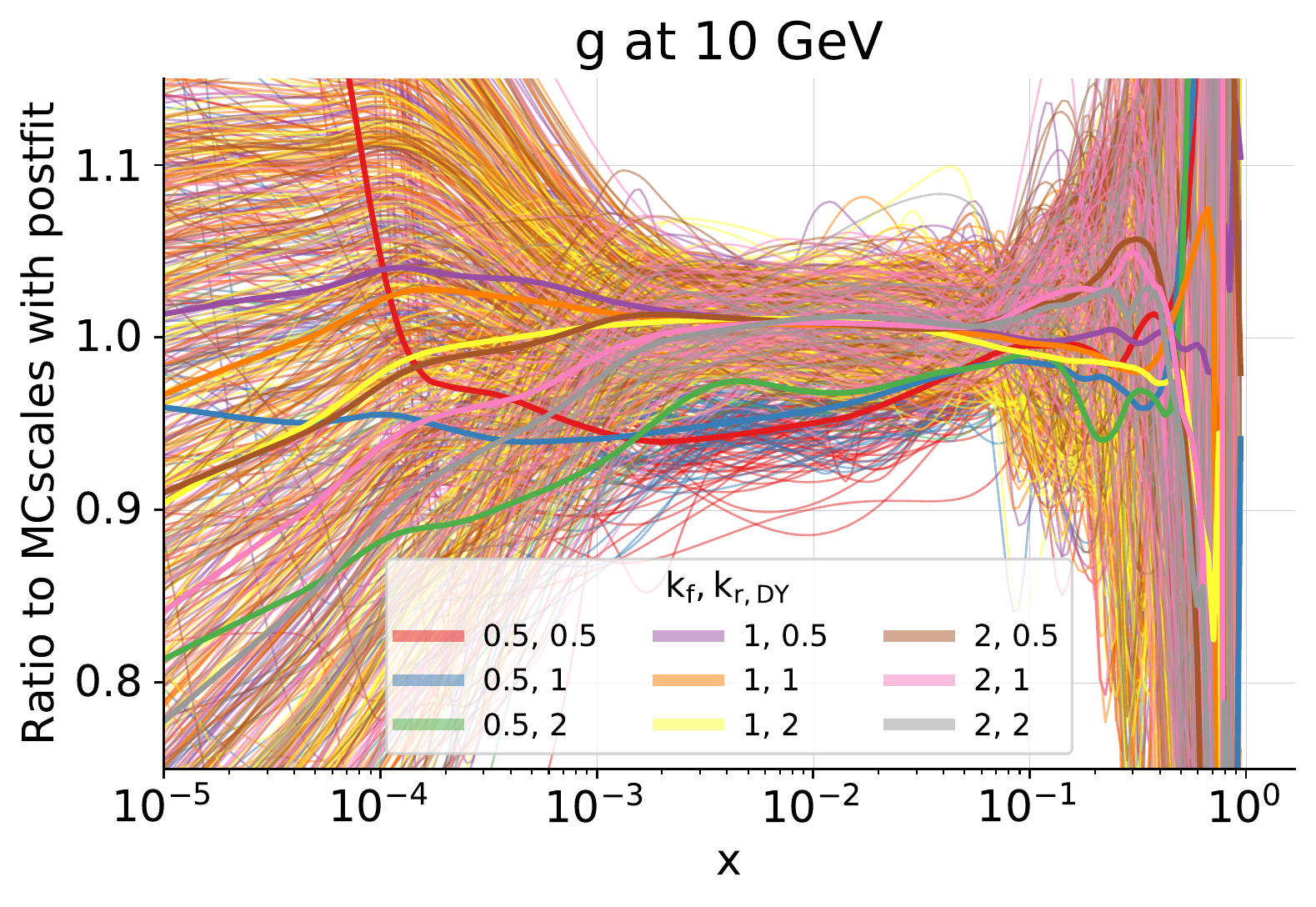}
\caption{PDF replicas of the {\tt MCscales with postfit} set, for the
  gluon at $Q=10$~GeV.
  Each replica is coloured according to its scale choice, where the numbers in the legend correspond to
($k_f, \, k_r$), where $k_r$ corresponds to the renormalisation scale
for the DIS CC predictions (left panel) or to the one for the DY
predictions (right panel). The colouring of the scales is done according to the scales of
two different processes included in the fit: the left-hand plot shows the values
of the scales for DIS CC and the right-hand plot shows them for DY. The bold
lines indicate the average over all replicas for the corresponding scale choice.
The plot is normalised to the central replica so that deviations from the central value are shown.}
\label{fig:pdf_replicas_coloured}
\end{figure}

Since we have information on the scales used to generate each replica, we can
study how different scale choices affect the PDFs and their fit quality.
Fig~\ref{fig:pdf_replicas_coloured} shows the 823 PDF replicas of the
{\tt MCscales with postfit} set for the gluon at
$Q=10$~GeV. Each plot shows the same replicas, while they are coloured
differently in each panel according to the combination of scales that
were used to generate them. This is shown for two different processes:
DIS CC (left panel) and DY (right panel). Here we see that for each case, a scale of $k_f=1/2$ (the red,
blue and green curves) leads to a diminution of the gluon in the region
$10^{-3} \lesssim x \lesssim 0.1$ and to an increase in the region
$x\gtrapprox 0.1$. On the other hand, the replicas generated with
$k_f=2$ (the brown, pink and grey curves) yield a similar gluon in the
central-$x$ region but increase it for $x\gtrapprox 0.1$.
A very similar behaviour is observed for all of the
five processes included in the fit. 
Interestingly, for DIS NC, the scale
combination $(k_f, \, k_r) = (2, \, 0.5)$ also leads to
diminution of the gluon, which is not seen for any other process.

%-------------------------------------------------------------------------------
%-------------------------------------------------------------------------------

\begin{figure}[!htbp]
\centering
\includegraphics[width=.49\textwidth]{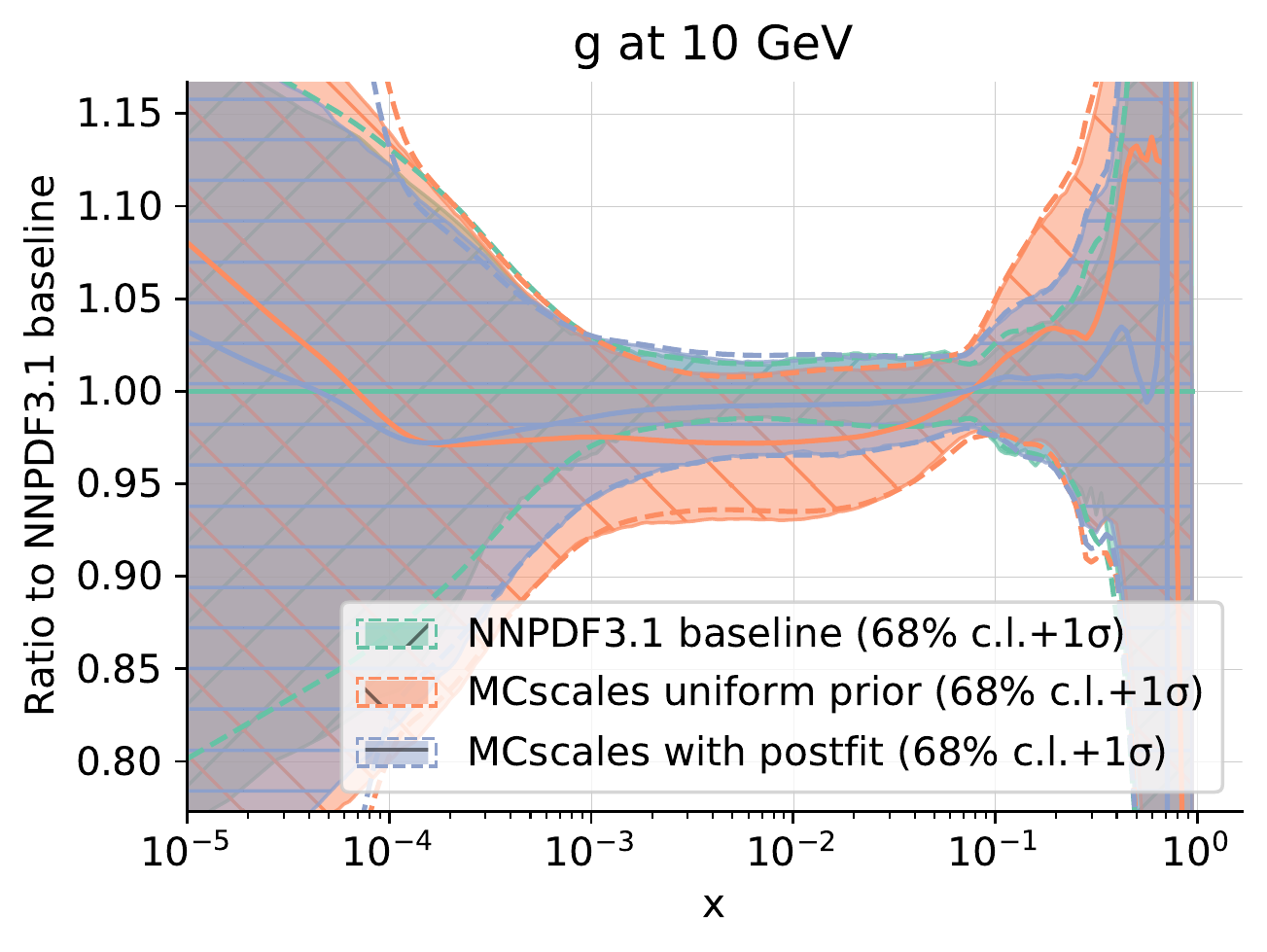}
\includegraphics[width=.49\textwidth]{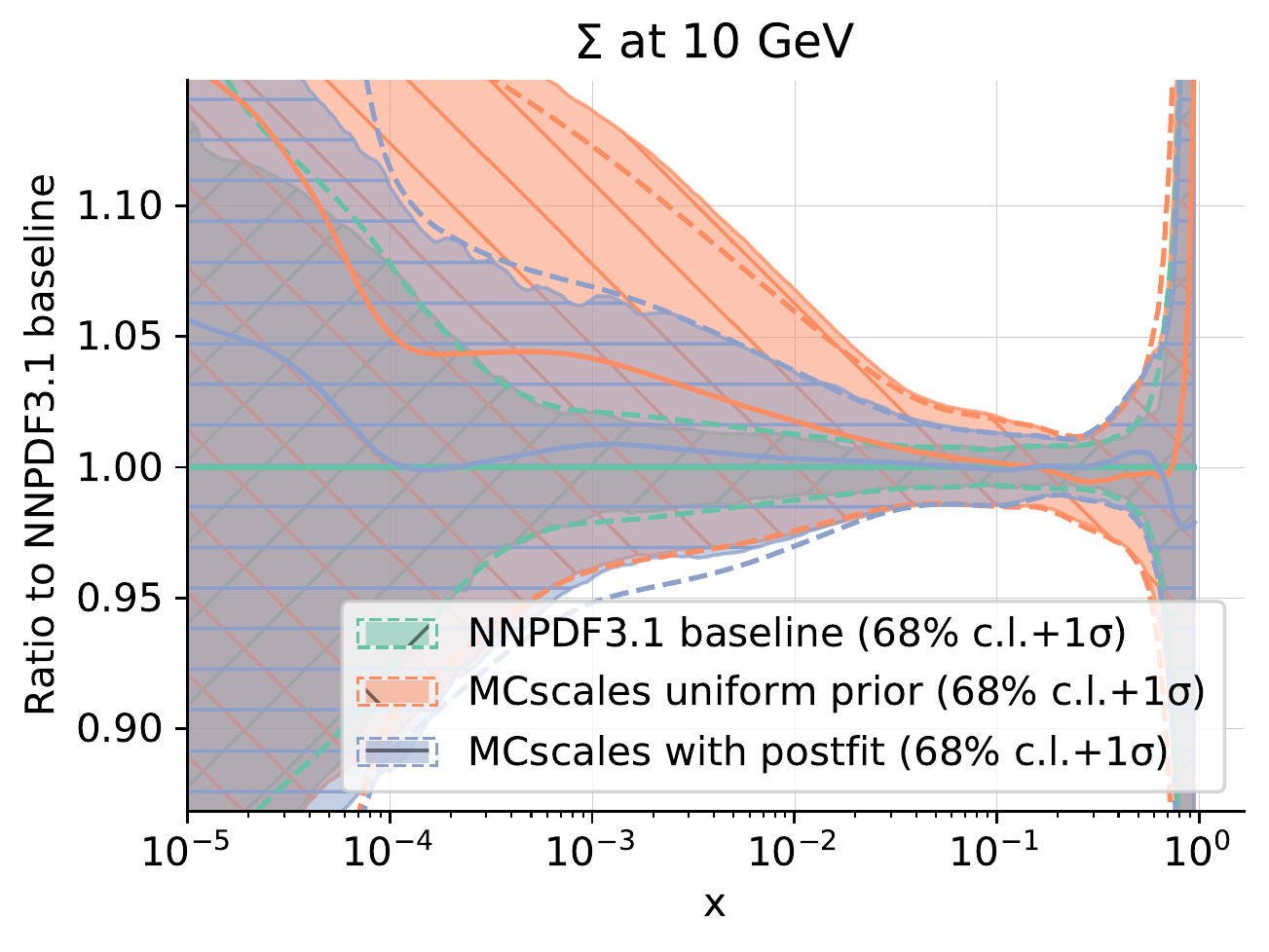}
\includegraphics[width=.49\textwidth]{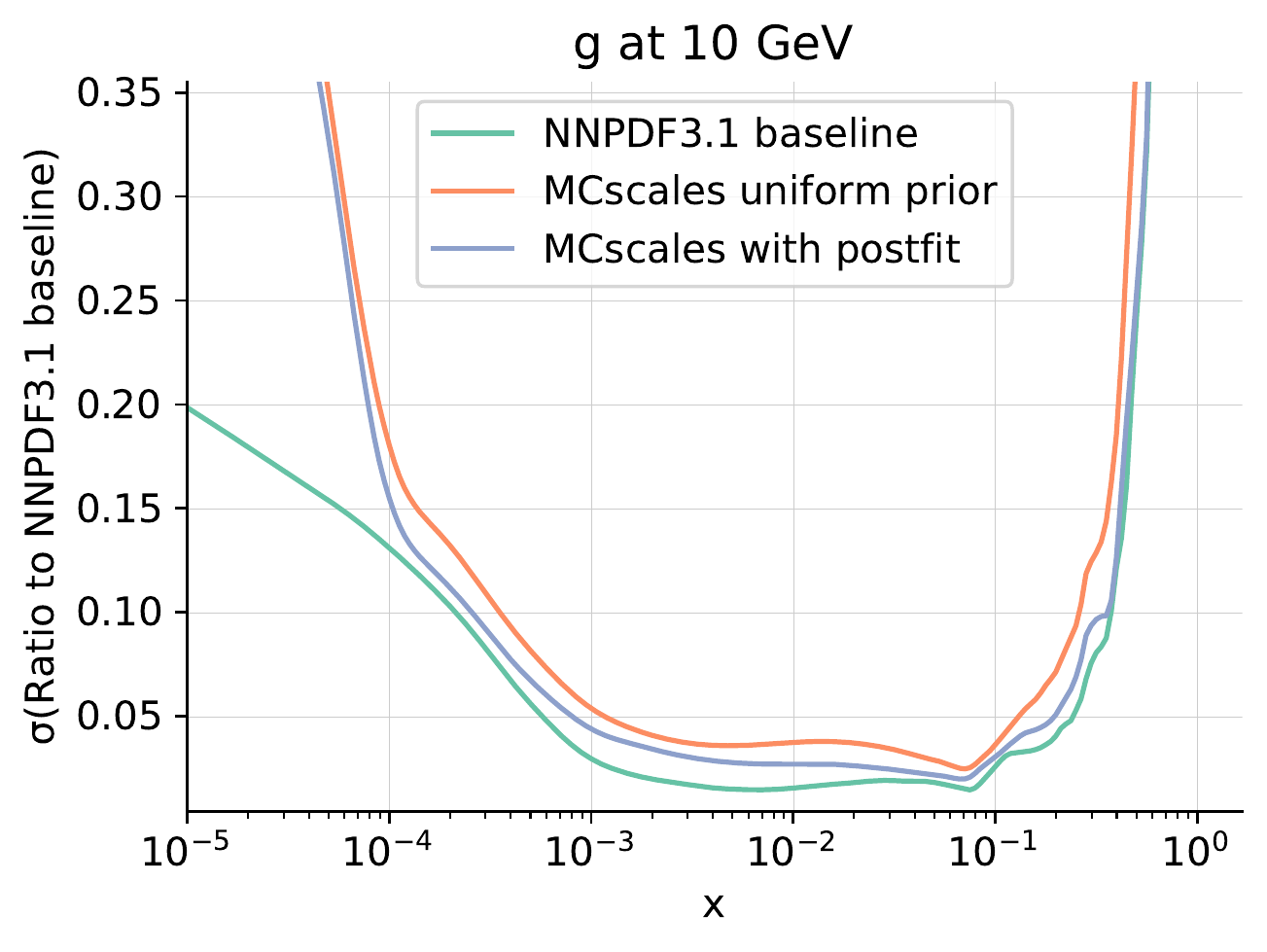}
\includegraphics[width=.49\textwidth]{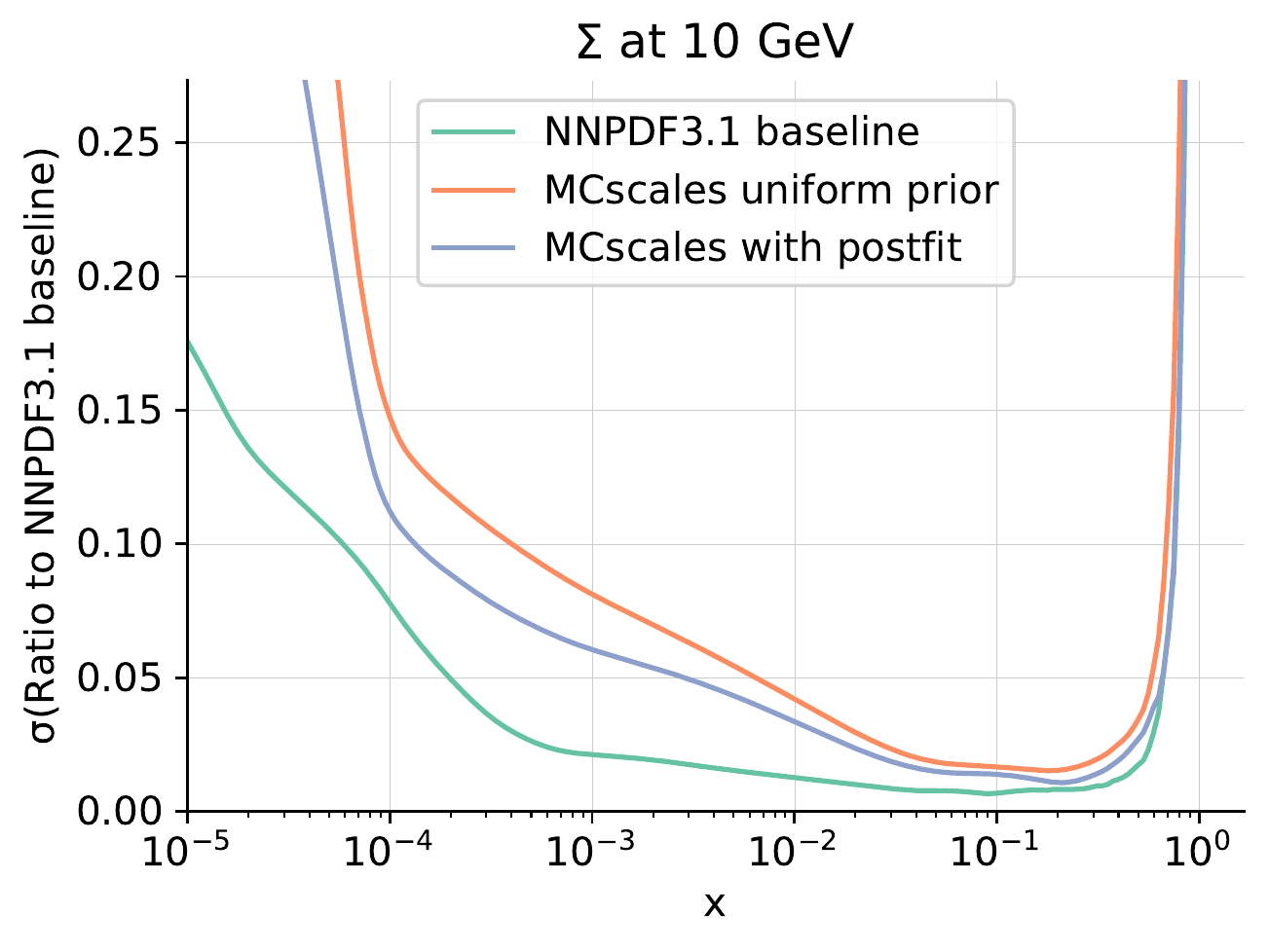}
\caption{Top: plots of the gluon (left) and singlet $\Sigma$ (right) PDFs at $Q = 10$~GeV, where
\texttt{NNPDF3.1 baseline} (green) is
compared to \texttt{MCscales uniform prior} (orange) and the same fit
with after postfit (blue).
Bottom: The 1$\sigma$
PDF uncertainties, normalised to the central value of the \texttt{NNPDF3.1 baseline} PDF.}
\label{fig:data-driven-pdf-comps-1}
\end{figure}

We now study the effect of the scale uncertainties at the level of the PDFs.
Fig~\ref{fig:data-driven-pdf-comps-1} %and~\ref{fig:data-driven-pdf-comps-2}
compares the {\tt NNPDF3.1} baseline
with the {\tt MCscales uniform prior} set and
the {\tt MCscales with postfit} set. The gluon and the singlet PDFs are displayed
along with their 68\% C.L. error bands at $Q =10$~GeV.
The plots are normalised to the {\tt NNPDF3.1} baseline so its central values
sit along the $x$-axes. We see that including scale variations in the PDF sets,
as is done in the {\tt MCscales} PDF sets, leads to increased PDF uncertainties, as we would
expect. In the $x\in [10^{-4}-10^{-2}]$ region the uncertainty
broadening is much stronger for the singlet
PDFs than for the gluon PDF. At very small $x$, $x\lessapprox 10^{-4}$
this effect seems to generally be most pronounced, especially
for the {\tt MCscales uniform prior}. We see further that the uniform
prior set leads to a general enhancement of the singlet PDFs, while
for the gluon PDF the effect is much less marked.
However in both cases, after postfit has been applied to the {\tt
MCscales} set based on an uniform prior, the general
enhancement is no longer present and the increase in PDF uncertainties tends to
become more symmetric. 
%-------------------------------------------------------------------------------
%-------------------------------------------------------------------------------
% \begin{figure}[!htbp]
% \centering
% \includegraphics[width=.49\textwidth]{Flavour_specs0_Norm_specs0_plot_pdfs_u.pdf}
% \includegraphics[width=.49\textwidth]{Flavour_specs0_Norm_specs0_plot_pdfs_d.pdf}
% \includegraphics[width=.49\textwidth]{Flavour_specs0_plot_pdf_uncertainties_u.pdf}
% \includegraphics[width=.49\textwidth]{Flavour_specs0_plot_pdf_uncertainties_d.pdf}\\
% \includegraphics[width=.49\textwidth]{Flavour_specs0_Norm_specs0_plot_pdfs_s.pdf}\\
% \includegraphics[width=.49\textwidth]{Flavour_specs0_plot_pdf_uncertainties_s.pdf}
% \caption{Same as Fig.~\ref{fig:data-driven-pdf-comps-1} for $u,\, d, \,s$ PDFs.}
% \label{fig:data-driven-pdf-comps-2}
% \end{figure}
% %---------

Comparing the {\tt NNPDF3.1} baseline and the {\tt MCscales with postfit} set, we see
the effect of scale variations once outliers have been sensibly removed. For
each flavour the PDFs are compatible within uncertainties, with any shift in the
central value induced by scale variations within the uncertainties of the
{\tt NNPDF3.1} baseline. The shift in the central value appears to be most prominent
in the data region for the gluon, where part of the reduction in the gluon PDF
observed for the {\tt MCscales} PDFs without postfit remains after their application.

A discussion concerning the comparison with the PDFs obtained in
Refs.~\cite{AbdulKhalek:2019bux,AbdulKhalek:2019ihb} using the theory covariance matrix approach is given in Appendix~\ref{sec:pdf-9pts}.
Note that the comparison at the level of PDFs does not necessarily reflect the comparison of
theory uncertainties at the level of observables. The predictions obtained by matching the scales in the
{\tt MCscales} PDF sets with those in the partonic cross section will have a comparable or even smaller
theory uncertainty due to the effect of the correlations between scales. This will be discussed explicitly
in Sect.~\ref{sec:pheno} and in Appendix~\ref{sec:pdf-9pts}.

\subsection{Non-uniform models for the prior sampling}  \label{sec:theory-driven}

So far we have illustrated the results that can be obtained with the
{\tt MCscales} methodology using the simplest option for the prior
probability, {\it i.~e.} a uniform probability distribution over all
scale multipliers contained in the set $\Omega$, Eq.~\eqref{eq:omega}.

Of course, more involved models for the prior probability can be considered.  In
Appendix~\ref{app:abc} we present a more general model for the prior probability
which can be tuned according to the user's assumptions on the correlation
between processes.

In this section, we limit ourselves to study the simplest non-uniform prior probabilities
based on the selection of a subset of scale multipliers in
$\Omega$. Those correspond to the scales multipliers that are used in the usual
the 7-points and 5-points envelope prescriptions. This study will allow us
to assess how the {\tt MCscales} PDF sets behave as the prior
probability spans a lower number of scale combinations.
%For reference we additionally present
%the results to those obtained using the theory covariance matrix
%approach that we presented in
%Refs.~\cite{AbdulKhalek:2019bux,AbdulKhalek:2019ihb}, while noting again the
%caveats discussed in Sec.~\ref{sec:pdf-9pts}.

We start by reminding the reader how the scale multipliers are chosen in the
n-point prescription. The 7-pts envelope prescription is the most commonly
adopted option in phenomenological studies. The multipliers $k_f$ and $k_{r_p}$
are  varied by a factor of two about the central choice while ensuring that 1/2
$\le k_f/k_{r_p}\le$ 2.
\begin{equation}
  \label{eq:omega7}
\Omega|_{\rm 7 pts}=\{(\xi_{f},\xi_{1},\ldots,\xi_{N_\text{p}})
\; \forall \; \xi_{f},\xi_{1},\ldots,\xi_{N_\text{p}}\in\Xi \Big|
\frac{1}{2}\le \frac{\xi_f}{\xi_p} \le 2 \}\ .
\end{equation}
We can also consider the 5-pts prescription where multipliers are obtained by varying $k_f$ up and down by a
factor of 2 while keeping $k_{r_p}=1$ and vice versa, namely
\begin{equation}
  \label{eq:omega5}
\Omega|_{\rm 5 pts}=\{(\xi_{f},\xi_{1},\ldots,\xi_{N_\text{p}})
\; \forall \; \xi_{f},\xi_{1},\ldots,\xi_{N_\text{p}}\in\Xi \Big|
\xi_f=1 \land \xi_p\in \Xi  \lor \xi_p = 1 \land  \xi_f\in \Xi \forall
p=1,\ldots,N_p\}\ .
\end{equation}
% Notation
The {\tt MCscales-7pts}, {\tt MCscales-5pts} and PDF sets presented in this
section are obtained by starting from the non-uniform priors described in
$\Omega|_{\rm 7 pts}$, $\Omega|_{\rm 5 pts}$ and applying the postfit criteria
spelled out in Sect.~\ref{sec:postfit}\footnote{We note that these sets can also be obtained by eliminating replicas from a
sufficiently big uniform prior set, by discarding replicas that do not match the
criteria in Eqs.~\ref{eq:omega7} and~\ref{eq:omega5}.  We provide the necessary
tools for this task, which are presented in Sect.~\ref{sec:delivery}.}

% Results
\begin{figure}[!htbp]
\centering
\includegraphics[width=.45\textwidth]{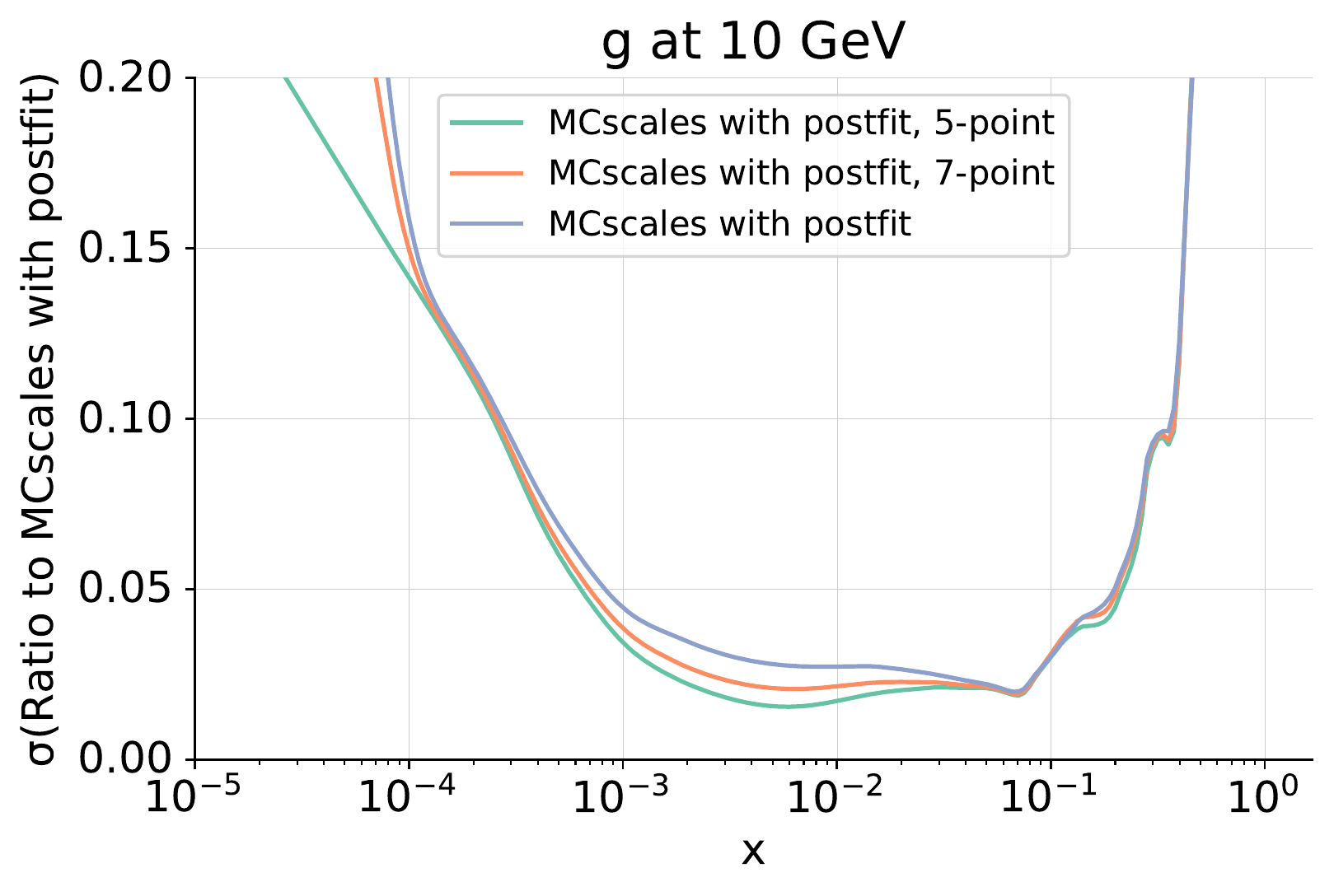}
\includegraphics[width=.45\textwidth]{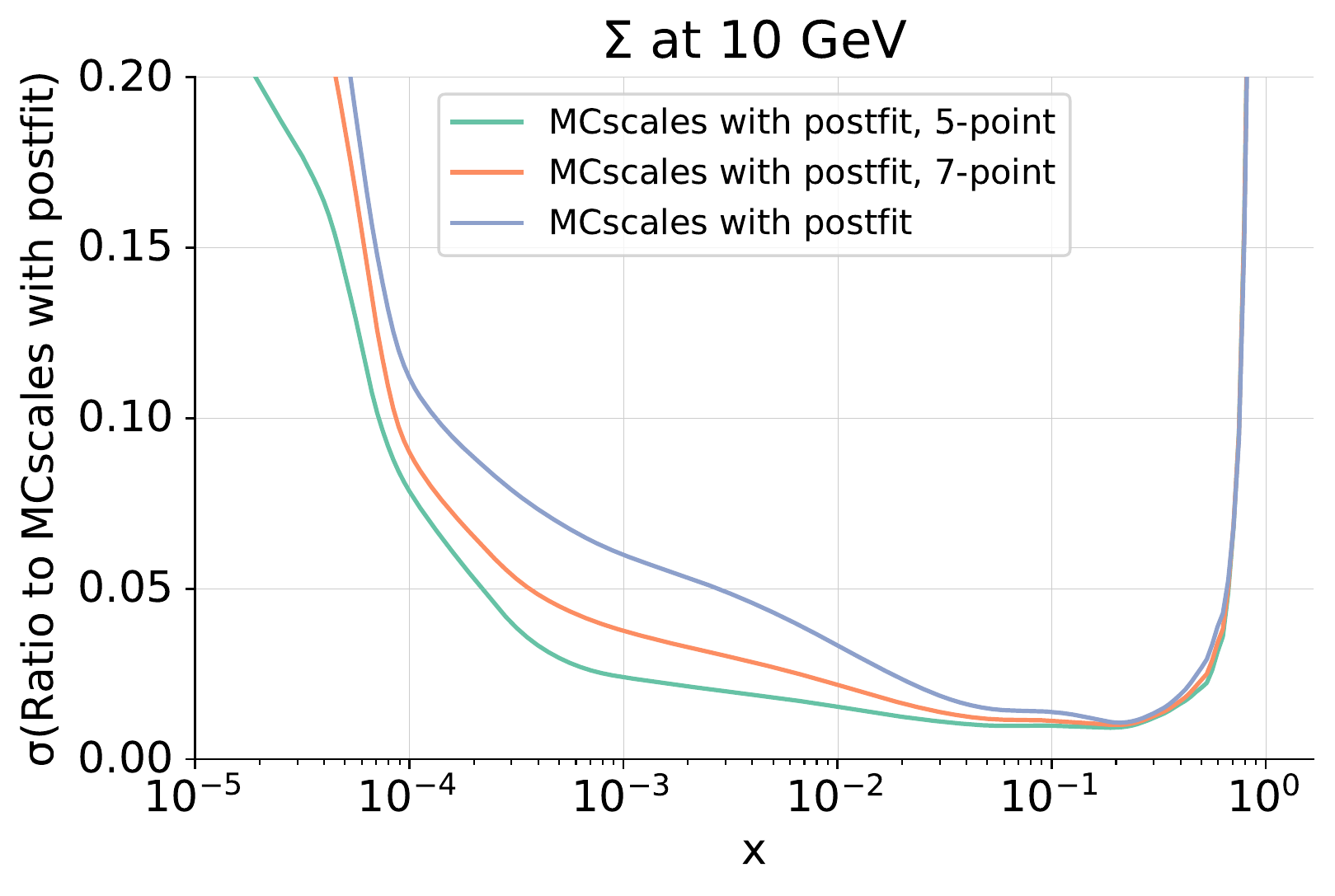}\\
\includegraphics[width=.45\textwidth]{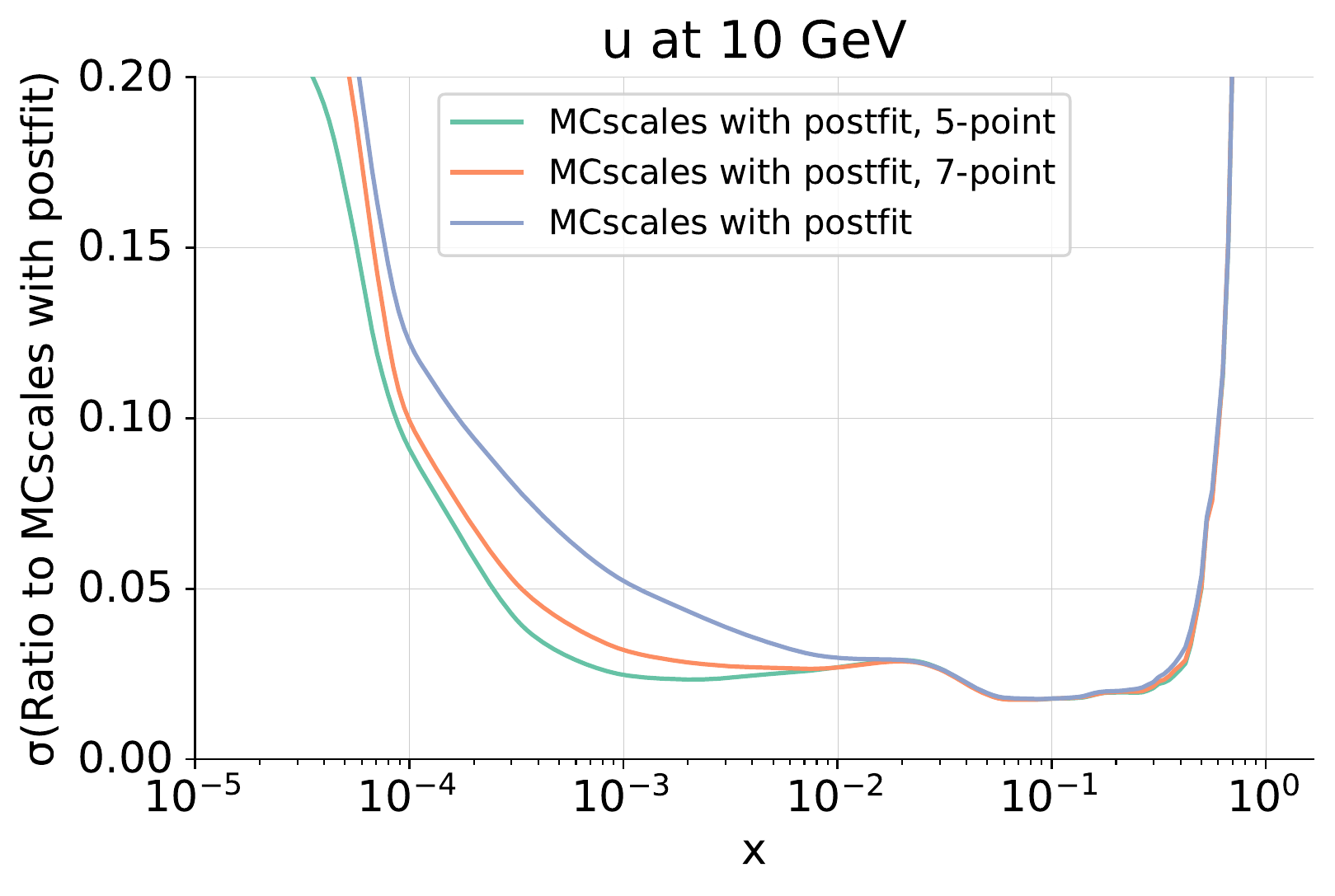}
\includegraphics[width=.45\textwidth]{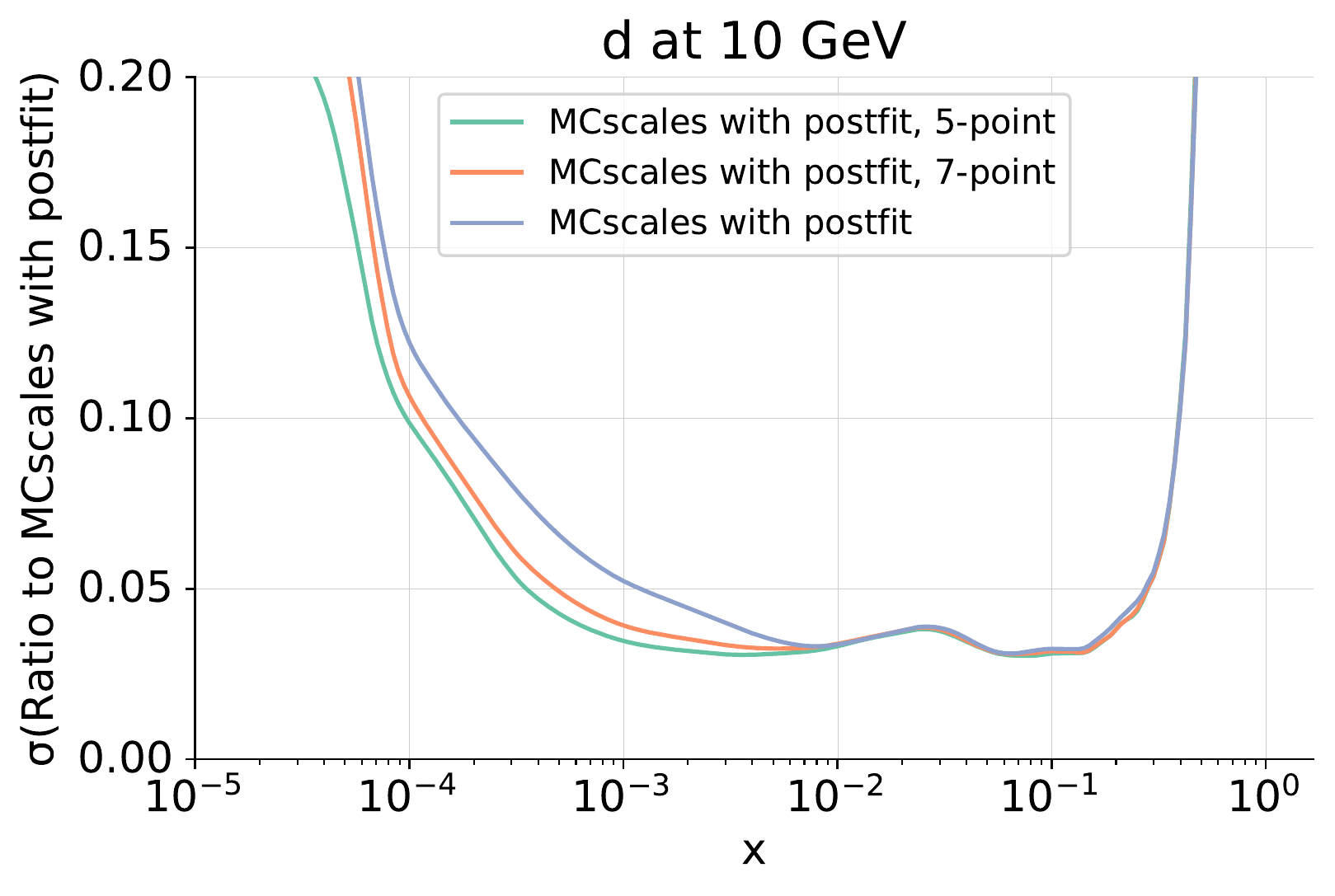}
\caption{Plots of the relative uncertainties of the gluon (top left), singlet $\Sigma$ (top right),
  up (bottom left) and down (bottom right) PDFs at $Q = 10$~GeV once
  the prior probability is chosen according to the subset of scale
  choices defined in Eq.~\eqref{eq:omega} (blue curves),
  \eqref{eq:omega7} (orange curves) or \eqref{eq:omega5} (green curves).}
\label{fig:5-7-9pts-unc}
\end{figure}
Given that the central value of the sets we explore in this section is very similar
to the central value of the {\tt MCscales} set obtained with a uniform prior, we focus on
the relative uncertainties of the PDFs
determined by starting from a prior that includes the subsets of
points defined in \eqref{eq:omega7} and \eqref{eq:omega5} behave. In Fig.~\ref{fig:5-7-9pts-unc} 
we observe that uncertainties are comparable in the medium to large-$x$ region, for $x\gsim 10^{-2}$,
while the moderate- to small-$x$ region is more affected, especially for the quarks.
Indeed the relative uncertainty decreases by about a factor of 2 for
$x\sim 10^{-3}$. This may be associated with the fact that the (2,1/2) and
(1/2,2) combinations have a stronger pull to the gluon and singlet
evolution, and thus yield replicas that span a broader range in the
small-$x$ region. 

\section{Cross section computation with {\tt MCscales} PDFs}
\label{sec:xsec}

Here we describe the machinery to produce phenomenological results
with {\tt MCscales} PDFs. Because the scale variation information used to
produce each PDF replica is stored, users have the ability to match the scale
variations in the partonic cross section of interest with those used in the theoretical predictions 
in the PDF fit, thus attaining theory predictions with fully consistent scale
variations. These steps are laid out in Sect.~\ref{sec:maincalc}.
Additionally, in Sect.~\ref{sec:other-ways} we define two other ways to
compute predictions which make use of a more limited amount of information,
for the purpose of assessing the effect of fully matching the scales.

\subsection{Theoretical predictions with matched scales}
\label{sec:maincalc}

The Monte Carlo formalism typically used in the NNPDF PDF sets requires 
computing a sample of theory predictions for a given 
partonic cross section $\hat{\sigma}$, from the set of $N$ PDF replicas $\{f_n\}$
with $n=1,\ldots, N$
\begin{equation}
  \label{eq:nnpdfsample}
  \left\{\sigma_n = \hat{\sigma} \otimes f_n\ \, \forall \, n=1, \ldots, N \right\} \,,
\end{equation}
where the symbol $\otimes$ schematically represents the Mellin convolution between each 
PDF replica set and the partonic cross section for a given
process\footnote{Note that for hadronic observable, one should think
  of $f_n$ as a convolution between two PDFs, defined as parton
  luminosity. However we leave this indicated as $f_k$ in order not to
unnecessarily complicate the notation. }.
Statistics over this sample is then applied to extract information
on the PDF dependence of the observable under inspection;
notably the central value of the theory prediction for that given observable is
taken to be the average of the $\{\sigma_n\}$ sample and the PDF uncertainty is taken to be
its standard deviation.

Our present work allows this method to be extended to include information on the
combined PDF and scale uncertainty to determine the total ({\it i~e.} the PDF+scale)
theoretical uncertainty associated with a given theoretical prediction.
As described in Sect.~\ref{sec:the-method}, we store the information on the
scale variations used in the theory
predictions for the input data of each replica,
%in the PDF sets described in this work. In
%Sect.~\ref{sec:the-sampling-model} we specified that we choose to
%define a single
namely the factorisation scale and five distinct renormalisation scales, one for
each of the five physical processes into which we split our input data set. This
makes it possible to match the factorisation and renormalisation scale
variations of a given cross section prediction with those used in the
determination of the PDF set. We distinguish between two cases, depending on the
nature of the process under inspection: the first case is when the process is
one of the five processes included in the PDF fit, and the second is when
instead we consider a new process not present in the fit. We now consider each
in turn.

In the first case, both the factorisation and renormalisation scale
variations can be matched: the partonic cross section for the process
$p$, $\hat{\sigma}_p(k_f, k_{r_p})$ (where we now note explicitly both the index
of the process and the dependence
on the scale multipliers) can be computed with the nine scale variations,
yielding the nine partonic cross sections,
\begin{equation}
  \label{eq:scalevar}
  \left\{\hat{\sigma}_p(k_f, k_{r_p}) \,\, \forall \, k_f, k_{r_p} \in \Xi \right\}\ .
\end{equation}
% Given that we will consider an individual process at the time
%from now on we will omit the index $p$ of the process we are
%considering.  
%
The scale variations of \eqref{eq:scalevar} can be associated with variation in
the replicas of the {\tt MCscales} PDF set to yield a sample that matches both
the factorization and the renormalization scale for a given process in the
input PDF set and in the computation of the partonic cross section.
The sample is built as
%contains the information of the combined PDF and scale uncertainty
%
\begin{equation}
  \label{eq:mcscalessample}
  \left\{\sigma_n = \hat{\sigma}_p(k_f^{(n)}, k_{r_p}^{(n)}) \otimes
    f_n(k_f^{(n)}, k_{r_p}^{(n)}) \ \,\, \forall \, n = 1,\ldots, N \right\}
  \ ,
\end{equation}
where $k_f^{(n)}$ is the factorisation scale multiplier associated with replica
$n$, and $k_{r_p}^{(n)}$ is the renormalisation scale multiplier associated
with replica $n$ for process $p$, and where we have left implicit the
dependence of $f_n$ on the scale multipliers associated with the
renormalization scale of the other $(N_p-1)$ processes. The sample of
replicas constructed as in Eq.~\eqref{eq:mcscalessample} now contains the
information both on the scales used in the fit yielding each
individual PDF replica and on the scales used in
the computation of the partonic cross section for each PDF replica.
%The importance of each scale variation will be determined by the data driven method described
%in Sect.~\ref{sec:data-driven} thereby ensuring that the scale choices lead
%to a good description of the experimental data included in the {\tt
%MCscales} fit.
Crucially, computing the standard deviation of the sample in
Eq.~\eqref{eq:mcscalessample} results in a total theory uncertainty, including
the experimental PDF uncertainty and the scale uncertainty. The scale
uncertainty comprises both the scale uncertainty in the partonic cross
section and the one in the
PDFs, with the correlations fully accounted.

We now address the situation in which a prediction is being made for a process
that is not among the input data sets in the PDF analysis, such as
Higgs production. In this case we indicate the renormalisation scale
associated with the Higgs production partonic cross section as $k_r$
(rather than $k_{r_p}$ as we do when the process is one of those
included in a PDF fit). 
Here only the factorisation scale variations can be matched, while there is no process to
match the renormalisation scale variations with.
In this case one can instead construct a sample of size $3N$ by first setting the
renormalization scale for the process not included in the PDF fit to 1/2 and
convoluting the partonic cross section with all 
replicas in the \mcscales set by matching the factorization scale in the computation
of the partonic cross section with those in the PDF set. Subsequently, the same
matched convolution is done by setting the renormalization scale for the process to 1
and then to 2. The set is built as 
%which results from matching the factorisation scales and considering the
%renormalisation scale variations,
%
\begin{gather}
  \label{eq:mcscalessample-noren}
  \begin{split}
    \left\{\sigma_n = \hat{\sigma}_p\left(k_f^{(n)}, k_r^{(n)}=\frac{1}{2}\right) \otimes f_n(k_f^{(n)}) \ \,\, \forall \,n= 1,\ldots, N \right\} \cup\\
    \left\{\sigma_n = \hat{\sigma}_p\left(k_f^{(n)}, k_r^{(n)}=1\right) \otimes f_n(k_f^{(n)}) \ \,\, \forall \, n= 1,\ldots, N \right\} \cup\\
    \left\{\sigma_n = \hat{\sigma}_p\left(k_f^{(n)}, k_r^{(n)}=2\right) \otimes f_n(k_f^{(n)}) \ \,\, \forall \,n=1,\ldots, N \right\} \, .
  \end{split}
\end{gather}
The resulting sample will weight the factorisation scale variation as in the
input PDF and will assume that the three renormalisation scale variations are
equally likely, as per our prior assumption in Sect.~\ref{sec:data-driven}\footnote{We
note that other choices are possible here, such as for example constructing a
sample of $N$ cross sections still matching the factorisation scales but
picking the renormalisation scale variation in the hard cross section
randomly, possibly with a non uniform weighting. However, the
results in Sect.~\ref{sec:data-driven} suggest that a uniform prior for the
renormalisation scale is a good guess, at least for the processes we
analysed.}.

Regardless of how the sample is constructed, the central value
$\bar{\sigma}$ and the total theory uncertainty $\Delta$ including
both the total PDF uncertainty and the scale uncertainty of the
partonic cross section can be obtained  %for instance
by computing the mean and standard deviation of the sample:
%Both for an in sample or an out of sample process, the central value and
%combined PDF and scale uncertainty is obtained by computing the mean and
%standard deviation over the sample, respectively
%
\begin{equation}
    \label{eq:sigmacorr}
  \bar{\sigma}^{\rm(corr)} = \left\langle \sigma_n \right\rangle\ ,
\end{equation}
\begin{equation}
  \label{eq:deltacorr}
%  \Delta ^{\rm(corr)}\equiv \Delta^{{\rm PDF}_{C+S}\, + \,{\rm scale}} = \text{std}(\sigma_n)\ ,
  \Delta ^{\rm(corr)} = \text{std}(\sigma_n)\ ,
\end{equation}
where the set $\{\sigma_n\}$ has been obtained with either 
Eq.~\eqref{eq:mcscalessample} or Eq.~\eqref{eq:mcscalessample-noren},
depending on whether the process is included or not in a PDF fit. 

\subsection{Theoretical predictions using less information}
\label{sec:other-ways}

In order to study the effect of employing {\tt MCscales} PDFs using the full
information on the scale correlations, as described in Sect.~\ref{sec:maincalc}, here we define two other
ways to compute cross sections, which utilise less of the available information.

The first method mimics what happens when one estimates theory
uncertainties using conventional PDFs; that is, PDFs that do not
include any scale variation uncertainty. 
In this case, rather than working with an {\tt MCscales} PDF set, we use the subset of the 
$N_c=192$ PDF replicas for which all factorisation and renormalisation scales are set to
their central values. To compute the PDF uncertainty, we follow the
standard approach of computing a set of $n$ replicas of the cross
sections computed for the central PDF set 
\begin{equation}
  \label{eq:dummysample}
  \left\{\sigma_n^{(c)} = \hat{\sigma}(k_f = 1, k_{r_p} = 1) \otimes f^{(c)}_n,
  \ \,\, \forall \, n= 1,\ldots, N_c 
  \right\}
  \ ,
\end{equation}
where $(c)$ stands for ``central'' and the replicas in $\left\{f^{(c)}_n\right\}$ are all obtained
with the central values of the scales. The central value of the theory prediction is given by
\begin{equation}
  \bar{\sigma}^{(c)} = \hat{\sigma}(k_f = 1, k_{r_p} = 1)\otimes f^{(c)}_0 ,
\end{equation}
where $f^{(c)}_0$ is the central replica (replica 0) of the $\{f_n^{(c)}\}$ subset of
the \mcscales set. The PDF-only uncertainty is given by
\begin{equation}
  \Delta^{{\rm PDF}_{\rm exp}} = {\rm std} (\sigma_n^{(c)}) \ ,
\end{equation}
where the suffix ${\tt exp}$ indicates a PDF uncertainty that only
propagates the experimental uncertainty of the data included in a fit.

Next we define the set of nine cross sections computed over all nine
combinations as in Eq.~\eqref{eq:scalevar} and we convolve them with $f^{(c)}_0$,
 yielding nine values of the cross section for the process under consideration 
%$(k_f,k_r)_i \in
%\{(1/2,1/2),(1/2,1),(1/2,2),(1,1/2),(1,1),(1,2),(2,1/2),(2,1),(2,2)\}$
%and convolved with the central replica of the PDF set
%
\begin{equation}
\sigma_{0,i}^{(c)} =  f^{(c)}_0 \otimes \hat{\sigma}(k_f, k_r)   \ ,\qquad i
= 1,\ldots,9,
\end{equation}
where $i$ indicates one of the 9 possible combinations of $\{k_f,k_r\}\in\Xi$,
which will be used to compute an estimate of the scale uncertainty. In order
to maintain an equivalence with the computation using the {\tt MCscales}
model as described in Sect.~\ref{sec:maincalc}, we weight each scale variation
by the same amount as in the original {\tt MCscales} PDF set (see
Fig.~\ref{fig:data-driven-dists-comp}). The weight for each of the nine contributions is
\begin{equation}
  w_i=\frac{N_i}{N} %=\frac{N_{\rm rep}^{(i)}}{N} \ ,
\end{equation}
where $N_i$ counts the number of replicas in the {\tt MCscales} with relevant
scale multipliers set to one of the 9 elements of ~\eqref{eq:scalevar}. 
%set consistent with the scale variation of $\sigma_i^{\rm th}$
% as discussed in Sect.~\ref{sec:maincalc}).
%is assumed to be in sample or out of sample as discussed in Sect.~\ref{sec:maincalc}).
The associated weighted variance is given by 
\begin{equation}
    \label{eq:thunc}
    (\Delta^{\rm scale})^2 = \frac{1}{1-\sum_{i=1}^9 w_i^2}
    \sum_{i=1}^9w_i (\sigma_{0,i} ^{\rm (c)}-\bar{\sigma}^{\rm (c)})^2 \, ,
\end{equation}
where
\begin{equation}
  \label{eq:sigmacentral}
  \bar{\sigma}^{\rm (c)}=\sum_{i=1}^9 w_i\sigma_{0,i} ^{\rm (c)} \, .
\end{equation}
Thus, the total theory uncertainty $\Delta^{\rm (c)}$ including
only PDF experimental uncertainty and the scale variation in the
partonic cross section in an uncorrelated way is given by
\begin{equation}
    \label{eq:deltacentral}
    \Delta^{\rm (c)}%\equiv \Delta^{{\rm PDF}_{\rm exp}\, \cup \, {\rm scale}}
    = \sqrt{(\Delta^{{\rm PDF}_{\rm exp}})^2 + (\Delta^{\rm scale})^2}
    \ .
\end{equation}

The second alternative method involves combining the scale variation in the
partonic cross section with that in the {\tt MCscales} PDF in an uncorrelated
way. This will allow us to gauge the loss of accuracy incurred in when adding
scale uncertainties in PDFs but no mechanism to match them to the partonic cross
sections, as was done in Ref.~\cite{AbdulKhalek:2019ihb}.  The PDF uncertainty
here is composed of both the experimental and the scale variation components, to
which we add a component to account for the scale uncertainty in the partonic
cross section.  We convolve all replicas in the {\tt MCscales} set with the hard
cross section using unmatched, central scales
\begin{equation}
  \label{eq:mcscalessample-central}
  \left\{\sigma_{n}^{\rm (uncorr)} = \hat{\sigma}_p(k_f=1, k_{r_p}=1) \otimes f_n(k_f^{(n)}, k_{r_p}^{(n)}) \ \,\, \forall \, n= 1,\ldots, N \right\}
  \ .
\end{equation}
The central value for the prediction is given by
\begin{equation}
  \label{eq:sigmauncorr}
  \bar\sigma^{\rm (uncorr)}=\,\left\langle\sigma_{n}^{\rm (uncorr)}\right\rangle
  \ .
\end{equation}
In order to obtain a component of the uncertainty accounting for the
experimental and scale uncertainties in the PDF fit, we compute
\begin{equation}
\Delta^{{\rm PDF}_{\rm exp+scale}} = {\rm std}(\sigma_{n}^{\rm (uncorr)})
\ ,
\end{equation}
to which we add a component to account for the scale uncertainty in the hard
cross section, $\Delta^{\rm scale}$ defined as in Eq.~\eqref{eq:thunc}.
Finally, the total uncertainty including both the PDF uncertainty
(experimental and scale) and the scale uncertainty of the partonic
cross section in an uncorrelated way is given by
\begin{equation}
  \label{eq:deltauncorr}
  \Delta^{\rm (uncorr)}%\equiv \Delta^{{\rm PDF}_{C+S}\, \cup \,{\rm scale}}
  = \sqrt{(\Delta^{{\rm PDF}_{\rm exp+scale}})^2 \,+\,
  (\Delta^{\rm scale})^2} \, .
\end{equation}
In Sect.~\ref{sec:pheno} we will explore in details how the estimate
of theory uncertainty in a NLO prediction changes depending on whether
one uses the full information on scale variations in the PDF input set
and its correlation with the scales in the partonic cross sections or not.
\section{Comparison with the theory covariance matrix approaches}
\label{sec:pdf-9pts}

As we mentioned in Sect.~\ref{introduction}, the methodology presented in this
work is based on a different principle as compared to the approach based on the
construction of a theoretical covariance matrix
~\cite{AbdulKhalek:2019bux,AbdulKhalek:2019ihb}.  In the latter, a covariance
matrix is  postulated, based on differences in predictions with varying scale
choices, and it is added up to the experimental covariance matrix, both in the
sampling of the Monte Carlo replicas and in the figure of merit used in the fit
itself. The predicted quantities, on which the PDF determination is based, are
not dependent on individual scales but rather a marginalisation of the
distribution of scales (see Eq.~(2.9) of Ref.~\cite{AbdulKhalek:2019ihb} and
surrounding discussion). In the notation of Sects.~\ref{sec:the-method}
and~\ref{sec:xsec}, this would roughly correspond to
\begin{equation}
  \label{eq:thcovmatmargi}
  \sum_{\omega\in \Omega}\hat{\sigma}_p(\omega) \otimes
  f(\omega) P(\omega)
\ ,
\end{equation}
where the dependence on $\hat{\sigma}$ is only through the relevant scale
variations $k_f$ and $k_{r_p}$ in $\omega$.

In the {\tt MCscales} approach presented in this paper,
the scales entering a PDF fit are treated as free
parameters of the fixed-order theory. Thus, the experimental
uncertainty of the data and of the scales uncertainty associated to
theory predictions are both propagated onto the PDF uncertainties via a Monte Carlo
sampling in the joint space of experimental data and scale variations
(the latter subject to the constraints spelled out in
Sect.~\ref{sec:the-method}), which implies producing independent determinations
for each $\hat{\sigma}_p(\omega) \otimes f(\omega), \forall \omega \in \Omega$.

We compare the results obtained starting from a uniform prior
probability for the scales, to those obtained with the theory covariance matrix
approach.
Among the various possibilities described in Ref.~\cite{AbdulKhalek:2019ihb}, we
compare our results to those obtained with the 9-point
prescription, in which all scale variations in $\Omega$ are considered.

The {\tt MCscales} approach might look slightly more similar to the approach
explored in Sect.~6 of Ref.~\cite{AbdulKhalek:2019ihb}, in which the theory
covariance matrix is included only in the data generation
but not in the $\chi^2$ used as a figure of merit in the fit.
The corresponding PDF set is called {\tt
NNPDF31\_nlo\_as\_0118\_scalecov\_9pt\_sampl}.
In that case, increased uncertainties and a worse fit quality are
obtained, since the data 
replica fluctuations are wider due to the MHOUs, but this is not
accounted for in the $\chi^2$ minimisation. As a result,
the increase in uncertainty due to the inclusion of MHOU in the
sampling 
is not compensated by a rebalancing of the datasets and a reduction of
the scale uncertainty obtained via the inclusion of MHOU in the fit.
However, one should be careful in comparing the two approaches. Indeed, while in
the theory covariance matrix approach the predictions $T_i^{(n)}$ that enter the
computation of the $\chi^2$, Eq.~\eqref{eq:chi2k}, are computed by using the
central scales ($k_f=k_{r_p}=1$), in the {\tt MCscales} approach, the
theoretical predictions are computed with the factorisation and renormalisation
scales matched to the element $\omega$ of the sampling of the replicas
($k_f=\xi_f, k_{r_p}=\xi_{r_p}, p=1,...,N_p$). This has the effect of
approximating the scale varied predictions of the data in the fit to the
corresponding experimental values.

%-------------------------------------------------------------------------------
\begin{figure}[!htbp]
\centering
\includegraphics[width=.45\textwidth]{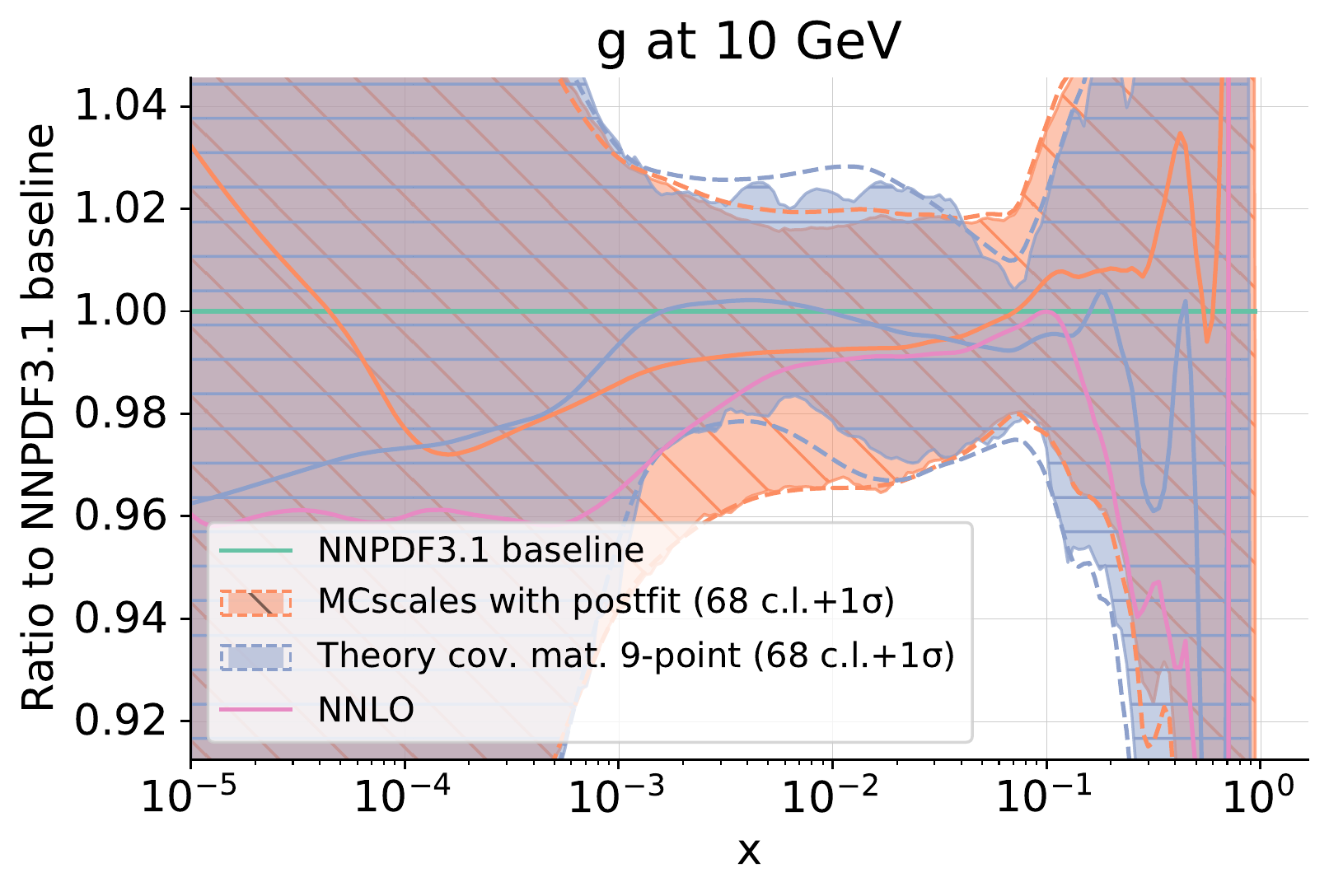}
\includegraphics[width=.45\textwidth]{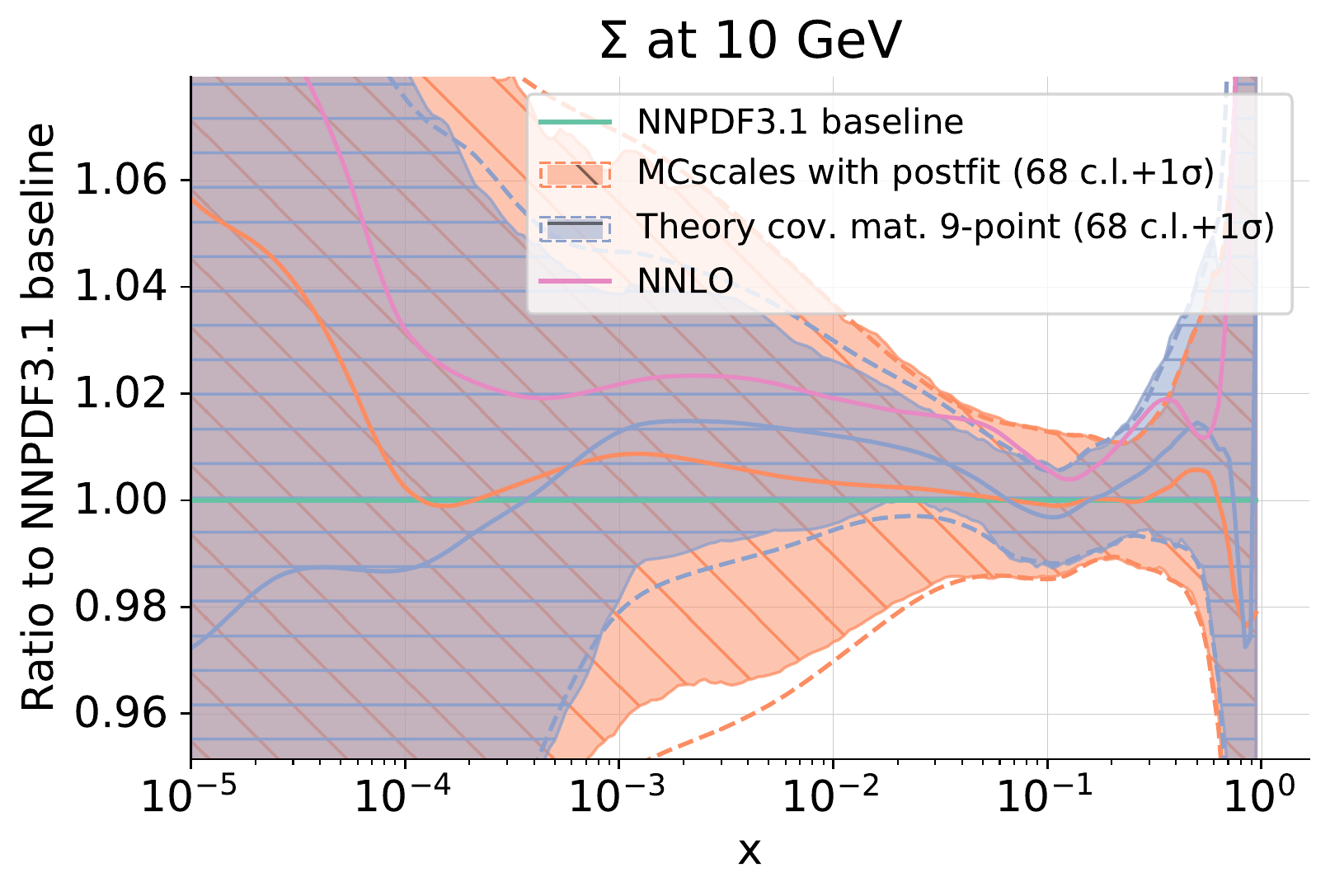}\\
\includegraphics[width=.45\textwidth]{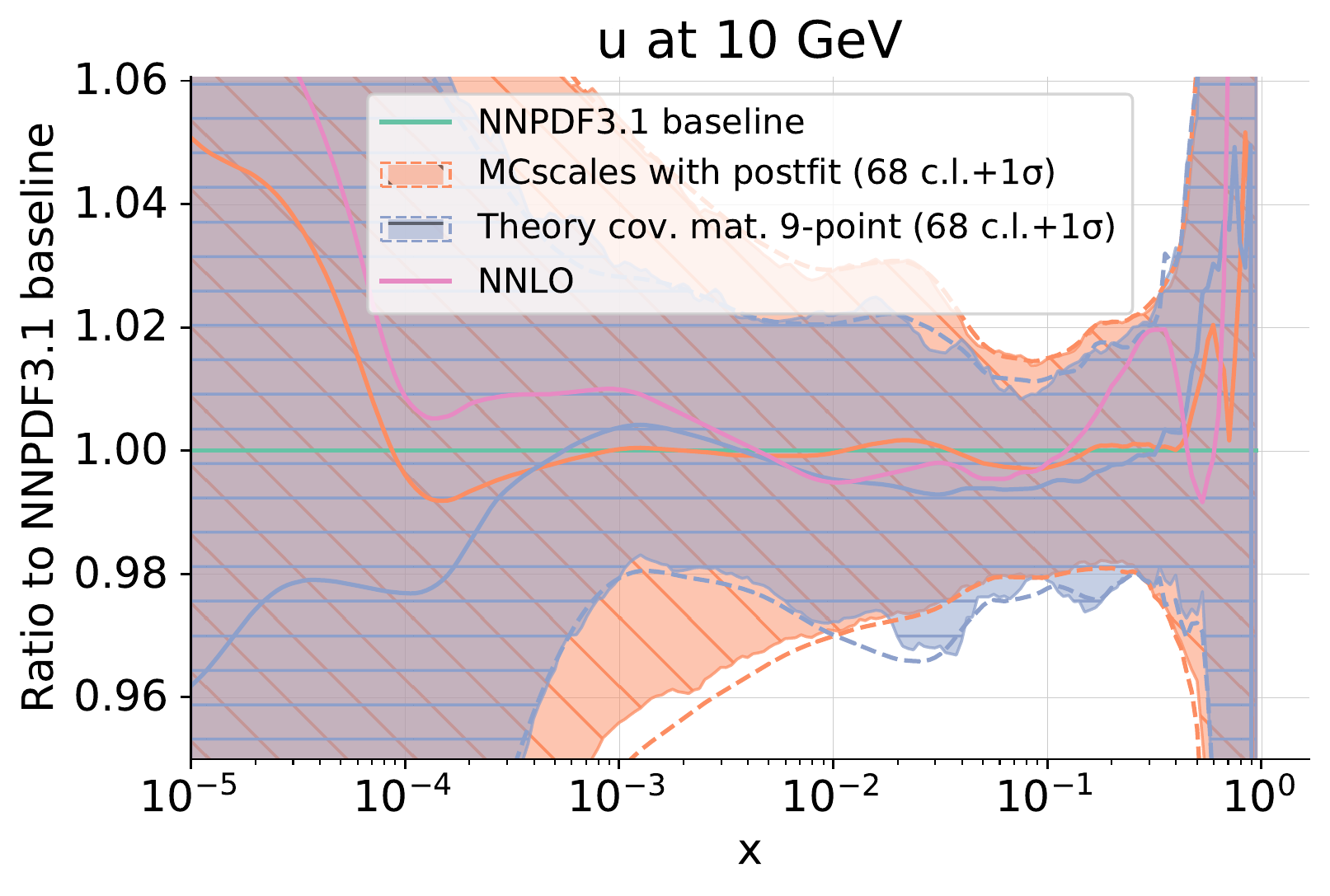}
\includegraphics[width=.45\textwidth]{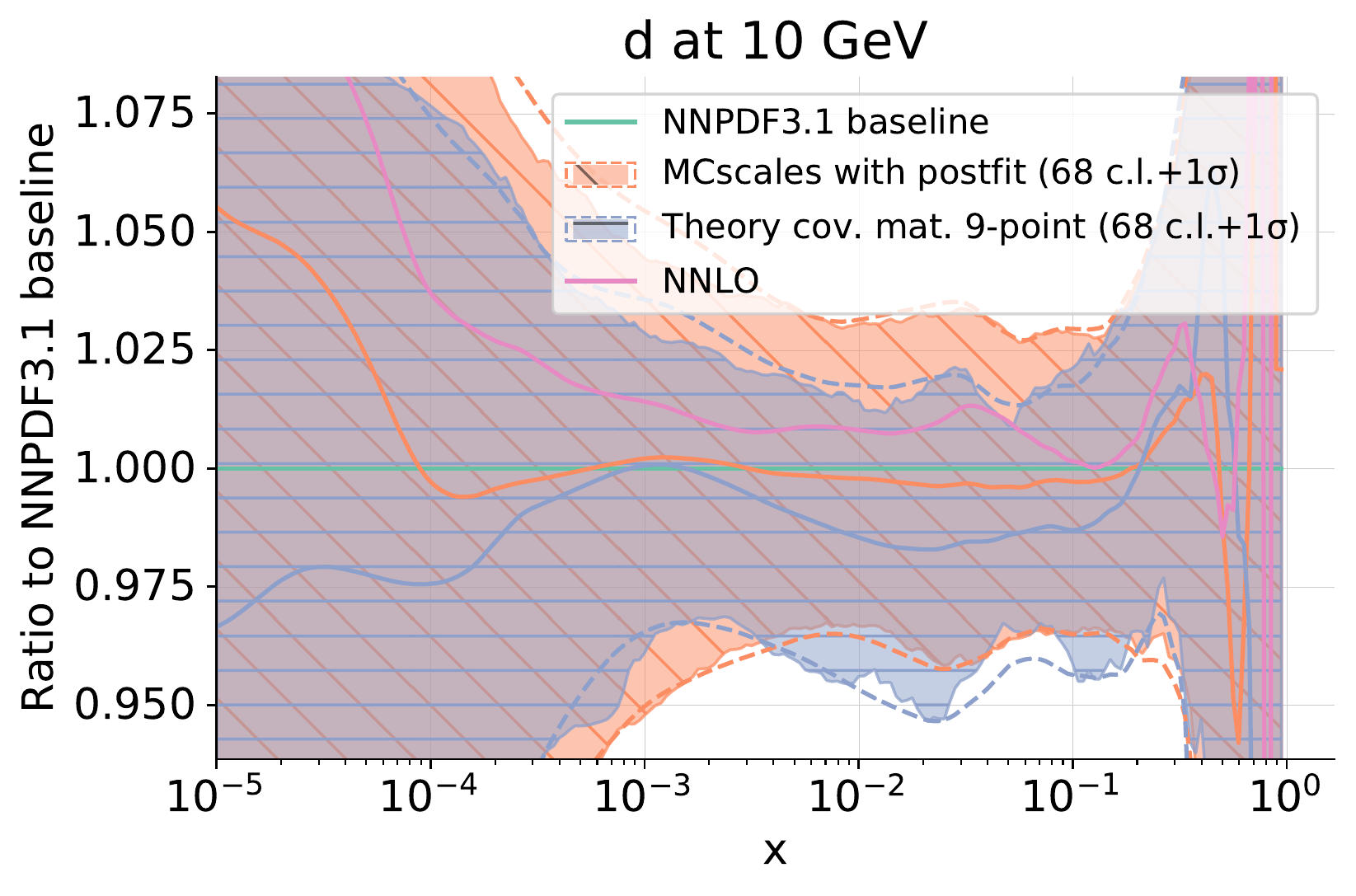}
\caption{Plots of the gluon (top left), singlet $\Sigma$ (top right),
  up (bottom left) and down (bottom right) PDFs at $Q = 10$~GeV, where
  the PDF set obtained in Ref.~\cite{AbdulKhalek:2019ihb} by using the theory
  covariance matrix built with a 9-point prescription
  (\texttt{NNPDF31\_nlo\_as\_0118\_scalecov\_9pt}) (blue) is compared to the PDF
  set obtained with the {\tt MCscales} approach starting from a uniform prior
  and after postfit (\texttt{MCscales with postfit}) (orange). The central NNLO
  baseline set (\texttt{NNPDF31\_nlo\_as\_0118}) is also displayed, in pink.
  The PDFs are normalised to the central value of the NLO baseline set
  (\texttt{NNPDF31\_nlo\_as\_0118}, green). We note that cross section
  computations using the {\tt MCscales} sets should use matched scale variations
  in the partonic cross section (see Sect.~\ref{sec:xsec}), which heavily limits
  the interpretability of this comparison.}
\label{fig:data-driven-pdf-9pts}
\end{figure}
%-------------------------------------------------------------------------------
We compare the default results of both approaches, namely the PDF set obtained
by using the 9-point theory covariance matrix both in the sampling and in the
fitting, the {\tt NNPDF31\_nlo\_as\_0118\_scalecov\_9pt} set, and the PDF set
that is obtained with the {\tt MCscales} approach starting from a uniform prior
and after postfit {\tt MCscales with postfit}. These sets are compared in
Fig.~\ref{fig:data-driven-pdf-9pts}, by displaying the gluon, singlet, up and
down-quark PDFs at the scale $Q=10$~GeV. The plots are normalised to the central
value of the {\tt NNPDF3.1} NLO baseline PDFs.  We observe that, as far as the
gluon and the singlet are concerned, the PDF uncertainties of the {\tt MCscales}
set are only moderately larger than those obtained with a 9-point theory
covariance matrix. The larger increase in uncertainty is observed in the
up-quark PDF, especially in the medium-small-$x$ region $x\in[10^{-4},10^{-2}]$.
We expect the uncertainty increase in PDFs to be compensated, when computing
partonic cross sections, and even even result in smaller uncertainties, when
correctly accounting for the correlations in scale variations in PDFs and
partonic cross sections, as we will show explicitly in what follows. 
The central values of the two sets are compatible within $1\sigma$ error
band, with the {\tt MCscales} set yielding harder PDFs at
small-$x$. It is interesting to observe that the small-$x$ behaviour reminds the one observed in
Ref.~\cite{McGowan:2022nag}, although our study is conducted at NLO
rather than NNLO. The shift in the 
central values that we observe here is much less dramatic than
in~\cite{McGowan:2022nag}, given the somewhat
different approach that we adopt: rather than having the data
determine the scales in the theoretical predictions and their
uncertainties by treating them as nuisance parameters to be constrained
by the fit, here we only use the data to discard the scale combination that yield a very poor
agreement between theory predictions and experimental data.

\begin{figure}[!htbp]
\centering
\includegraphics[width=.49\textwidth]{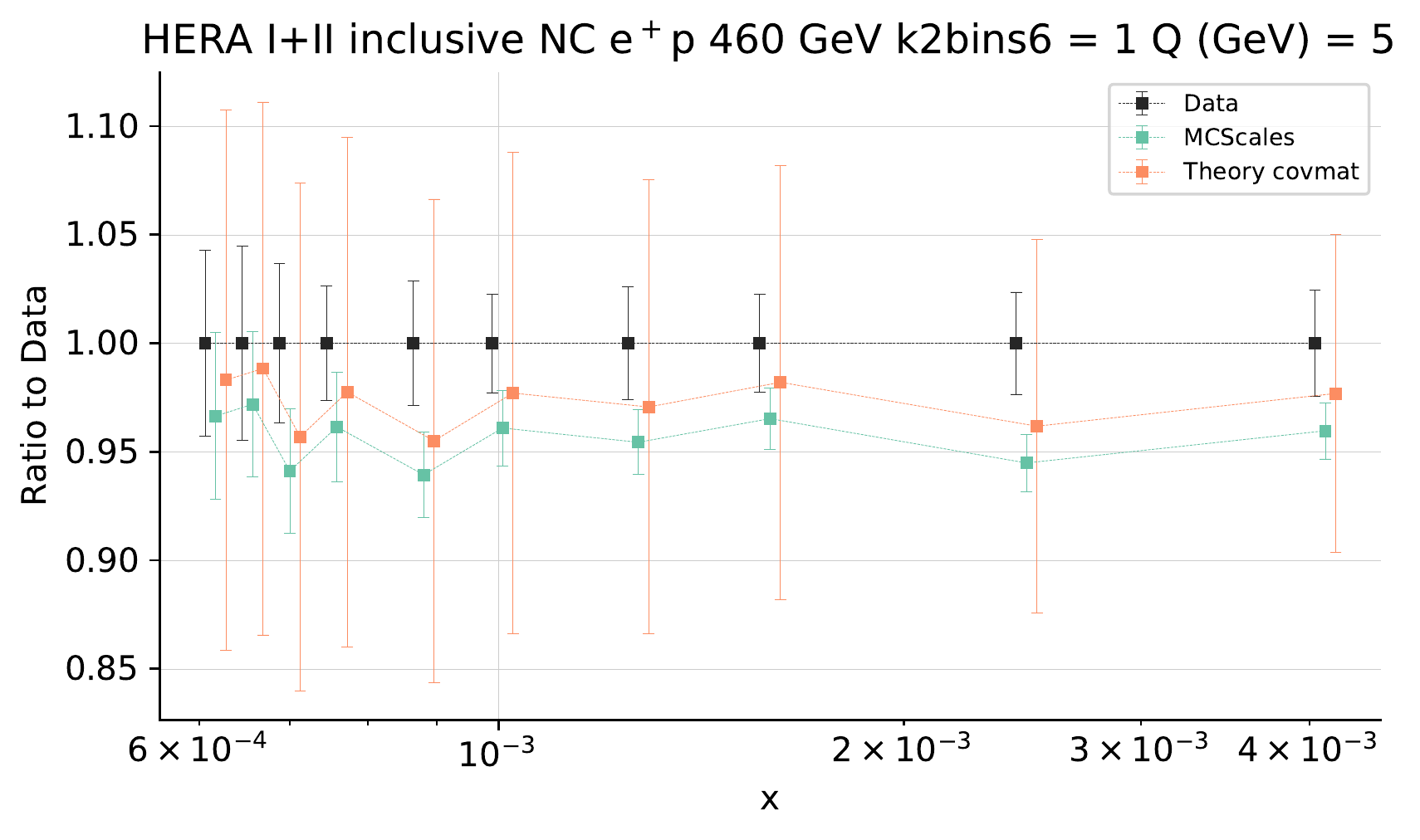}
\includegraphics[width=.49\textwidth]{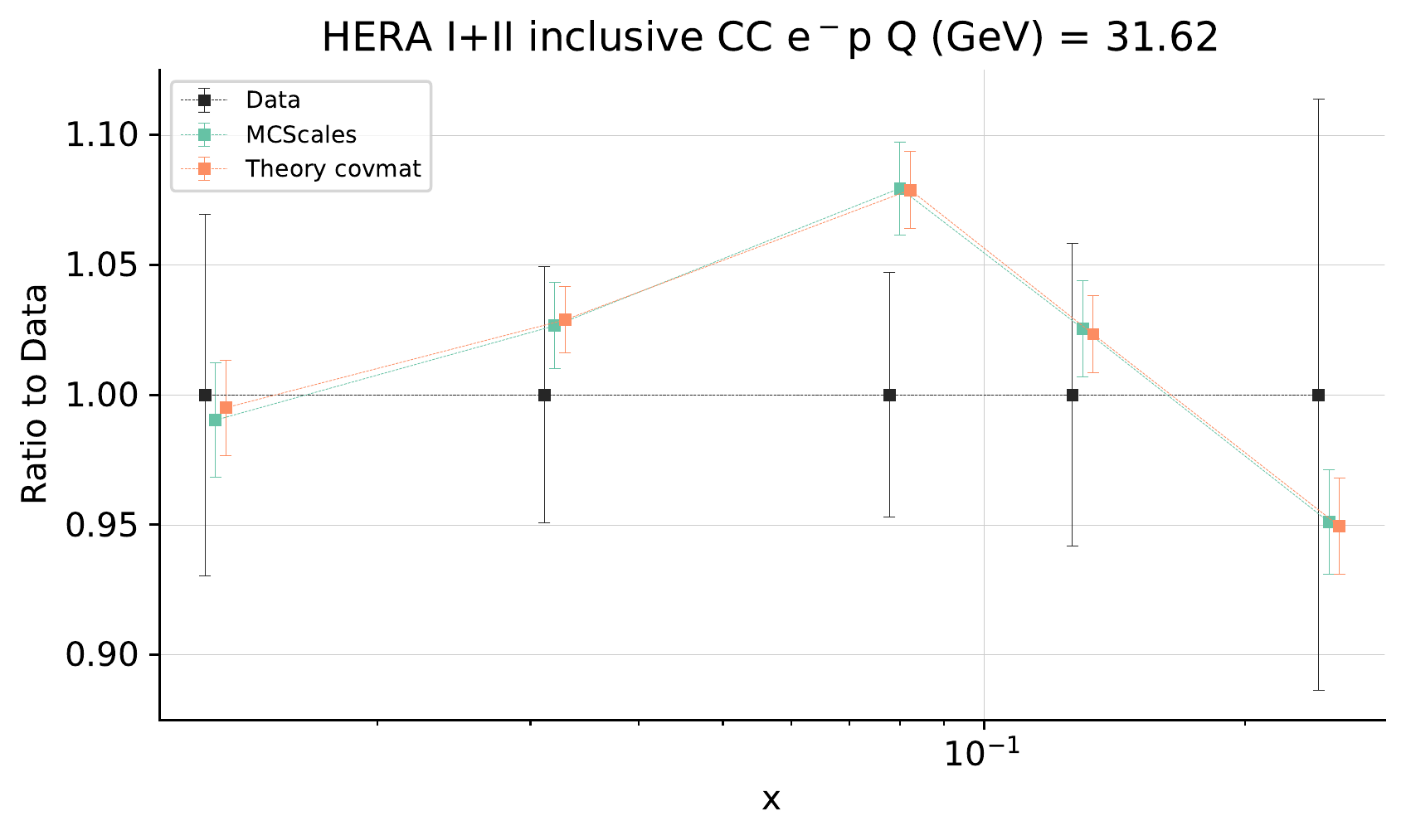}
\caption{Ratio to HERA inclusive neutral current positron-electron
  cross section data~\cite{Abramowicz:1900rp} (left) and HERA inclusive charged current
  data~\cite{H1:2018flt} (right) of the theoretical predictions obtained
  with \mcscales and theory covariance matrix approaches to the respective data.}
\label{fig:obs-9pts}
\end{figure}

We emphasise that the cross section predictions that use {\tt MCscales} PDFs are
intended to use matched scale variations in the partonic cross sections. As a
result the PDFs are not directly comparable to those in the theory covariance
matrix approach, where PDFs are to be convoluted with partonic cross sections
using central scales only.  This is an important caveat to the PDF plots
presented so far. To illustrate this point, in Fig.~\ref{fig:obs-9pts} we show
the comparison between the experimental data and the theoretical predictions for
a reduced neutral current DIS cross section measured at
HERA~\cite{Abramowicz:1900rp} and for the charged current cross
section~\cite{H1:2018flt}, where the theoretical predictions and their total
theoretical uncertainty (including both the scale uncertainty in the partonic
cross sections and the PDF uncertainty - in turn including both the experimental
and the scale uncertainties in a PDF fit) have been computed
using the PDF set obtained with a 9-pts theory covariance matrix and with the
\mcscales set. The uncertainty bands for the theory covariance matrix have been
obtained using Eq.~(8.3) of Ref.~\cite{AbdulKhalek:2019ihb}, which is analogous to
the uncorrelated prescription presented in Sect.~\ref{sec:other-ways} while the
\mcscales predictions use Eq.~\ref{eq:mcscalessample}.
%
%-------------------------------------------------------------------------------
We observe that the \mcscales uncertainties on the cross sections can indeed be
smaller than those computed with the theory covariance matrix approach, as they
account for the correlations of the scales and also due to the postfit
selection, but also bigger in cases where the underlying assumptions differ.
This illustrates how the comparison at the level of PDFs in
Fig.~\ref{fig:data-driven-pdf-9pts} might lead to incorrect conclusions with
regard to the relative uncertainty in the two approaches.

An alternative procedure to disentangle the effect of individual scale
variations and study the correlation between theory uncertainties in PDF fits
and predictions made with such PDF sets was presented in
Ref.~\cite{Ball:2021icz}.
The method is based on the theory covariance matrix formalism. However, rather than producing a
PDF fit by computing the marginalised predictions (as it is done in
Eq.~\ref{eq:thcovmatmargi}) it interprets the difference between data and theory
predictions computed with the central scales as the shifts in the predictions
derived from the theory covariance matrix (insofar those are projected into the
relevant subspace, as explained in detail in Ref.~\cite{Ball:2021icz}). By
contrast, the \mcscales procedure discussed here operates directly on the scale
parameters, which are determined for each replica by the prior distribution and
remain unchanged during the fit. As a result, the post-fit replica
selection in \mcscales has a weaker effect
in terms of constraining the scales. On the other hand the \mcscales
procedure does not assume that the differences between data and predictions come
exclusively from the theory uncertainties.
Another difference in the two approaches comes from the fact that in
the \mcscales procedure both the predictions used in the fit and those computed with the resulting PDFs
are matched exactly, thereby avoiding any assumptions on the Gaussianity and
smallness of the theory corrections. Such hypotheses do not
necessarily apply to all data. Indeed, in 
Sect.~\ref{sec:scale-dists}, we show that scale variations have a large impact
on the fit quality and are therefore comparable to experimental uncertainties.
Secondly, the \mcscales procedure provides direct access to the scale
parameters, allowing external users to study and even tune the scale dependence of the
resulting fit.
Furthermore, given that the procedure in Ref.~\cite{Ball:2021icz} is based on
the results of a pre-existing PDF fit scale shifts, it might be subject to
the problems pointed out in
Ref.~\cite{Forte:2020pyp} for the equivalent case of fitting the strong coupling
constant.  The result in this case would not be the same as
those of a simultaneous fit of PDFs and scale shifts.

In order to understand how the differences in the approaches affect
the results, it would be very interesting to implement the two
procedures in the computation of the LHC observables that we will present in
Sect.~\ref{sec:pheno}, and compare the results in all cases. We leave
this to a future study, in which both methodologies may take advantage
of the more modern {\tt NNPDF4.0} PDF set~\cite{nnpdf40} and of the theory
pipeline implementation presented in~\cite{Barontini:2023vmr}. 

\section{Application to phenomenology}
\label{sec:pheno}

In this section we compute reference LHC standard candles with the NLO PDF
sets from our Monte Carlo samplings presented in
Sect.~\ref{sec:data-driven} and the definitions given in
Sect.~\ref{sec:xsec}. 
Most importantly, for the first time we will be able to assess the importance
of the correlation between the scale variation in the computation of
the partonic cross sections and the scale variation in the PDF fit as well as the impact
of their double counting. 

For our assessment to be meaningful, it is important that the partonic cross
sections are computed at the same perturbative order as the observables
included in a PDF fit. While this requirement does not hold if correlations
are neglected, as in Ref.~\cite{AbdulKhalek:2019ihb}, once correlations are
fully taken into account according to Eq.~\eqref{eq:mcscalessample} it is
important that the scale variations are performed at the same perturbative
order. %, because the matching of scales is not meaningful otherwise.
For this reason, we will only consider the NLO computaton of key processes at the LHC.
Once a PDF set including scale variation will be available at NNLO, our
analysis can be straightforwardly extended.

In our two-fold investigation, we will first consider the Higgs boson production
in gluon-fusion and in vector-boson fusion, as examples of
processes that are not included in a PDF fit, where the effect of correlations is
expected to be milder.
We then consider top quark pair production and the $Z$ and $W$ electroweak gauge boson production,
as examples of processes included in a PDF fit, where the effect of correlations
is expected to be stronger.

\subsection{Higgs production}
\label{sec:higgscorrelations}
We start by discussing Higgs production in gluon fusion (ggF) and in vector boson fusion (VBF).
These two processes are of direct relevance for the characterisation of the Higgs sector and
are both currently known at N$^3$LO
accuracy~\cite{Anastasiou:2015vya,Anastasiou:2016cez,Mistlberger:2018etf,Dreyer:2016oyx}. The
processes are complementary in two respects. On the one hand, Higgs production in
gluon fusion is driven by the gluon-gluon luminosity and its perturbative
expansion converges slowly. On the other hand vector boson fusion is driven by the quark-
anti-quark luminosity and it exhibits fast perturbative convergence.

In Table~\ref{tab:higgs-gluon-fusion} we present predictions for Higgs production in ggF 
at the LHC for $\sqrt{s}=$ 13 TeV. Given that we focus on the effect of correlations between the
factorisation scale used in the PDF replicas and the factorisation scale used in the partonic cross
section, it is important to perform them at the same order, i.e. at NLO. We perform the
calculation of ggF at NLO in the rescaled effective theory
approximation using {\tt ggHiggs}~\cite{Bonvini:2014jma,Bonvini:2016frm} with
$\mu_f^0=\mu_r^0=m_H/2$, with the {\tt MCscales with postfit} set obtained in this
paper. In the first line, we use the fully correlated
prescription, Eqs.~\eqref{eq:sigmacorr} and \eqref{eq:deltacorr}, while
in the second and third lines we use the prescriptions that use less information,
namely the uncorrelated prescription, Eqs.~\eqref{eq:sigmauncorr} and
~\eqref{eq:deltauncorr}, in which the scale uncertainty in
the PDFs is combined with the scale uncertainty in the partonic cross section in an uncorrelated
way, and finally the prescription in which we ignore scale uncertainty
in PDFs, as in Eqs.~\eqref{eq:sigmacentral} and~\eqref{eq:deltacentral}. 
The comparison between the first and the
second lines gives us an estimate of the size of correlations between
scales and the double-counting of scale variation in the PDFs and in
the partonic partonic cross sections, 
while the comparison between the first two lines and the third gives us an idea on the effect
of neglecting scale uncertainty in PDF fits.
%
%-------------------------------------------------------------------------------
\begin{table}[!htbp] \centering
\begin{tabular}{l|c}
\multicolumn{2}{c}{Higgs production by gluon fusion at 13 TeV [pb]}          \\
\hline
                                  & NLO  \\%& $\hat{\sigma}$ at NNLO, PDF at NLO               \\
\hline
  $\bar{\sigma}^{\rm(corr)}\,\pm\,\Delta ^{\rm(corr)}$    & 38.10 $\pm$ 19.5\% \\
  % & 46.91 $\pm$ 9.50\% \\
\hline
  $\bar{\sigma}^{\rm (uncorr) }\,\pm\,\Delta^{\rm (uncorr)}$      & 37.41 $\pm$ 17.9\% \\
  %(1.99\% (PDF), 17.78\% (scale))\\%& 47.12 $\pm$ 8.74\% \\
\hline
  $\bar{\sigma}^{\rm (c) }\,\pm\,\Delta^{\rm (c)}$   & 37.64 $\pm$ 17.7\% \\
  % (1.18\% (PDF), 17.67\% (scale)) \\%& 47.41 $\pm$ 8.52\% \\
\hline
\end{tabular}
\caption{The central value and the total theoretical uncertainty for the Higgs production
  cross section by gluon fusion at NLO for $\sqrt{s}=13$ TeV. The settings are indicated in the
  text. All results are produced with NLO PDFs, and the central values and total theoretical
  uncertainty are computed according to three different prescriptions. $\Delta ^{\rm(corr)}$ is the
  correlated theoretical uncertainty in which the scale variation in the PDF fit are correlated
  with the scale variations in the computation of the partonic cross
  section, Eqs.~\eqref{eq:sigmacorr} and \eqref{eq:deltacorr}.
  $\Delta^{\rm (uncorr)}$ is the theoretical uncertainty
  in which the scale variations in the PDF fit are uncorrelated
  with the scale variations in the computation of the partonic cross
  section, Eqs.~\eqref{eq:sigmauncorr} and \eqref{eq:deltauncorr}.
  $\Delta^{\rm (c)}$ is the total theory uncertainty in which scale uncertainty is included
  only in the partonic cross section, 
  Eqs.~\eqref{eq:sigmacentral} and \eqref{eq:deltacentral}.
  The NNLO central value is $46.53$ pb. }
\label{tab:higgs-gluon-fusion}
\end{table}
%-------------------------------------------------------------------------------
%

\begin{figure}[!htbp] \centering
\includegraphics[width=0.8\textwidth]{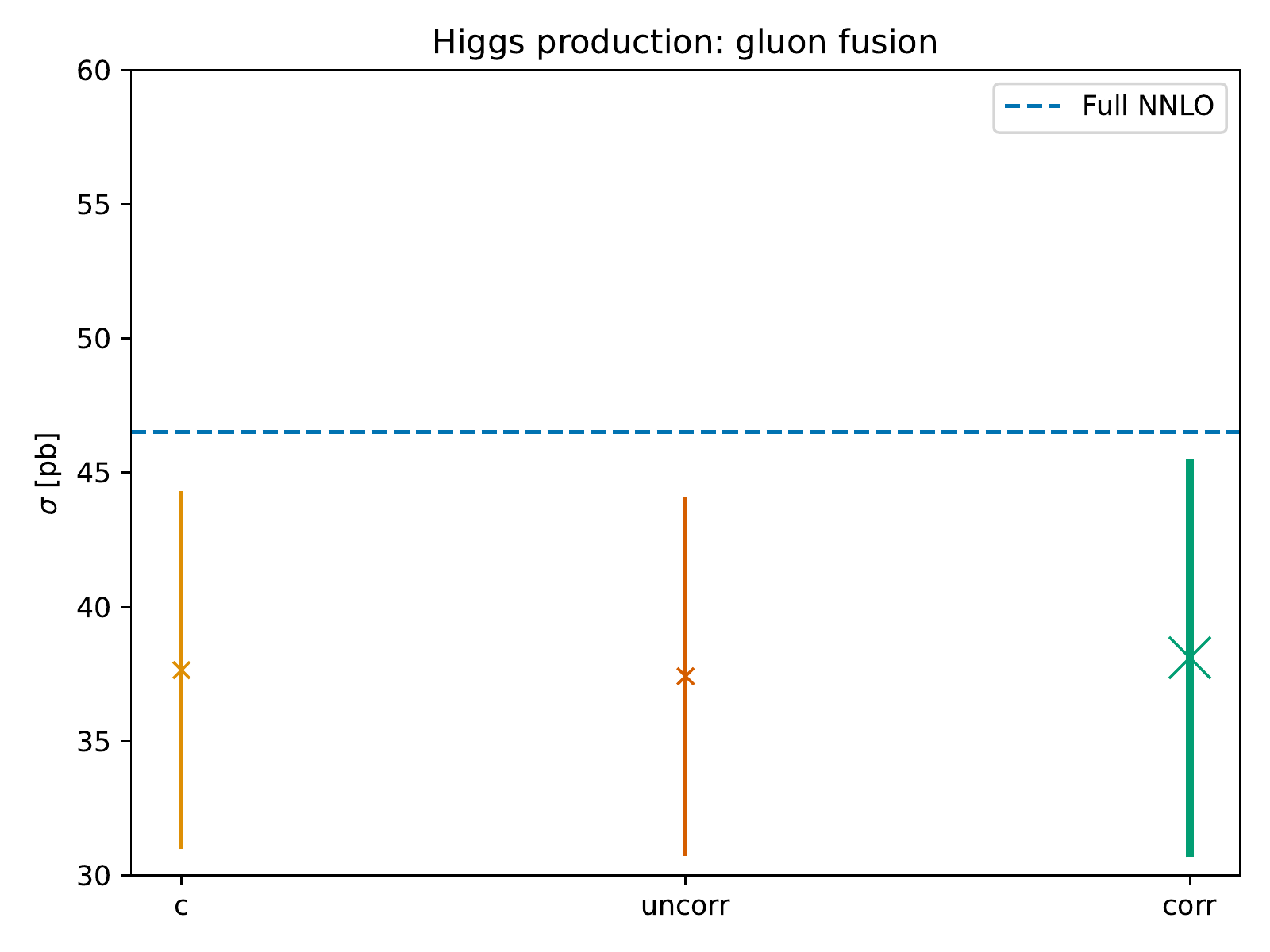}
\caption{NLO total cross sections for the Higgs production
  cross section by gluon fusion at NLO for $\sqrt{s}=13$ TeV.
  The uncertainty on each result is the
  total  theoretical uncertainty, which includes both PDF and scale
  uncertainties according to the three different prescriptions
  described in the text and in Table~\ref{tab:higgs-gluon-fusion}. The
  NNLO result, which uses the NNLO {\tt NNPDF3.1} baseline PDFs, is also displayed. }
\label{fig:higgs-gluon-fusion}
\end{figure}

We observe that in this case not accounting for the correlation between the factorisation
scale variation in the PDFs and the factorisation scale variation in the computation
of the Higgs cross section underestimates the theory uncertainty by a non-negligible amount.
The total theory uncertainty increases by nearly 2\% in absolute value. This finding
contradicts the somewhat intuitive statement according to which the computation of the
total theory uncertainty in an uncorrelated way is a conservative estimate of theory
uncertainties. In this case the correlation between the factorisation scale
variation in the PDF fit and those in the computation of ggF Higgs production cross section
does increase the total uncertainty, rather than decreasing it.
Results are displayed graphically in Fig.~\ref{fig:higgs-gluon-fusion} where,
for the NNLO computation, we also show the central value found using NNLO PDFs.
%-------------------------------------------------------------------------------
%-------------------------------------------------------------------------------
We observe that the central value of the NLO cross section computed over the \mcscales set
according to the correlated prescription gets closer to the next perturbative order.

\begin{figure}[!htbp] \centering
\includegraphics[width=0.8\textwidth]{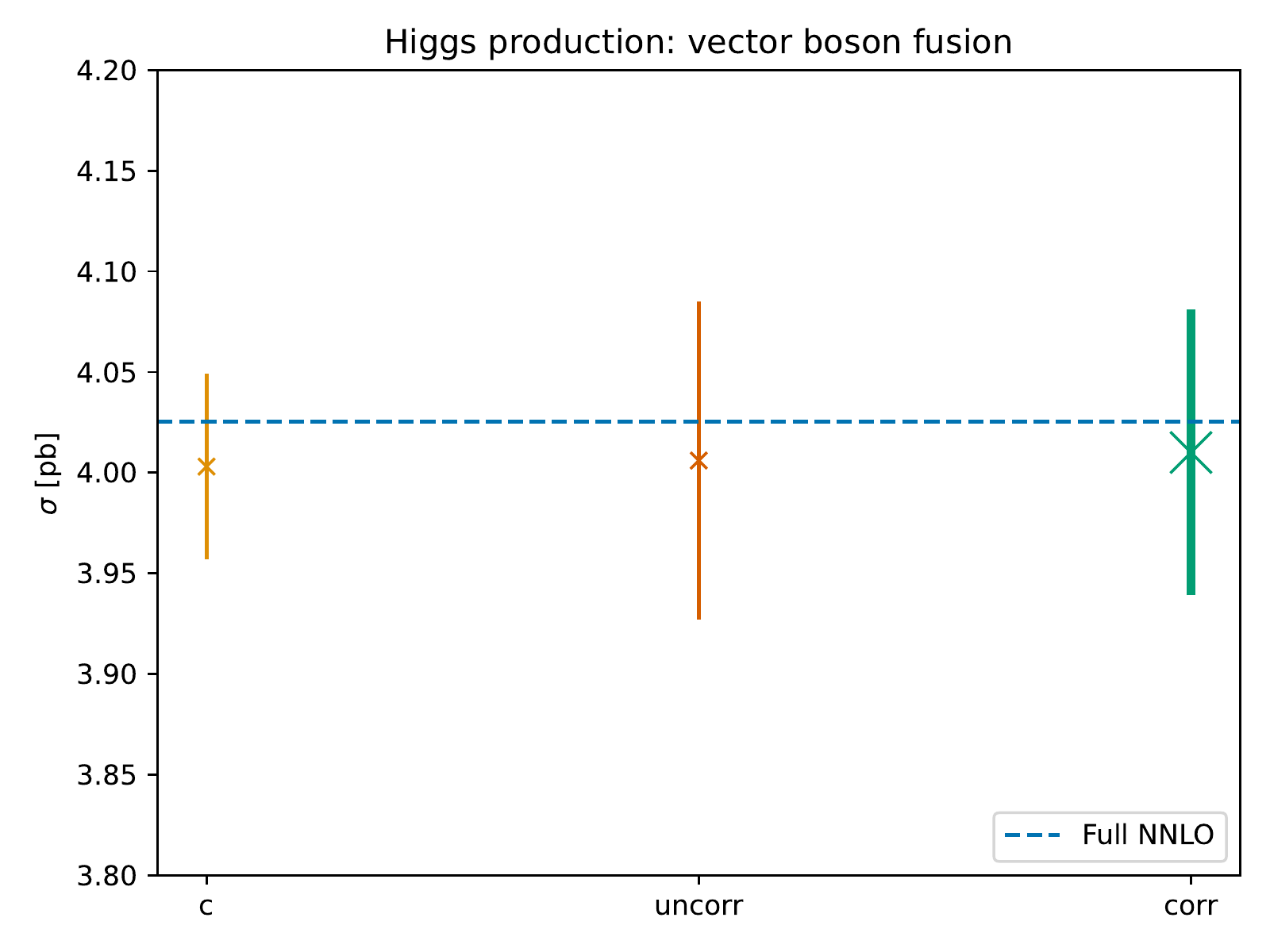}
\caption{The same as Fig.~\ref{fig:higgs-gluon-fusion} but for Higgs production by vector boson
  fusion at 13 TeV.}
\label{fig:higgs-vector-boson-fusion}
\end{figure}

We now turn to Higgs production in vector boson fusion. We perform the NLO calculation
using proVBFH-inclusive~\cite{Cacciari:2015jma} with central factorisation and normalisation
scales set equal to the squared four-momentum of the vector boson. Results are collected in
Table~\ref{tab:higgs-vbf-fusion} and shown in Fig.~\ref{fig:higgs-vector-boson-fusion}.
We observe that in this case the inclusion of the correlation between the factorisation
scale variations in the Higgs production partonic cross section and
those in the PDFs does 
decrease the scale uncertainty by a non-negligible amount. On the other hand, we also observe that
not including the effect of theory uncertainties in PDFs would significantly
underestimate theory uncertainties in this process.
%-------------------------------------------------------------------------------
\begin{table}[!htbp] \centering
\begin{tabular}{l|c}
\multicolumn{2}{c}{Higgs production by virtual boson fusion at 13 TeV [pb]}          \\
\hline
                                  & NLO    \\%             & NNLO               \\
\hline
$\bar{\sigma}^{\rm(corr)}\,\pm\,\Delta ^{\rm(corr)}$      & 4.010 $\pm$ 1.77\% \\%& 3.962 $\pm$ 2.12\% \\
\hline
$\bar{\sigma}^{\rm (uncorr) }\,\pm\,\Delta^{\rm (uncorr)}$     & 4.006 $\pm$ 1.97\% \\%& 3.957 $\pm$ 2.0\% \\
  \hline
$\bar{\sigma}^{\rm (c) }\,\pm\,\Delta^{\rm (c)}$               & 4.003 $\pm$ 1.15\% \\%& 3.950 $\pm$ 1.14\% \\
  \hline
\end{tabular}
\caption{Same as Table~\ref{tab:higgs-gluon-fusion} for the NLO total cross sections for the Higgs production
  cross section by vector boson fusion at NLO for $\sqrt{s}=13$
  TeV. In this case, the NNLO central value is $4.025$ pb.}
\label{tab:higgs-vbf-fusion}
\end{table}
%-------------------------------------------------------------------------------

Results are displayed graphically in Fig.~\ref{fig:higgs-vector-boson-fusion}, in which we see that
the uncertainty bands increases once scale uncertainty in PDFs is accounted for, and slightly
decreases once correlations are fully taken into account. Also in this case the central value of the NLO cross section computed over the \mcscales set
according to the correlated prescription gets closer to the next perturbative order. 
%-------------------------------------------------------------------------------
%-------------------------------------------------------------------------------

\subsection{Top pair production}
We now turn to the impact of PDF-related scale uncertainty and the effect of
the correlations between scale variations in the PDFs and in the partonic
cross section by computing the total top-pair production cross section at the LHC at
$\sqrt{s}=$ 13 TeV.

\begin{figure}[!htbp] \centering
\includegraphics[width=0.8\textwidth]{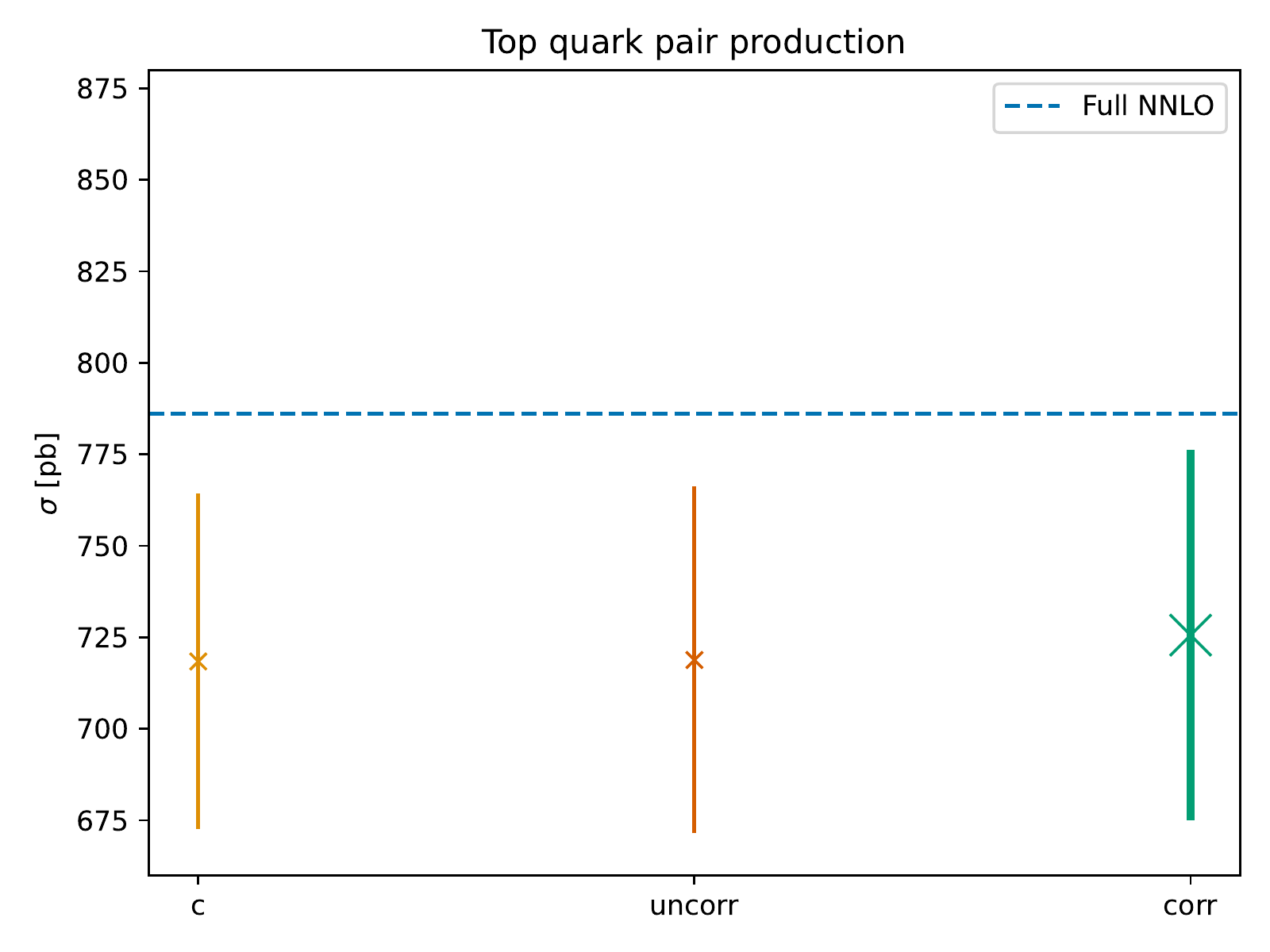}
\caption{The same as Fig.~\ref{tab:higgs-gluon-fusion} but for
  the top quark pair production at $\sqrt{s}=13$ TeV.}
\label{fig:top-pair}
\end{figure}

\begin{table}[!htbp]
  \centering
  \begin{tabular}{l|c}
    \multicolumn{2}{c}{Top pair production at 13 TeV [pb]}          \\
\hline
&  NLO \\
\hline
$\bar{\sigma}^{\rm(corr)}\,\pm\,\Delta ^{\rm(corr)}$      & 725.6 $\pm$ 6.98\%  \\ %& 802.2  $\pm$ 2.78\%           \\
\hline
$\bar{\sigma}^{\rm (uncorr) }\,\pm\,\Delta^{\rm (uncorr)}$ & 718.8 $\pm$ 6.59\%  \\ %&  800.1 $\pm$ 3.24\%         \\
\hline
$\bar{\sigma}^{\rm (c) }\,\pm\,\Delta^{\rm (c)}$   & 718.4 $\pm$ 6.38\% \\ %&  799.7  $\pm$  2.78\%        \\
\hline
\end{tabular}
\caption{Same as Table~\ref{tab:higgs-gluon-fusion} for the NLO total cross sections for
  the top pair production cross section NLO for $\sqrt{s}=13$
  TeV. In this case, the NNLO central value is $786.0$ pb.}
\label{tab:ttbar}
\end{table}
In Table~\ref{tab:ttbar} we collect, using
the same format as Table~\ref{tab:higgs-gluon-fusion} , the predictions for the top-quark
pair-production cross-sections at $\sqrt{s}$ = 13 TeV obtained using the {\tt top++}
code~\cite{Czakon:2011xx} and setting the central scales
to $\mu_f^0 = \mu_f^0= m_t=$ 173.3 GeV. The results are also displayed in
Fig.~\ref{fig:top-pair}, where again for at NNLO we also show the
result obtained using {\tt NNPDF3.1} NNLO baseline PDFs. 

%-------------------------------------------------------------------------------
%-------------------------------------------------------------------------------
We observe that in this case the effect of the correlation is almost
negligible. As in the case of Higg production via gluon fusion, the
correlation is negative and its inclusion increases the total
theoretical uncertainty, although by a small amount in this case.
This is in contrast with what is observed in Ref.~\cite{Ball:2021icz}, possibly
due to a different methodology to compute the matched predictions. In
this paper we simply match the subsets of PDF replicas to the partonic
cross section that set matching values of scale, while in
~\cite{Ball:2021icz} the theoretical predictions are computed using
different definitions of the theory covariance matrix and a direct
comparison is not straightforward.

\subsection{Vector boson production}
\begin{figure}[!htbp] \centering
\includegraphics[width=0.6\textwidth]{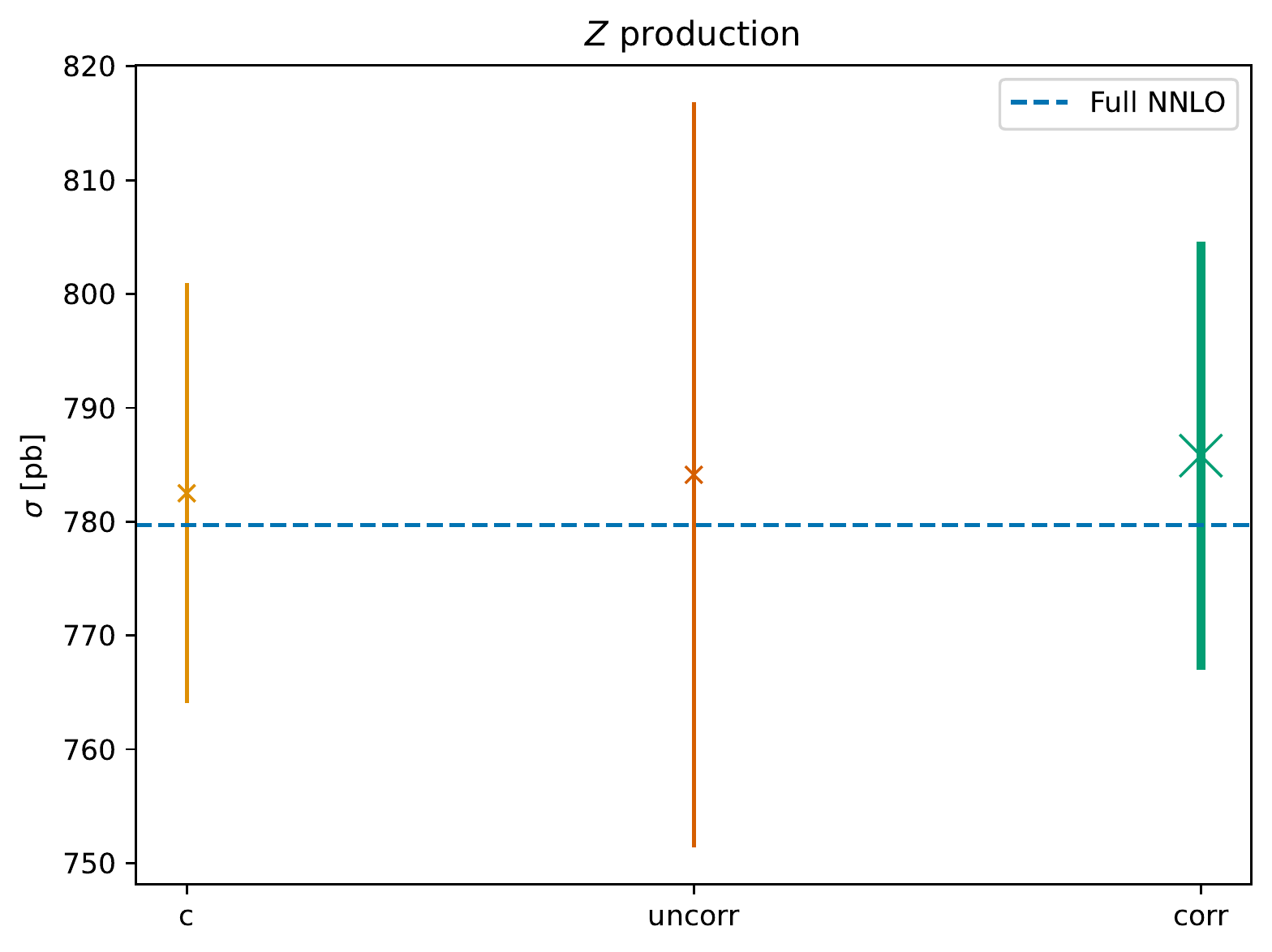}\\
\includegraphics[width=0.6\textwidth]{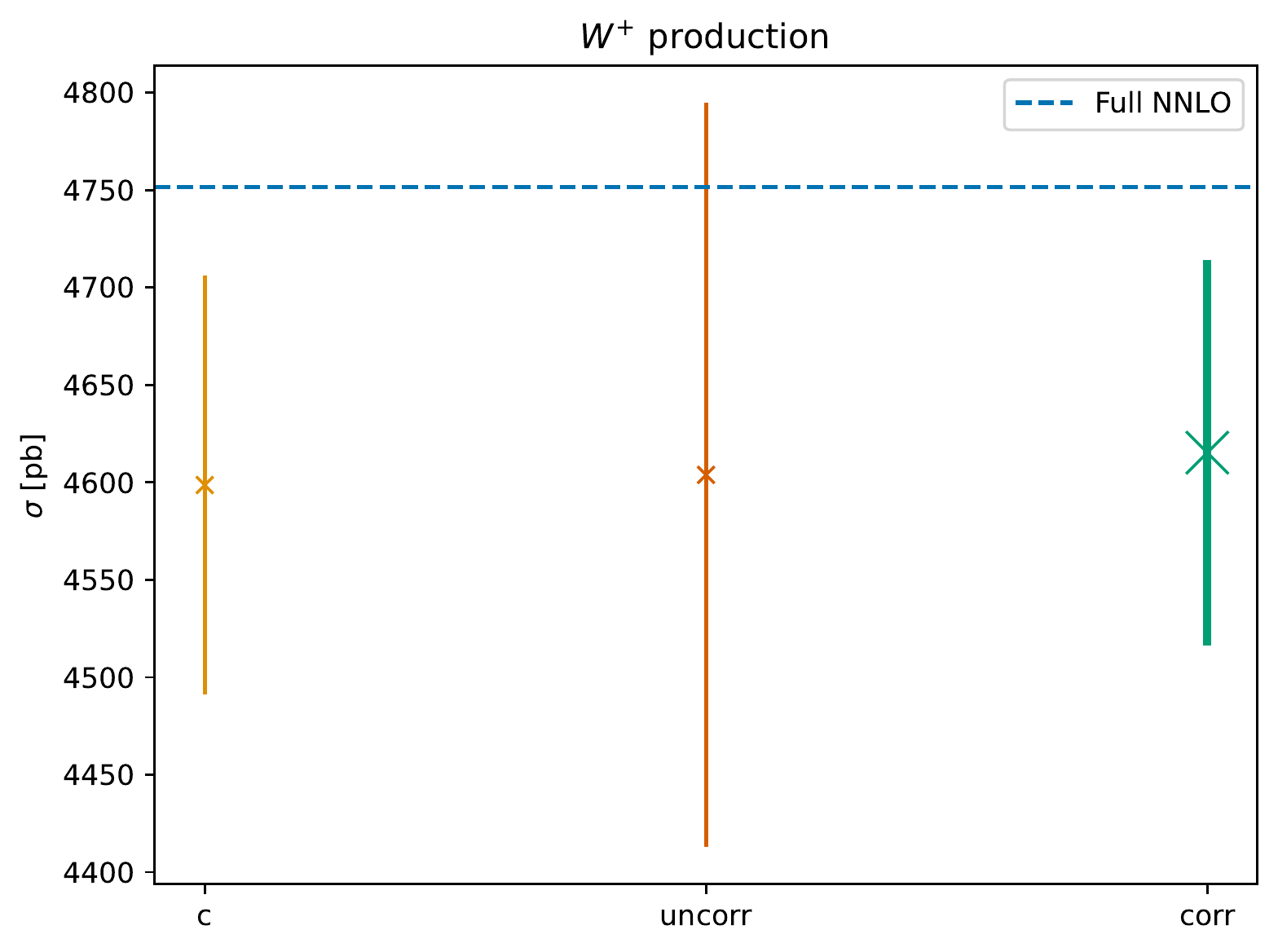}\\
\includegraphics[width=0.6\textwidth]{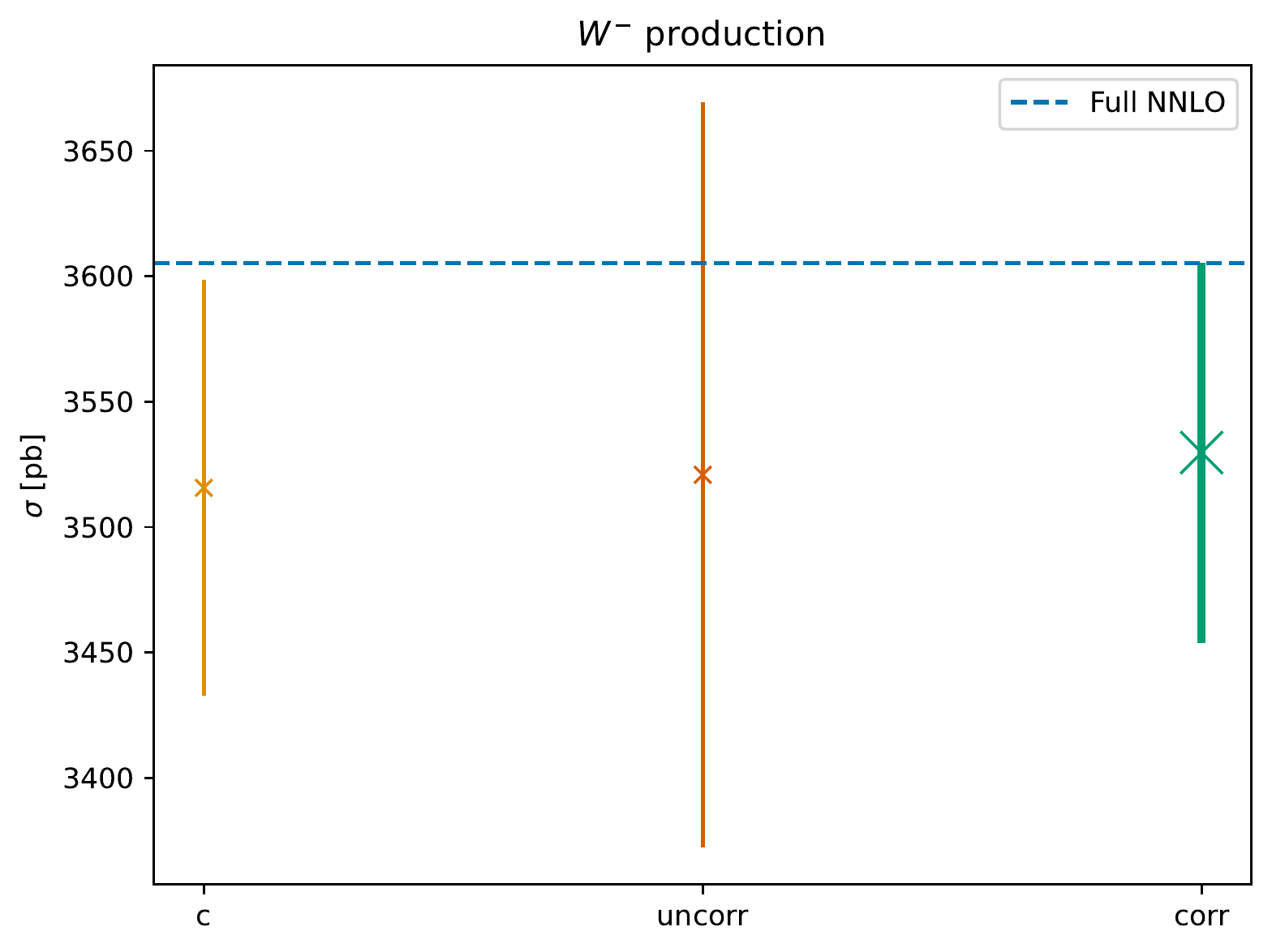}
  \caption{The same as Fig.~\ref{tab:higgs-gluon-fusion} but for $Z$ (top), $W^{+}$ (middle) and
  $W^-$ (bottom) production at $\sqrt{s}=$13 TeV.}
\label{fig:wz-prod}
\end{figure}

We finally turn to gauge boson production, for which we obtain predictions using
{\tt FEWZ}~\cite{Li:2012wna,Gavin:2012sy,Gavin:2010az}.
For the theoretical predictions for inclusive $W$ and $Z$ production cross sections at
$\sqrt{s}$ = 13 TeV, we adopt realistic kinematic cuts similar to those applied
by ATLAS and CMS. The fiducial phase space for the $W^\pm$ cross-section is defined by requiring
$p_T^l \ge$ 25 GeV and $\eta_l\le$ 2.5 for the charged lepton transverse momentum and pseudo-rapidity
and a missing energy from the neutrino of $p_T^\nu\ge $ 25 GeV. In the case of $Z$ production, we
require $p_T^l\ge$ 25 GeV and $|\eta_l|\le$ 2.5 for the charged leptons transverse momentum and
rapidity and 66 $\le m_{ll}\le$ 116 GeV for the di-lepton invariant mass.
\begin{table}[!htbp]
  \centering
  \begin{tabular}{l|c|c|c}
    \multicolumn{4}{c}{Vector boson production at LHC 13 TeV [pb]}          \\
\hline
&  Z (NLO) & $W^+$ (NLO)  & $W^-$ (NLO) \\
\hline
$\bar{\sigma}^{\rm(corr)}\,\pm\,\Delta ^{\rm(corr)}$       & 785.8 $\pm$2.40\% & 4615.3$\pm$2.14\% &  3529.7$\pm$2.15\%         \\
\hline
$\bar{\sigma}^{\rm (uncorr) }\,\pm\,\Delta^{\rm (uncorr)}$  & 784.1$\pm$4.17\%  & 4603.86$\pm$4.15\% &3520.9$\pm$4.21\%\\
\hline
$\bar{\sigma}^{\rm (c) }\,\pm\,\Delta^{\rm (c)}$ & 782.5 $\pm$2.36\%  & 4598.6$\pm$2.34\% &  3515.7$\pm$2.36\%\\
\hline
  \end{tabular}
\caption{Same as Table~\ref{tab:higgs-gluon-fusion} for the NLO total cross sections for
  vector boson production cross section NLO for $\sqrt{s}=13$
  TeV. The NNLO result are  779.7 pb, 4751.6 pb and 3605.2 pb for $Z$, $W^+$, $W^-$
  production respectively.}
\label{tab:WZ}
\end{table}
In Table~\ref{tab:WZ} we display a similar comparison as in Table~\ref{tab:higgs-gluon-fusion}
now for $W$ and $Z$ gauge boson production at $\sqrt{s}$ = 13 TeV.
The corresponding graphical representation of the results is provided in Fig.~\ref{fig:wz-prod},
again using the same conventions as in Fig.~\ref{fig:higgs-gluon-fusion} and
again showing the NNLO result with NNLO PDFs.

% -------------------------------------------------------------------------------
It is very interesting  to observe how in this case the correlation
between the scale variation in the PDF set and the scale variation in the
partonic cross section decreases the total theoretical uncertainty by about a factor of 2,
while bringing the result closer to the next perturbative order. This
happens both for neutral and charged current Drell-Yan. 
In this case not including the correlation between the scale variation in the PDF
set and the scale variation in the partonic cross section grossly overestimates the
theoretical uncertainty of a precision observable. 
This is a very interesting result as the positive correlation between
the scale variation in the PDFs and the scale variation in the
computation of the partonic cross section makes the theoretical
uncertainty smaller than in the case in which the scale uncertainty in
PDFs is completely neglected.

\section{Delivery}
\label{sec:delivery}

In this section we present the deliverables of this work. They consist of the
{\tt MCscales} PDF, which is delivered as a {\tt LHAPDF} grid with enhanced
metadata containing information on scale variations, as well as a set of tools
to operate on it.  The main deliverable of this work is PDF set obtained as
described in Sect.~\ref{sec:data-driven}, dubbed
\begin{center}
    {\tt mcscales\_v1} .
\end{center}
It can be downloaded from the following URL:
\begin{center}
    \href{https://data.nnpdf.science/pdfs/mcscales_v1.tar.gz}{\tt https://data.nnpdf.science/pdfs/mcscales\_v1.tar.gz} .
\end{center}

This PDF set is based on the NNPDF 3.1~\cite{Ball:2017nwa} global analysis
with the same minor changes in dataset selection made in
Ref.~\cite{AbdulKhalek:2019ihb}. The PDF set contains 823 replicas, which have
been fitted with a different choice of factorisation and renormalisation
scales for the input data. This allows computing predictions with matched
scale variations, as discussed in Sec.~\ref{sec:maincalc}, such as those used
in the phenomenology studies in Sec.~\ref{sec:pheno}. These computations
could be implemented directly on codes, with the help of the metadata
information stored in the {\tt LHAPDF} grids discussed next, or executed manually
on any code that allows choosing PDFs and scales, with the help of a script
to split the grids, which we also provide and describe briefly below.

The metadata of the {\tt LHAPDF} grid of the {\tt mcscales\_v1} set has been
enhanced with the following keys in the global {\tt .info} file affecting the
full PDF set:
\begin{description}
    \item[{\tt mcscales\_processes}] A list of strings containing the
    processes for which separate renormalisation scales are considered, as
    discussed in Sec.~\ref{sec:the-sampling-model}. Its current value is [DIS
    CC, DIS NC, DY, JETS, TOP].
    \item[{\tt mcscales\_scale\_multipliers}] A list of numbers with the
    possible scale multiplier choices, Eq.~\ref{eq:scalemultchoices}. Its current value
    is [0.5, 1, 2].
\end{description}
The metadata header for each of replica grids (that is, all the members of
the {\tt LHAPDF} set except the data member 0, which corresponds to the central
value) has been enhanced with the following keys:
\begin{description}
    \item[{\tt mcscales\_ren\_multiplier\_<PROCESS>}]: One key for each of
    the values of {\tt mcscales\_processes} (i.e. {\tt
    mcscales\_ren\_multiplier\_DIS CC},
    {\tt mcscales\_ren\_multiplier\_DIS NC},\break {\tt mcscales\_ren\_multiplier\_DY}, {\tt mcscales\_ren\_multiplier\_JETS} and \break{\tt
    mcscales\_ren\_multiplier\_TOP}) with the value of the renormalisation
    scale multiplier for a given process and the current replica. The
    possible values are one of the entries in {\tt
    mcscales\_scale\_multipliers}.
    \item[{\tt mcscales\_fac\_multiplier}] The factorisation scale multiplier for the current replica. The
    possible values are one of the entries in {\tt
    mcscales\_scale\_multipliers}.
\end{description}
This metadata, accessible with the LHAPDF library, should allow for
incorporating support for the matched scale convolution described in
Sect.~\ref{sec:maincalc}: Users can split the replicas by scale multipliers as
appropriate for the computation and match the scale variation for the hard
cross section with each group, in order to implement either
Eq.~\ref{eq:mcscalessample} or Eq.~\ref{eq:mcscalessample-noren}.

In addition, we provide a set of tools to manipulate {\tt MCscales} PDFs. These are available from 
\begin{center}
    \href{https://github.com/Zaharid/mcscales_tools/}{\tt https://github.com/Zaharid/mcscales\_tools/}
    \ .
\end{center}

The repository contains two scripts, along with usage instructions. Here we
provide a brief overview.

The first one {\tt mcscales-partition-pdf} is used partition one {\tt
MCscales} PDF into several PDF sets with identical renormalisation or
factorisation scale choices across all replicas in a given set. These can be
either three sets, split by factorization scale variation, and with arbitrary
renormalisation scales, or nine sets, split by factorisation or
renormalisation scale variations for a given process, which is given as
input. This allows implementing matched convolution computations with existing
tools. The computation of a given cross section is repeated for all PDF sets
in the split, taking care to compute the hard cross section with the matched
scale variation for each set. Eqs.~\ref{eq:mcscalessample}
and~\ref{eq:mcscalessample-noren} can then be implemented with a nine way
(factorization and renormalisation scale for a given process) and a tree way
(factorisation scale) split respectively. This strategy was used to obtain
the results in Sec.~\ref{sec:pheno} with a variety of existing codes.

The second script, {\tt mcscales-resample-pdf}, allows to experiment with
simple variatons of the prior scale choices. It generates a new {\tt
MCscales} PDF set by vetoing particular scale combinations from an original
{\tt MCscales} PDF set. We note that owing to the particularly simple form of
the choice of postfit likelihood in Eq.~\ref{eq:postfit}, this is identical
to modifying the prior distribution of scales (as long as central scale
choices are kept). The script allows reproducing the PDF sets shown in
Sect.~\ref{sec:theory-driven}, with different variations on the prior. Users
are encouraged to modify the source code of this script in order to
investigate more general scale priors, such as those discussed in
Appendix~\ref{app:abc}.

%The final deliverable is a Monte Carlo set, much like a typical NNPDF set,
%where each replica contains information not only on the experimental
%uncertainty and the degeneracy of the PDF parametrization, but also on
%the scale variation uncertainty. These replicas can be used to compute
%statistics over ensembles of observables, accounting for the combined
%PDF and scale uncertainty. For example, given a hard cross section of
%the process type \(p\), which is also included in the PDF fit, we would
%partition the replica sample into subsets with identical values for
%\(k_{r_p}\), and convolve those with the appropriately matched
%renormalisation scale in the hard cross section \(\hat{\sigma}(k_{r_p})\).
%We would construct the Monte Carlo sample of observables
%
%TODO: Deduplicate with sec 5
%
%\begin{equation}
%\label{eq:ensembles}
%\bigcup_{\xi=\{\frac{1}{2},1,2\}}\left\{
%\hat{\sigma}(k_{r_p}=\xi)\otimes f_{k\left(k_f,k_{r_{1}}\ldots k_{r_p}\ldots s_{N_{\text{p}}}\right)} \, | \, k_{r_{p}}=\xi
%\right\}
%\end{equation} and apply the usual statistical estimators over the
%resulting ensemble.
%
%{\bf What will be made public? Where can it be found? Where can instructions be
%found?}
%

\section{Conclusion} \label{sec:conclusion}

%Prospect of NNLO PDFs with this method.
In this work we have presented a PDF determination that includes 
scale uncertainties by expanding the Monte Carlo sampling method for
propagating experimental uncertainties to PDFs to the space of factorisation and
renormalisation scales.
In this approach, dubbed {\tt MCscales}, the scales are
viewed as free parameters of the fixed-order theory that should be chosen
jointly so as to produce theory predictions that are compatible with
experimental data included the PDF fit.

This method has several key benefits. First of all, the information
provided in LHAPDF sets is extended to include the scales used to
produce each PDF replica. As a result, external users are able to
modify the PDF set based on their preferences. Second, this added scale
information can be utilised to correlate scale variations in partonic cross
sections with those in the {\tt MCscales} PDF set that we provide.

Having explicitly set out the assumptions used to construct a prior distribution of scales in the
{\tt MCscales} approach, we have demonstrated that both experimental data and theoretical considerations can
be used to constrain the prior. 

Our procedure provides a way to benchmark the suitability of different scales
variations across all the data in PDF fits, allowing to educe a quantitative
underpinning from this traditional procedure. We observed that the customary
variation range of a factor of two around a central scale performs differently
when applied to factorisation and renormalisation scales. For factorisation
scales, a variation of a factor 2 adequately covers the set of values that are
likely to yield a satisfactory fit to the data.  This conclusion however relies on the
fitting of the PDF replicas and post selection procedure described in
Sect.~\ref{sec:the-method}. Indeed, performing such scale variations without our
procedure implies PDFs that are not in agreement with the experimental data,
which is the only source of information on precise constraints of PDFs, and
therefore, unphysical predictions.
We observed that this is not the case for renormalisation scale
variations of a factor of two, which instead exhibit a rather uniform fit
quality for the three choices. This suggests that the theoretical uncertainty
on the renormalisation scale might be underestimated, at least at NLO and with
appropriate matching between PDF and hard cross section. We leave for
future work to study scale variation ranges in more detail.
Furthermore we observed that the behaviour of scale choices was
highly variable across different processes, suggesting that a method that treats each process in the same way (as in the theory
covariance matrix approach) may be undesirable. At the PDF level, the
{\tt MCscales} approach appears to lead to rather larger PDF
uncertainties. However, since in the {\tt MCscales} approach each PDF replica uses theoretical predictions
with matched scale combinations, comparisons at the PDF level
are misleading. Indeed we noticed that convolutions with matched scales
can both increase and decrease the uncertainty on cross sections as
compared to the uncertainties on PDF themselves. 

We exploited the possibility of modifying the prior distribution of scales by
simply choosing to omit certain scale combinations, producing {\tt MCscales}
variants based on priors that include only the 7-points in the
renormalisation and factorisation scale space that are included in the widely
adopted 7-points envelope. This led to us observing that excluding the scale
combinations with ratios of four or one-quarter (i.e. those excluded in
7-point) drives a significant reduction in the MCscales PDF uncertainties.

Finally, we moved to phenomenology by, firstly, setting out recipes for making predictions with
MCscales PDF sets, and then applying these to processes relevant to the LHC. We demonstrated
that it is possible to include correlations between scale variations in the hard cross section and those
in the PDF, whether or not the process in question is included in the PDF fit. 
As with the theory
covariance matrix approach, the inclusion of scale uncertainties with {\tt MCscales} leads generally to more accurate
predictions. The increase in uncertainty induced by their inclusion (if indeed there is an increase) is
typically slightly larger when using {\tt MCscales} than the theory covariance matrix, though they remain
mild, at around the 1\% level.
Many developments are possible on the common assumptions of the two methods, including a more refined
factorisation scale variations or more nuanced process categorisations. Such extensions are possible
avenues for future research.
That said, the most pressing phenomenological development will be
the extension of both methodologies to the determination of MHOUs in a
state-of-the-art NNLO PDF set. Indeed, the inclusion of different sources of theoretical
uncertainties at NNLO is expected to lead to the most precise and accurate
PDF sets currently attainable.
The {\tt MCscales} methodology could be naturally extended to handle these uncertainties.
\appendix
\section{The $\alpha\beta\gamma$ model}
\label{app:abc}

As it was mentioned in the paper, the \mcscales approach allows the
users to select a specific {\it prior} probability in the space of
factorisation and renormalisation scales. Any user could apply any
symmetric or asymmetric choice of scales depending on their
theoretical preference. In this Appendix we outline a particular
theoretical model, dubbed the $\alpha\beta\gamma$ model, that a user
could choose to
construct a more convolved prior probability than  the uniform priors
presented in the main
paper and tune it according to the user's theoretical assumptions on the
correlation between processes. The model can be seen as a generalisation for
the prior choices considered in Sect.~\ref{sec:data-driven}, which can be
recovered for certain values of the parameters. In the second part of the
Appendix, we determine the value of the $\alpha\beta\gamma$ parameters that the
data included in the fit tend to prefer.

\subsection{Prior assumptions}
\label{assumptions}

We construct the model for the prior probability by imposing a
set of theoretically-motivated symmetries. As we will show below, we can
characterise the prior probability for
each theory hypothesis in terms of a set of total probabilities for each
scale, defined as the sum of the probabilities of all the hypotheses
where a particular scale variation \(k_i\) has a given value, \(x\),
\begin{equation}
P(k_{i}=x)=
\sum_{\{\omega \in \Omega | \xi_i=x\}}
P(\omega)\ ,
\end{equation}
and set of conditional probabilities
\begin{equation}
P(k_i = x | k_j = y) = \frac{1}{P(k_j=y)}
\sum_{\{\omega \in \Omega|\xi_i=x, \xi_j=y\}}
P(\omega)\ .
\end{equation}

\begin{description}
\item[Symmetry between renormalisation and factorisation scales]
When considering only one process, we assume that the probability is
invariant under the exchange of the factorisation and the renormalisation scales.
Thus, when sampling independently, we have,
\begin{equation}
  P(k_f=\xi) = P(k_r=\xi) \ ,
\end{equation}
with the conditional probabilities also being symmetric,
\begin{equation}
  P(k_f=\xi_x | k_r=\xi_y) = P(k_r=\xi_x|k_f=\xi_y) \ . \label{eq:dacrensymm}
\end{equation}
\item[Symmetry between upper and lower variations]
We assume that the probabilities are unchanged if we flip all of the upper
and lower scale variations. That is, given
\begin{equation}
  \text{flip}(\xi)=\begin{cases}
  2 & \xi=\frac{1}{2}\\
  1 & \xi=1\\
  \frac{1}{2} & \xi=2
  \end{cases} \ ,
\end{equation}
we assume
\begin{equation}
  \label{eq:flipsymm}
  P\left(\xi_{f},\xi_{1},\ldots, \xi_{N_\text{p}}\right) =
  P\left(\text{flip}(\xi_{f}),\text{flip}(\xi_{1}),\ldots, \text{flip}(\xi_{N_\text{p}})\right) \ .
\end{equation}
\item[Conditional independence between renormalisation scales]
We assume that the probabilities of sampling different renormalisation
scales are conditionally independent given the factorisation scale. That is, for $i \neq j$,
\begin{equation}
  \label{eq:independence}
  P(k_{r_i}=\xi_i | k_f=\xi_f, k_{r_j}=\xi_j) = P(k_{r_i}=\xi_i | k_f=\xi_f)\ .
\end{equation}
Note that this does not imply
\(P(k_{r_i}=\xi_i | k_{r_j}=\xi_j) = P(k_{r_i}=\xi_i)\) since knowledge of \(k_{r_j}\)
provides knowledge on \(k_f\) and thus renormalisation scale variations
are in principle correlated with each other.
\item[Symmetry between renormalisation scales]
We assume that the probabilities are unchanged under permutations
of the values of the renormalisation scales. In particular, this implies
\begin{equation}
  \label{eq:rensym}
  P(k_{r_i}=\xi) = P(k_{r_j}=\xi)
\end{equation}
and
\begin{equation}
  P(k_{r_i}=\xi|k_f=\xi_f) = P(k_{r_j}=\xi|k_f=\xi_f)
\end{equation} for all \(i,j\in1,\ldots,N_\text{p}\).
\end{description}

%%%%%%%%%%%%%%%%%%%%%%%%%%%%%%%%%
%\subsubsection{Prior model parametrization}
%\label{model-parametrization}

Because we assume that renormalisation scale variations are independent
(Eq.~\eqref{eq:independence}), we can give the probability of any set of
scale variations in terms of the total probability of sampling a given
value of the factorisation scale and conditional probabilities for
sampling a renormalisation scale given the factorisation scale:
\begin{equation}
  P(k_f=\xi_{f},\ldots,k_{r_{N_\text{p}}}=\xi_{N_\text{p}})=P(k_f=\xi_{f})
  \prod_{i=1}^{N_\text{p}}P(k_{r_i}=\xi_{i}|k_f=\xi_{f})
  \ .
  \label{eq:modelprob}
\end{equation}
Let us now count how many parameters we must specify to define
all of the probabilities. To begin, we have to specify three values for \(P(k_f)\). Further, because
we assume that the model acts the same for all renormalisation scales
(Eq.~\eqref{eq:rensym}), we have to specify nine values for the conditional
probabilities, one for each combination of renormalisation and
factorisation scale variation. These 12 total values are not independent, as
they are constrained by one normalisation relation for the total
factorisation scale variation probabilities,
\begin{equation}
\sum_{\xi\in\Xi}P(k_f=\xi)=1\label{eq:totalnorm} \, ,
\end{equation}
and three relations for the conditional probabilities,
one for each possible value of \(\xi_f\),
\begin{equation}
\sum_{\xi\in\Xi}P(k_r=\xi|k_f=\xi_{f})=1\label{eq:condnorm} \qquad  \forall \, \xi_f \in \Xi\, .
\end{equation}
We have four additional constraints from the symmetry of upper and
lower variations (Eq.~\eqref{eq:flipsymm}),
\begin{align*}
\label{eq:explicitflipsymm}
P\left(k_f=\frac{1}{2}\right)= &P(k_f=2)\\
P\left(k_r=\frac{1}{2}|k_f=1\right)= & P(k_r=2|k_f=1)\\
P\left(k_r=1|k_f=\frac{1}{2}\right)= & P(k_r=1|k_f=2)\\
P\left(k_r=\frac{1}{2}|k_f=2\right)= & P\left(k_r=2|k_f=\frac{1}{2}\right) \, .
\end{align*}
Finally, the symmetry between factorisation and
renormalisation scales (Eq.~\eqref{eq:dacrensymm}) adds another constrain,
\begin{equation}
P(k_r=1|k_f=2)=\frac{P(k_f=2|k_r=1)P(k_r=1)}{P(k_f=2)}=\frac{P(k_r=2|k_f=1)P(k_f=1)}{P(k_f=2)} \ , \label{eq:bayescons}
\end{equation}
where in the first equality we have used the law of conditional probabilities (Bayes' theorem).

It follows that with our assumptions, the sampling model is completely
specified with three free parameters, which we call $\alpha$, $\beta$ and $\gamma$. We can choose them to be
probability ratios; we define \(\alpha\) as the probability of sampling a
central factorisation scale, over the probability to sample a non-central
scale,
\begin{equation}
\label{eq:adef}
\alpha \equiv \frac{P(k_f=1)}{P(k_f=2)} = \frac{P(k_f=1)}{P(k_f=\frac{1}{2})} \, ,
\end{equation}
\(\beta\) to be the ratio of probabilities for central and
non-central variations given that a central factorisation scale is
chosen,
\begin{equation}
\label{eq:bdef}
\beta \equiv \frac{P(k_r=1|k_f=1)}{P(k_r=2|k_f=1)} = \frac{P(k_r=1|k_f=1)}{P(k_r=\frac{1}{2}|k_f=1)} \, ,
\end{equation}
and \(\gamma\) as the probability of sampling the same upper or
lower renormalisation scale as the factorisation scale, over the probability of
sampling the opposite renormalisation scale given the same factorisation scale
as in the numerator,
\begin{equation}\gamma\equiv\frac{P(k_r=2|k_f=2)}{P(k_r=\frac{1}{2}|k_f=2)}=\frac{P(k_r=\frac{1}{2}|k_f=\frac{1}{2})}{P(k_r=2|k_f=\frac{1}{2})}\label{eq:cdef} \, .
\end{equation}

In terms of the parameters, the probabilities are
\begin{align} \label{eq:sampling-model}
P(k_f=1) & =\frac{\alpha}{2+\alpha} \nonumber \\
P\left(k_f=\frac{1}{2}\right)=P(k_f=2) & =\frac{1}{2+\alpha} \nonumber \\
P(k_r=1|k_f=1) & =\frac{\beta}{2+\beta} \nonumber \\
P\left(k_r=\frac{1}{2}|k_f=1\right)=P(k_r=2|k_f=1) & =\frac{1}{2+\beta}\\
P\left(k_r=1|k_f=\frac{1}{2}\right)=P(k_r=1|k_f=2) & =\frac{\alpha}{2+\beta} \nonumber \\
P\left(k_r=\frac{1}{2}|k_f=\frac{1}{2}\right)=P(k_r=2|k_f=2) & =\frac{c(\beta+2-\alpha)}{(2+\beta)(1+\gamma)}\nonumber \\
P\left(k_r=\frac{1}{2}|k_f=2\right)=P\left(k_r=2|k_f=\frac{1}{2}\right) & =\frac{(\beta+2-\alpha)}{(2+\beta)(1+\gamma)} \nonumber
\end{align}
Note that these parameters have a clear physical interpretation. For
example, a model where factorisation and renormalisation variations are
completely independent corresponds to \(\alpha=\beta\) and \(\gamma=1\), a model
where they are completely correlated corresponds to \(\beta\to \infty\) and
\(\gamma \to \infty\), while $\alpha=\beta=1, \gamma\to\infty$ corresponds to the prescription
where the extreme asymmetric variations are discarded while the rest have the
same probability (sometimes referred to as the 7-point prescription).

\subsection{Data-driven determination of the $\alpha\beta\gamma$ coefficients}

Thus far our analysis of the replica distributions over scale choices has been
largely qualitative. To make it more quantitative we now define some effective
values of $\alpha$, $\beta$ and $\gamma$, which can be computed from the aforementioned
distributions to give an indication of the particular $\alpha\beta\gamma$ model that the
resulting PDF set corresponds to. Here we define effective model parameters by
modifying Eqs~\eqref{eq:adef}, \eqref{eq:bdef} and~\eqref{eq:cdef}, where the
modifications take into account the fact that the symmetries of the model are
broken after postfit vetoes are imposed.

Beginning with $\alpha$, we see that in Eq.~\eqref{eq:adef} it is defined in two ways,
which are of course equal due to the model assumptions of the $\alpha\beta\gamma$ model, where
$P(k_f=\frac{1}{2}) = P(k_f=2)$. However, after the postfit vetoes have been
applied this equality is unlikely to hold. This breaking of the model symmetries
can be seen for example in Fig.~\ref{fig:data-driven-dists-comp}.
Therefore, if we evaluate $\frac{P(k_f=1)}{P(k_f=\frac{1}{2})}$ in the hope of
defining an effective value for $a$ and then compare it to
$\frac{P(k_f=1)}{P(k_f=2)}$, we will get different answers. In order to give
a rough estimate of the effective value of $\alpha$, we therefore symmetrise the
distribution over $k_f$ and define
\begin{equation}
    \alpha_{\text{eff}} = \frac{P(k_f=1)}{(P(k_f=\frac{1}{2})+P(k_f=2))/2} \, .
\end{equation}
Similarly,
\begin{equation}
    \beta_{\text{eff}} = \frac{1}{N_{\text{p}}} \sum_{p = 1}^{N_{\text{p}}} \frac{P(k_{r_p}=1|k_f=1)}{(P(k_{r_p}=\frac{1}{2}|k_f=1)+P(k_{r_p}=2|k_f=1))/2} \, ,
\end{equation}
where we average over processes because the summand will in general be different
for each process. This is because the postfit vetoes act differently on
different processes.

Finally,
\begin{align}
    \gamma_{\text{eff}} & = \frac{1}{N_{\text{p}}}
         \sum_{p = 1}^{N_{\text{p}}}
             \frac{(P(k_{r_p}=\frac{1}{2}|k_f=\frac{1}{2})+P(k_{r_p}=2|{k_f}=2))/2}
                  {(P({k_{r_p}}=\frac{1}{2}|k_f=2)+P(k_{r_p}=2|k_f=\frac{1}{2}))/2} \\
    & = \frac{1}{N_{\text{p}}} 
        \sum_{p = 1}^{N_{\text{p}}}
        \frac{P(k_{r_p}=\frac{1}{2}|k_f=\frac{1}{2})+P(k_{r_p}=2|k_f=2)}
        {P(k_{r_p}=\frac{1}{2}|k_f=2)+P(k_{r_p}=2|k_f=\frac{1}{2})}
    \, .
\end{align}

In Table~\ref{tab:eff-params} we show these effective  model parameters for the
two {\tt MCscales} PDF sets that we are studying in this section. Note that in the
case that all of the replicas do satisfy the $\alpha\beta\gamma$ model symmetries, which is
the case for the {\tt MCscales} replicas without postfit applied, the effective model
parameters coincide with the model parameters that were chosen as input, namely
$a=b=c=1$, as we would expect.

%-------------------------------------------------------------------------------
\begin{table} [!htbp]\centering
\begin{tabular}{ |c|c|c|c| }
\hline
Parameter        & MCscales without postfit & MCscales with postfit \\
\hline
$\alpha_{\text{eff}}$ & 1.0   & 3.2                    \\
$\beta_{\text{eff}}$ & 1.0   & 1.1                    \\
$\gamma_{\text{eff}}$ & 1.0   & 1.4                    \\
\hline
\end{tabular}
\caption{Estimates of the effective model parameters before and after applying
postfit vetoes, where for $\beta_{\text{eff}}$ and $\gamma_{\text{eff}}$ the values are
averaged over processes.}
\label{tab:eff-params}
\end{table}
%-------------------------------------------------------------------------------

These effective parameters for the {\tt MCscales} PDF set with postfit indicate
the $\alpha\beta\gamma$ model that best fits the experimental data. We see that a value of
$\alpha_{\text{eff}} \sim 3$ is preferred, which indicates that $P(k_f = 1) \approx 3
P(k_f = \frac{1}{2}) \approx 3 P(k_f = 2)$ and therefore the degree to which the
central factorisation scale fits the data better than the non-central
factorisation scales. Next we see that $\beta_{\text{eff}} \sim 1$, which tells us
that given a replica with $k_f = 1$, the fit quality is roughly independent of
the processes renormalisation scale, since
$P(k_r = 1 | k_f = 1) \approx P(k_r = 1 | k_f = \frac{1}{2}) \approx P(k_r = 1 | k_f = 2)$
is implied. Finally, we see that $\gamma_{\text{eff}} = 1.4$, which tells us that for
the non-central factorisation scales, there is a mild asymmetry in the
renormalisation scales, with a slight preference for the same renormalisation
scale multipliers as the factorisation scale multiplier.

The values of $\beta_{\text{eff}}$ and $c_{\text{eff}}$ discussed above have been
averaged over processes. We can also compute these effective parameters per
process, which we show in Table~\ref{tab:eff-params-by-proc}. For
$\beta_{\text{eff}}$ we see a small variance of the values for the {\tt MCscales} set with
postfit, with the numbers exhibiting small fluctuations around the mean
value of 1.1. For $\gamma_{\text{eff}}$ the fluctuations are more pronounced, with
values of around one being preferred for the DY, jet and top data, and
significantly higher values being preferred for the DIS data. This suggests that
the DY, jet and top data are more stable upon variations of the renormalisation
scale from a given non-central factorisation scale, and indeed the goodness of
fit is almost independent of these variations. This is in fact true for the jet
and top data, but misleading for the DY data, where we see from
Fig.~\ref{fig:data-driven-dists-comp} that the distribution is in fact highly
skewed, which is something not captured by $\gamma_{\text{eff}}$. The fact that the
DIS data correspond to higher values of $\gamma_{\text{eff}}$, particularly for the
DIS NC data, matches the observations drawn from
Fig.~\ref{fig:data-driven-dists-comp}. This is true especially for the DIS NC
data, for which we recall that the $(k_f, k_r) = (2, \frac{1}{2})$ bin is
entirely unfilled after postfit has been applied.

%-------------------------------------------------------------------------------
\begin{table}[!htbp] \centering
\begin{tabular}{ |c|c|c|c|c| }
\hline
Parameter        & Process & MCscales without postfit & MCscales with postfit \\
\hline
$\beta_{\text{eff}}$ & DIS CC  & 1.0   & 1.2                    \\
                 & DIS NC  & 0.9   & 1.0                    \\
                 & DY      & 1.0   & 1.3                    \\
                 & Jet     & 0.9   & 0.9                    \\
                 & Top     & 1.0   & 1.2                    \\
\hline
$\gamma_{\text{eff}}$ & DIS CC  & 1.0   & 1.6                    \\
                 & DIS NC  & 1.1   & 2.3                    \\
                 & DY      & 1.0   & 1.1                    \\
                 & Jet     & 0.9   & 1.0                    \\
                 & Top     & 1.0   & 0.9                    \\
\hline
\end{tabular}
\caption{Estimates of the effective values of $\beta$ and $\gamma$ per process. Note
that $\alpha$ is not shown because it does not differ between processes. Note also
  that the results for the MCscales set without postfit should each be equal to one
in the limit of infinite statistics, and that the reason they are not is due
to insufficient PDF replicas.}
\label{tab:eff-params-by-proc}
\end{table}
%-------------------------------------------------------------------------------

The variability of these parameters across processes suggests that using a
model that treats these five processes on the same footing is perhaps ill
advised. For example, it is perhaps not preferable to include results for
$(k_f, k_r) = (2, \frac{1}{2})$ for DIS NC on the same footing as all other
combinations, when this particular combination leads to such poor agreement
with the data. Indeed, the $\alpha\beta\gamma$ model withut any post selection does exactly
this, as do the current implementations of the theoretical covariance matrix.
However, importantly, after postfit has been applied, this is no longer the
case for the $\alpha\beta\gamma$ model. We consider this to be an advantage of the {\tt
MCscales} PDF set with postfit applied over other PDF sets with scale
variations taken into account. As can be seen from
Fig.~\ref{fig:data-driven-dists-comp}, despite the fact that the processes
are initially treated equally, they very much are not after postfit has been
applied.

\section*{Acknowledgements}

We thank Ben Allanach and the Cambridge Pheno Working Group for useful
discussions that inspired the original idea. We thank Luigi Del
Debbio, Stefano Forte and Richard D.~Ball for their precious comments
and the NNPDF collaboration for their invaluable input.
M.~U. and Z.~K. are supported by the European Research Council under the
European Union’s Horizon 2020 research and innovation Programme (grant agreement n.950246).
M.~U., Z.~K. and C.~V. are partially supported by STFC consolidated grants ST/P000681/1, ST/T000694/1.
M.~U. is also supported by the Royal Society grant RGF/EA/180148.
The work of M.~U. is also funded by the Royal Society grant DH150088.
C.V. is thankful to the Science and Technology Facilities Council for their support through the
grant ST/R504671/1.

\renewcommand{\em}{}
\bibliographystyle{UTPstyle}
\bibliography{references}

\providecommand{\href}[2]{#2}\begingroup\raggedright\begin{thebibliography}{10}

\bibitem{Ball:2017nwa}
{\bf NNPDF} Collaboration, R.~D. Ball et~al., {\it {Parton distributions from
  high-precision collider data}},  {\em Eur. Phys. J.} {\bf C77} (2017), no.~10
  663, [\href{http://arxiv.org/abs/1706.00428}{{\tt arXiv:1706.00428}}].

\bibitem{Bailey:2020ooq}
S.~Bailey, T.~Cridge, L.~A. Harland-Lang, A.~D. Martin, and R.~S. Thorne, {\it
  {Parton distributions from LHC, HERA, Tevatron and fixed target data: MSHT20
  PDFs}},  \href{http://arxiv.org/abs/2012.04684}{{\tt arXiv:2012.04684}}.

\bibitem{Hou:2019efy}
T.-J. Hou et~al., {\it {New CTEQ global analysis of quantum chromodynamics with
  high-precision data from the LHC}},  {\em Phys. Rev. D} {\bf 103} (2021),
  no.~1 014013, [\href{http://arxiv.org/abs/1912.10053}{{\tt
  arXiv:1912.10053}}].

\bibitem{nnpdf40}
{\bf NNPDF} Collaboration, R.~D. Ball et~al., {\it {The path to proton
  structure at 1\% accuracy}},  {\em Eur. Phys. J. C} {\bf 82} (2022), no.~5
  428, [\href{http://arxiv.org/abs/2109.02653}{{\tt arXiv:2109.02653}}].

\bibitem{Ball:2022hsh}
R.~D. Ball et~al., {\it {The PDF4LHC21 combination of global PDF fits for the
  LHC Run III}},  \href{http://arxiv.org/abs/2203.05506}{{\tt
  arXiv:2203.05506}}.

\bibitem{Amoroso:2022eow}
S.~Amoroso et~al., {\it {Snowmass 2021 whitepaper: Proton structure at the
  precision frontier}},  \href{http://arxiv.org/abs/2203.13923}{{\tt
  arXiv:2203.13923}}.

\bibitem{Khalek:2018mdn}
R.~Abdul~Khalek, S.~Bailey, J.~Gao, L.~Harland-Lang, and J.~Rojo, {\it {Towards
  Ultimate Parton Distributions at the High-Luminosity LHC}},  {\em Eur. Phys.
  J.} {\bf C78} (2018), no.~11 962,
  [\href{http://arxiv.org/abs/1810.03639}{{\tt arXiv:1810.03639}}].

\bibitem{Ball:2014uwa}
{\bf NNPDF} Collaboration, R.~D. Ball et~al., {\it {Parton distributions for
  the LHC Run II}},  {\em JHEP} {\bf 04} (2015) 040,
  [\href{http://arxiv.org/abs/1410.8849}{{\tt arXiv:1410.8849}}].

\bibitem{DelDebbio:2021whr}
L.~Del~Debbio, T.~Giani, and M.~Wilson, {\it {Bayesian approach to inverse
  problems: an application to NNPDF closure testing}},  {\em Eur. Phys. J. C}
  {\bf 82} (2022), no.~4 330, [\href{http://arxiv.org/abs/2111.05787}{{\tt
  arXiv:2111.05787}}].

\bibitem{Cruz-Martinez:2021rgy}
J.~Cruz-Martinez, S.~Forte, and E.~R. Nocera, {\it {Future tests of parton
  distributions}},  {\em Acta Phys. Polon. B} {\bf 52} (2021) 243,
  [\href{http://arxiv.org/abs/2103.08606}{{\tt arXiv:2103.08606}}].

\bibitem{Carli2010}
T.~Carli, D.~Clements, A.~Cooper-Sarkar, C.~Gwenlan, G.~P. Salam, F.~Siegert,
  P.~Starovoitov, and M.~Sutton, {\it A posteriori inclusion of parton density
  functions in nlo qcd final-state calculations at hadron colliders:
  theÂ applgridÂ project},  {\em The European Physical Journal C} {\bf 66}
  (Apr, 2010) 503--524.

\bibitem{Bertone:2016lga}
V.~Bertone, S.~Carrazza, and N.~P. Hartland, {\it {APFELgrid: a high
  performance tool for parton density determinations}},  {\em Comput. Phys.
  Commun.} {\bf 212} (2017) 205--209,
  [\href{http://arxiv.org/abs/1605.02070}{{\tt arXiv:1605.02070}}].

\bibitem{Kluge:2006xs}
T.~Kluge, K.~Rabbertz, and M.~Wobisch, {\it {FastNLO: Fast pQCD calculations
  for PDF fits}},  in {\em {14th International Workshop on Deep Inelastic
  Scattering}}, 9, 2006.
\newblock \href{http://arxiv.org/abs/hep-ph/0609285}{{\tt hep-ph/0609285}}.

\bibitem{Frederix:2018nkq}
R.~Frederix, S.~Frixione, V.~Hirschi, D.~Pagani, H.~S. Shao, and M.~Zaro, {\it
  {The automation of next-to-leading order electroweak calculations}},  {\em
  JHEP} {\bf 07} (2018) 185, [\href{http://arxiv.org/abs/1804.10017}{{\tt
  arXiv:1804.10017}}]. [Erratum: JHEP 11, 085 (2021)].

\bibitem{Carrazza:2020gss}
S.~Carrazza, E.~R. Nocera, C.~Schwan, and M.~Zaro, {\it {PineAPPL: combining EW
  and QCD corrections for fast evaluation of LHC processes}},  {\em JHEP} {\bf
  12} (2020) 108, [\href{http://arxiv.org/abs/2008.12789}{{\tt
  arXiv:2008.12789}}].

\bibitem{McGowan:2022nag}
J.~McGowan, T.~Cridge, L.~A. Harland-Lang, and R.~S. Thorne, {\it {Approximate
  N$^{3}$LO Parton Distribution Functions with Theoretical Uncertainties:
  MSHT20aN$^3$LO PDFs}},  \href{http://arxiv.org/abs/2207.04739}{{\tt
  arXiv:2207.04739}}.

\bibitem{Cacciari:2011ze}
M.~Cacciari and N.~Houdeau, {\it {Meaningful characterisation of perturbative
  theoretical uncertainties}},  {\em JHEP} {\bf 09} (2011) 039,
  [\href{http://arxiv.org/abs/1105.5152}{{\tt arXiv:1105.5152}}].

\bibitem{David:2013gaa}
A.~David and G.~Passarino, {\it {How well can we guess theoretical
  uncertainties?}},  {\em Phys. Lett. B} {\bf 726} (2013) 266--272,
  [\href{http://arxiv.org/abs/1307.1843}{{\tt arXiv:1307.1843}}].

\bibitem{Bonvini:2020xeo}
M.~Bonvini, {\it {Probabilistic definition of the perturbative theoretical
  uncertainty from missing higher orders}},  {\em Eur. Phys. J. C} {\bf 80}
  (2020), no.~10 989, [\href{http://arxiv.org/abs/2006.16293}{{\tt
  arXiv:2006.16293}}].

\bibitem{Duhr:2021mfd}
C.~Duhr, A.~Huss, A.~Mazeliauskas, and R.~Szafron, {\it {An analysis of
  Bayesian estimates for missing higher orders in perturbative calculations}},
  \href{http://arxiv.org/abs/2106.04585}{{\tt arXiv:2106.04585}}.

\bibitem{AbdulKhalek:2019bux}
{\bf NNPDF} Collaboration, R.~Abdul~Khalek et~al., {\it {A first determination
  of parton distributions with theoretical uncertainties}},  {\em Eur. Phys.
  J.} {\bf C} (2019) 79:838, [\href{http://arxiv.org/abs/1905.04311}{{\tt
  arXiv:1905.04311}}].

\bibitem{AbdulKhalek:2019ihb}
{\bf NNPDF} Collaboration, R.~Abdul~Khalek et~al., {\it {Parton Distributions
  with Theory Uncertainties: General Formalism and First Phenomenological
  Studies}},  {\em Eur. Phys. J. C} {\bf 79} (2019), no.~11 931,
  [\href{http://arxiv.org/abs/1906.10698}{{\tt arXiv:1906.10698}}].

\bibitem{Ball:2020xqw}
R.~D. Ball, E.~R. Nocera, and R.~L. Pearson, {\it {Deuteron Uncertainties in
  the Determination of Proton PDFs}},  {\em Eur. Phys. J. C} {\bf 81} (2021),
  no.~1 37, [\href{http://arxiv.org/abs/2011.00009}{{\tt arXiv:2011.00009}}].

\bibitem{Pearson:2019upi}
R.~Pearson, R.~Ball, and E.~R. Nocera, {\it {Uncertainties due to Nuclear Data
  in Proton PDF Fits}},  {\em PoS} {\bf DIS2019} (2019) 027.

\bibitem{Harland-Lang:2018bxd}
L.~A. Harland-Lang and R.~S. Thorne, {\it {On the Consistent Use of Scale
  Variations in PDF Fits and Predictions}},  {\em Eur. Phys. J.} {\bf C79}
  (2019), no.~3 225, [\href{http://arxiv.org/abs/1811.08434}{{\tt
  arXiv:1811.08434}}].

\bibitem{Ball:2021icz}
R.~D. Ball and R.~L. Pearson, {\it {Decorrelation of Theoretical Uncertainties
  in PDF Fits and Theoretical Uncertainties in Predictions}},
  \href{http://arxiv.org/abs/2105.05114}{{\tt arXiv:2105.05114}}.

\bibitem{Buckley:2014ana}
A.~Buckley, J.~Ferrando, S.~Lloyd, K.~Nordstr\"om, B.~Page, M.~R\"ufenacht,
  M.~Sch\"onherr, and G.~Watt, {\it {LHAPDF6: parton density access in the LHC
  precision era}},  {\em Eur. Phys. J. C} {\bf 75} (2015) 132,
  [\href{http://arxiv.org/abs/1412.7420}{{\tt arXiv:1412.7420}}].

\bibitem{Iranipour:2022iak}
S.~Iranipour and M.~Ubiali, {\it {A new generation of simultaneous fits to LHC
  data using deep learning}},  {\em JHEP} {\bf 05} (2022) 032,
  [\href{http://arxiv.org/abs/2201.07240}{{\tt arXiv:2201.07240}}].

\bibitem{McCullough:2022hzr}
M.~McCullough, J.~Moore, and M.~Ubiali, {\it {The dark side of the proton}},
  \href{http://arxiv.org/abs/2203.12628}{{\tt arXiv:2203.12628}}.

\bibitem{Greljo:2021kvv}
A.~Greljo, S.~Iranipour, Z.~Kassabov, M.~Madigan, J.~Moore, J.~Rojo, M.~Ubiali,
  and C.~Voisey, {\it {Parton distributions in the SMEFT from high-energy
  Drell-Yan tails}},  {\em JHEP} {\bf 07} (2021) 122,
  [\href{http://arxiv.org/abs/2104.02723}{{\tt arXiv:2104.02723}}].

\bibitem{Giele:2001mr}
W.~T. Giele, S.~A. Keller, and D.~A. Kosower, {\it {Parton Distribution
  Function Uncertainties}},  \href{http://arxiv.org/abs/hep-ph/0104052}{{\tt
  hep-ph/0104052}}.

\bibitem{Ball:2008by}
{\bf NNPDF} Collaboration, R.~D. Ball, L.~Del~Debbio, S.~Forte, A.~Guffanti,
  J.~I. Latorre, A.~Piccione, J.~Rojo, and M.~Ubiali, {\it {A Determination of
  parton distributions with faithful uncertainty estimation}},  {\em Nucl.
  Phys. B} {\bf 809} (2009) 1--63, [\href{http://arxiv.org/abs/0808.1231}{{\tt
  arXiv:0808.1231}}]. [Erratum: Nucl.Phys.B 816, 293 (2009)].

\bibitem{Ball:2010de}
R.~D. Ball, L.~Del~Debbio, S.~Forte, A.~Guffanti, J.~I. Latorre, J.~Rojo, and
  M.~Ubiali, {\it {A first unbiased global NLO determination of parton
  distributions and their uncertainties}},  {\em Nucl. Phys. B} {\bf 838}
  (2010) 136--206, [\href{http://arxiv.org/abs/1002.4407}{{\tt
  arXiv:1002.4407}}].

\bibitem{Ball:2012cx}
R.~D. Ball et~al., {\it {Parton distributions with LHC data}},  {\em Nucl.
  Phys. B} {\bf 867} (2013) 244--289,
  [\href{http://arxiv.org/abs/1207.1303}{{\tt arXiv:1207.1303}}].

\bibitem{DelDebbio:2009zz}
{\bf NNPDF} Collaboration, L.~Del~Debbio, R.~D. Ball, S.~Forte, A.~Guffanti,
  J.~I. Latorre, A.~Piccione, J.~Rojo, and M.~Ubiali, {\it {Monte Carlo methods
  for robust error estimates for PDFs}},  in {\em {44th Rencontres de Moriond
  on QCD and High Energy Interactions}}, 3, 2009.

\bibitem{Ball:2011eq}
{\bf NNPDF} Collaboration, R.~D. Ball, V.~Bertone, F.~Cerutti, L.~Del~Debbio,
  S.~Forte, A.~Guffanti, J.~I. Latorre, J.~Rojo, and M.~Ubiali, {\it {Parton
  Distributions: Determining Probabilities in a Space of Functions}},  in {\em
  {PHYSTAT 2011}}, 10, 2011.
\newblock \href{http://arxiv.org/abs/1110.1863}{{\tt arXiv:1110.1863}}.

\bibitem{deFlorian:2016spz}
{\bf LHC Higgs Cross Section Working Group} Collaboration, D.~de~Florian
  et~al., {\it {Handbook of LHC Higgs Cross Sections: 4. Deciphering the Nature
  of the Higgs Sector}},  \href{http://arxiv.org/abs/1610.07922}{{\tt
  arXiv:1610.07922}}.

\bibitem{Ball:2009qv}
{\bf NNPDF} Collaboration, R.~D. Ball, L.~Del~Debbio, S.~Forte, A.~Guffanti,
  J.~I. Latorre, J.~Rojo, and M.~Ubiali, {\it {Fitting Parton Distribution Data
  with Multiplicative Normalization Uncertainties}},  {\em JHEP} {\bf 05}
  (2010) 075, [\href{http://arxiv.org/abs/0912.2276}{{\tt arXiv:0912.2276}}].

\bibitem{Ball:2012wy}
R.~D. Ball et~al., {\it {Parton Distribution Benchmarking with LHC Data}},
  {\em JHEP} {\bf 04} (2013) 125, [\href{http://arxiv.org/abs/1211.5142}{{\tt
  arXiv:1211.5142}}].

\bibitem{Abramowicz:1900rp}
{\bf H1, ZEUS} Collaboration, H.~Abramowicz et~al., {\it {Combination and QCD
  Analysis of Charm Production Cross Section Measurements in Deep-Inelastic ep
  Scattering at HERA}},  {\em Eur. Phys. J.} {\bf C73} (2013), no.~2 2311,
  [\href{http://arxiv.org/abs/1211.1182}{{\tt arXiv:1211.1182}}].

\bibitem{H1:2018flt}
{\bf H1, ZEUS} Collaboration, H.~Abramowicz et~al., {\it {Combination and QCD
  analysis of charm and beauty production cross-section measurements in deep
  inelastic $ep$ scattering at HERA}},  {\em Eur. Phys. J. C} {\bf 78} (2018),
  no.~6 473, [\href{http://arxiv.org/abs/1804.01019}{{\tt arXiv:1804.01019}}].

\bibitem{Forte:2020pyp}
S.~Forte and Z.~Kassabov, {\it {Why $\alpha _s$ cannot be determined from
  hadronic processes without simultaneously determining the parton
  distributions}},  {\em Eur. Phys. J. C} {\bf 80} (2020), no.~3 182,
  [\href{http://arxiv.org/abs/2001.04986}{{\tt arXiv:2001.04986}}].

\bibitem{Barontini:2023vmr}
A.~Barontini, A.~Candido, J.~M. Cruz-Martinez, F.~Hekhorn, and C.~Schwan, {\it
  {Pineline: Industrialization of High-Energy Theory Predictions}},
  \href{http://arxiv.org/abs/2302.12124}{{\tt arXiv:2302.12124}}.

\bibitem{Anastasiou:2015vya}
C.~Anastasiou, C.~Duhr, F.~Dulat, F.~Herzog, and B.~Mistlberger, {\it {Higgs
  Boson Gluon-Fusion Production in QCD at Three Loops}},  {\em Phys. Rev.
  Lett.} {\bf 114} (2015) 212001, [\href{http://arxiv.org/abs/1503.06056}{{\tt
  arXiv:1503.06056}}].

\bibitem{Anastasiou:2016cez}
C.~Anastasiou, C.~Duhr, F.~Dulat, E.~Furlan, T.~Gehrmann, F.~Herzog,
  A.~Lazopoulos, and B.~Mistlberger, {\it {High precision determination of the
  gluon fusion Higgs boson cross-section at the LHC}},  {\em JHEP} {\bf 05}
  (2016) 058, [\href{http://arxiv.org/abs/1602.00695}{{\tt arXiv:1602.00695}}].

\bibitem{Mistlberger:2018etf}
B.~Mistlberger, {\it {Higgs boson production at hadron colliders at N$^{3}$LO
  in QCD}},  {\em JHEP} {\bf 05} (2018) 028,
  [\href{http://arxiv.org/abs/1802.00833}{{\tt arXiv:1802.00833}}].

\bibitem{Dreyer:2016oyx}
F.~A. Dreyer and A.~Karlberg, {\it {Vector-Boson Fusion Higgs Production at
  Three Loops in QCD}},  {\em Phys. Rev. Lett.} {\bf 117} (2016), no.~7 072001,
  [\href{http://arxiv.org/abs/1606.00840}{{\tt arXiv:1606.00840}}].

\bibitem{Bonvini:2014jma}
M.~Bonvini, R.~D. Ball, S.~Forte, S.~Marzani, and G.~Ridolfi, {\it {Updated
  Higgs cross section at approximate N$^3$LO}},  {\em J. Phys. G} {\bf 41}
  (2014) 095002, [\href{http://arxiv.org/abs/1404.3204}{{\tt
  arXiv:1404.3204}}].

\bibitem{Bonvini:2016frm}
M.~Bonvini, S.~Marzani, C.~Muselli, and L.~Rottoli, {\it {On the Higgs cross
  section at N$^{3}$LO+N$^{3}$LL and its uncertainty}},  {\em JHEP} {\bf 08}
  (2016) 105, [\href{http://arxiv.org/abs/1603.08000}{{\tt arXiv:1603.08000}}].

\bibitem{Cacciari:2015jma}
M.~Cacciari, F.~A. Dreyer, A.~Karlberg, G.~P. Salam, and G.~Zanderighi, {\it
  {Fully Differential Vector-Boson-Fusion Higgs Production at
  Next-to-Next-to-Leading Order}},  {\em Phys. Rev. Lett.} {\bf 115} (2015),
  no.~8 082002, [\href{http://arxiv.org/abs/1506.02660}{{\tt
  arXiv:1506.02660}}]. [Erratum: Phys.Rev.Lett. 120, 139901 (2018)].

\bibitem{Czakon:2011xx}
M.~Czakon and A.~Mitov, {\it {Top++: A Program for the Calculation of the
  Top-Pair Cross-Section at Hadron Colliders}},  {\em Comput. Phys. Commun.}
  {\bf 185} (2014) 2930, [\href{http://arxiv.org/abs/1112.5675}{{\tt
  arXiv:1112.5675}}].

\bibitem{Li:2012wna}
Y.~Li and F.~Petriello, {\it {Combining QCD and electroweak corrections to
  dilepton production in FEWZ}},  {\em Phys. Rev. D} {\bf 86} (2012) 094034,
  [\href{http://arxiv.org/abs/1208.5967}{{\tt arXiv:1208.5967}}].

\bibitem{Gavin:2012sy}
R.~Gavin, Y.~Li, F.~Petriello, and S.~Quackenbush, {\it {W Physics at the LHC
  with FEWZ 2.1}},  {\em Comput. Phys. Commun.} {\bf 184} (2013) 208--214,
  [\href{http://arxiv.org/abs/1201.5896}{{\tt arXiv:1201.5896}}].

\bibitem{Gavin:2010az}
R.~Gavin, Y.~Li, F.~Petriello, and S.~Quackenbush, {\it {FEWZ 2.0: A code for
  hadronic Z production at next-to-next-to-leading order}},  {\em Comput. Phys.
  Commun.} {\bf 182} (2011) 2388--2403,
  [\href{http://arxiv.org/abs/1011.3540}{{\tt arXiv:1011.3540}}].

\end{thebibliography}\endgroup

\end{document}